\magnification1200

\def\lag{{\cal G}}
\def\lah{{\cal H}}
\def\lax{{\cal X}}

\input epsf

\rightline{DAMTP 2003-145}
\rightline{KCL-MTH-03-18}
\rightline{hep-th/0312247}

\vskip .5cm
\centerline
{\bf Representations of $\lag^{+++}$ and the role of space-time }
\vskip 1cm
\centerline{Axel Kleinschmidt}
\centerline{Department of Applied Mathematics and Theoretical Physics}
\centerline{Centre for Mathematical Sciences}
\centerline{Wilberforce Road, Cambridge CB3 0WA, UK}
\vskip .5cm
\centerline{Peter West}
\centerline{Department of Mathematics}
\centerline{King's College, London WC2R 2LS, UK}
\vskip 2cm

\leftline{\sl Abstract}
\vskip .2cm
\noindent 
We consider the decomposition of the adjoint and fundamental
representations of very extended Kac-Moody algebras
$\lag^{+++}$ with respect to their regular $A$ type subalgebra which,
in the corresponding non-linear realisation, is associated with gravity. We
find that for many very extended algebras 
almost all the  $A$ type representations
that occur in  the decomposition of the fundamental representations 
also occur in 
 the adjoint
representation of $\lag^{+++}$. In particular,  for
$E_8^{+++}$,  this applies to all its fundamental representations. 
However, there are some important
examples, such as $A_{N-3}^{+++}$,  where this is not true and indeed the
adjoint representation contains no generator that can be identified with a
space-time translation. We comment on the significance of these results 
for how space-time can occur in the non-linear realisation based on
$\lag^{+++}$. Finally we show that there is a correspondence between the  $A$
representations that occur in the fundamental representation associated
with the very extended node  and the adjoint representation of $\lag^{+++}$
which is consistent with the interpretation of the former as charges
associated with brane solutions. 
 
\vskip .5cm

\vfill

\vskip 1cm
email:  a.kleinschmidt@damtp.cam.ac.uk, pwest@mth.kcl.ac.uk

\eject

\medskip
{\bf{1. Introduction }}
\medskip
It is a consequence of supersymmetry that the scalars in supergravity
multiplets belong to  non-linear realisations [1].  One of the most
celebrated examples concerns  the four dimensional maximal supergravity
where the scalars belong to a non-linear realisation of  $E_7$ [2]. The
eleven dimensional supergravity theory does not possess any scalars and it
was widely believed that these symmetry algebras were not present in this
theory.  However, it was found that the eleven
dimensional supergravity theory could be formulated as a non-linear
realisation [3]. The infinite dimensional algebra involved in this
construction was  the closure of a finite dimensional algebra, denoted
$G_{11}$, with the eleven dimensional conformal algebra. The non-linear
realisation was carried out by ensuring that the equations of motion were
invariant under both finite dimensional algebras, taking into
account that some of their generators were in common.  
The algebra
$G_{11}$ involved the space-time translations together with an algebra
$\hat G_{11}$ 
which contained $A_{10}$ and the Borel subalgebra of $E_7$
as subalgebras. The algebra  $\hat G_{11}$  was not a Kac-Moody algebra,
however,  it was conjectured [4] that the theory could be extended so that
the algebra $\hat G_{11}$  was promoted to  a Kac-Moody algebra.  It was
shown that this symmetry would have to contain a certain rank eleven
Kac-Moody algebra denoted 
$E_{11}$ [4]. 
\par
Consequently, it was argued  [4] that an extension of eleven dimensional
supergravity should possess an $E_{11}$ symmetry that was non-linearly
realised. In particular,  the symmetries found
when  the  eleven dimensional  supergravity theory was dimensionally
reduced would  be  present in this eleven dimensional theory.
One of the advantages of a   non-linear realisation is that the dynamics 
is largely specified by the algebra if the chosen local subalgebra is
sufficiently large. This was not the case for the $G_{11}$ considered
in [3] when taken in isolation as  the
local subalgebra was choosen to be just the Lorentz algebra, but it  is
the case for 
$E_{11}$ with the local subalgebra that was specified in [4].  However,
the way the space-time generators and the conformal algebra should relate
to the
$E_{11}$ non-linear realisation  was  not considered in reference [4].
\par
A similar picture emerged for the IIA and IIB supergravity theories
in ten dimensions and it was conjectured that these theories could be
extended such that they were invariant under $E_{11}$ [4,5]. As a
result, the different type II theories arose by taking different
local subalgebras of $E_{11}$ [4,5]. 
\par
Similar  ideas  to that advocated  in [4] were adopted in [6] where
it was argued  that eleven dimensional supergravity was
invariant under a non-linear realisation of the $E_{10}$ subalgebra of
$E_{11}$. However, guided by earlier work on cosmological billiards [7], 
this work treated space-time in a way that was different to that
considered in [3], but consistent with the conjecture of [4]. The fields
in the non-linear realisation were presumed to depend only on time and any
spatial dependence was to arise in the equations of motion resulting from
the non-linear realisation. At the level at which this construction was
carried out, the fields were almost independent of spatial coordinates, 
but it was observed that the $E_{10}$ adjoint representation contained fields
with  the correct $A_9$  representations to be the spatial derivatives
of some of the fields that occcur in eleven dimensional supergravity. Put
another way, it was supposed that the spatial dependence would arise from
the structure of
$E_{10}$. However,   the space-time translation operator was not
thought to appear as one of the generators of $E_{11}$.  For the
non-linear realisation of a  Kac-Moody algebra with a 
 local subalgebra which is chosen to be  that which is invariant
under the Cartan involution there   exists an invariant 
Lagrangian [8]. Hence, regarding the spatial
dependence to arise  in  $E_{10}$ itself had the additional technical
advantage that one could use this   Lagrangian formulation. 
\par
An alternative  proposal to incorporate the space-time generators into
the context of a non-linear realisation of 
$E_{11}$ was made in reference [9]. The space-time translations naturally
occur as the lowest level object in   the fundamental representation
associated with the very extended node, denoted $l_1$, of
$E_{11}$.    It was proposed to take the non-linear
realisation of the algebra  formed from the $E_{11}$
semi-direct product of $E_{11}$ with generators corresponding to the 
$l_1$ representation. This had the advantage that at low levels this
construction was that carried out in reference [3] and so one did
obtain a theory that described correctly eleven dimensional supergravity
in terms of fields that had a full space-time dependence.  
 One not anticipated result in this approach was
that the 
$l_1$  representation contains at its first few levels 
the space-time translations as well as the central charges
found in the supersymmetry algebra together with an infinite number of
other states. 
\par
Very recently an alternative approach was advocated [10];  as in [4]
the theory was assumed to be a non-linear realisation of $E_{11}$,
but the fields were assumed  to depend on an auxiliary parameter and, in
the spirit of [6] all the space-time dependence was assumed   to
arise from the equations of motion of the non-linear realisation in some
way.  It was also noted, by examing the tables of [11],  that  at level
seven  a generator occurred with the correct 
$A_{10}$ representation to be identified as a space-time generator
and it was supposed that this might provide a mechanism for
introducing space-time into the fields. 
\par
A similar scenario to that advocated to eleven dimensional supergravity
in [4] was supposed to occur for gravity [12] in D dimensions and the
effective action of the closed bosonic string [4] generalised to  D
dimensions. It was realised that the algebras that arose in all these
theories were of a special kind and were called very extended Kac-Moody
algebras [13]. Indeed,  for any finite  dimensional semi-simple Lie algebra
$\lag$ one can systematically extend its Dynkin diagram by adding three more
nodes to obtain an indefinite 
Kac-Moody algebra denoted $\lag^{+++}$. The algebras for 
gravity and the closed bosonic string being $D^{+++}_{D-2}$ [12] and
$A^{+++}_{D-3}$ [5] respectively. This general approach was extended in
reference [14] where the non-linear realisation of $\lag^{+++}$ restricted
to its Cartan subalgebra was constructed and the Weyl transformations
resulting were shown to transform the Kasner solutions into each 
other and, for
$E^{+++}_8=E_{11}$ and $D_{24}^{+++}$, were shown to be related to U duality
transformations in the corresponding string theories. In  [15]  it was
shown that the theories associated with $\lag^{+++}$ admit BPS
intersecting solutions. Finally, it was shown [16] that the low level
content of the adjoint representation of $\lag^{+++}$ predicted a field
content for a non-linear realisation of $\lag^{+++}$ consistent with
oxidation theory. As a  result of [4,12,13] and [14,15,16,10], 
 one was lead to suppose, as was
explicitly stated in the latter papers, that  
the non-linear realisation of any  very extended algebra $\lag^{+++}$
leads to a theory, called ${\cal V}_\lag$ in [16],
that at low levels includes gravity and the other fields
found in the corresponding oxidised theory and it was  hoped that this
non-linear realisation contains an infinite number of propagating fields
that ensures the consistency of these theories. 
\par
As the above explains, there is growing evidence for the conjecture
set out in [4], namely that an extension of eleven dimensional supergravity
possessing an $E_{11}$ (or a  subalgebra of this) symmetry which is
non-linearly realised. However,  there are different proposals as to how
space-time should arise in this construction. To help resolve these
differences in this paper we investigate the content of the fundamental
and adjoint representations of 
$\lag^{+++}$. In particular we will, following [17,16], analyse
$\lag^{+++}$ and its representations
in terms
of the most physically relevant $A$ subalgebra which in the
non-linear realisation is associated with gravity and  find relations
between the content of the fundamental representations and adjoint
representation of
$\lag^{+++}$. We will show how the analogues of the representation
proposed for $E_8^{+++}$ in [9] contain precisely the right fields at
lowest levels to be interpreted as central charges. In addition, we analyse
the different proposals for how spacetime enters in the light of our
findings. 
\par
The structure of this paper is as follows. Section 2 contains the
decomposition strategy for $\lag^{+++}$ and its representations and
the results obtained by this method. In section 3 we show how the
so-called $l_1$ representation contains the correct fields in correspondence 
to the generalisations of the
central charges of supersymmetry algebras. Section 4 discusses the
consequences of the results in the earlier sections for the 
role of space-time. Appendix
A lists all the Dynkin diagrams of the very extended algebras,
appendix B tabulates a number of $\bar l_1$ representations 
of various $\lag^{+++}$
analysed to a
greater height. In appendix C we also list the $\bar l_1$ decomposition for
$E_{10}=E_8^{++}$.

\medskip
{\bf 2. Representations and analysis of indefinite Kac-Moody algebras}
\medskip
{\bf{2.1 Relationship between fundamental and adjoint representations of
$\lag^{+++}$}}
\medskip
Given a semi-simple finite dimensional Lie algebra $\lag$ we can construct
its very extension $\lag^{+++}$ [13].  We label the Dynkin
diagram of $\lag^{+++}$ by assigning to the very, over and affine nodes
the labels 1,2,3 and continue this labelling along the horizontal line of
the Dynkin diagram going from  left to right and then label the nodes
above the horizontal working from right to left.
The Dynkin diagrams with this labelling
  are given in appendix A. We are interested in
the fundamental representations of
$\lag^{+++}$. These are the standard highest weight representations 
of $\lag^{+++}$ with highest weights given by the fundamental weights 
of $\lag^{+++}$. We will study this problem by considering the 
fundamental representations embedded in an
extended algebra denoted $\lag^{+++a}$ whose Dynkin diagram is 
obtained from that of $\lag^{+++}$ by adding one more node attached 
by a single line to the node
labelled $a$ of $\lag^{+++}$. We label this additional node by
$0$. The representations of Kac-Moody algebras could also be
calculated from the Weyl-Kac character formula or the Freudenthal
formula [18] but using this auxiliary extended diagram is more suited to
our analysis of the relations between the algebra and its
representations. 
In order to reveal parts of the structure of the infinite dimensional 
$\lag^{+++}$ we consider a decomposition of $\lag^{+++}$ with
respect 
to a finite dimensional subalgebra as in reference [13]. There is 
(at least) one node $c$ of the Dynkin diagram of $\lag^{+++}$ whose 
deletion leaves the diagram of a finite dimensional semi-simple Lie 
algebra. We denote this subalgebra by $\lah$. 
For example, for the case of 
$E_8^{+++}$,
this can be the node labelled $11$ and the  algebra remaining after its
deletion is $A_{10}$.
\par
The roots
of $\lag^{+++a}$ can be written in terms of simple roots in the form
$$\beta = n_0\alpha_0+n_c\alpha_c+\sum_i n_i\alpha_i
\eqno(2.1)$$
Our notation here is that Latin indices from the middle of the
alphabet 
run over the subalgebra $\lah$, whereas letters from the beginning run 
over $\lag^{+++}$.
The concept of level was
implicit in reference [13], given explicitly for
$E_{10}$ in reference [6] and worked out in general in [17,16]. In our case,
each root $\beta$ has two levels, $(n_c,n_0)$, which are the numbers of
times the simple roots
$\alpha_c$ and $\alpha_0$ occur in the root expansion.
The roots of level $(\star,0)$,
where
$\star$ is any level,  are those of
$\lag^{+++}$ and amongst these the roots of level
$(0,0)$ generate the finite dimensional 
subalgebra $\lah$ in terms of which we will analyse 
the
algebra $\lag^{+++a}$. (There are also additional scalars from the 
Cartan subalgebra generators of the deleted nodes.)
As the notion of level
is preserved by the commutators, the roots of level
$(\star,1)$  form a representation of $\lag^{+++}$. It is  just the
standard lowest weight representation with lowest weight $\bar l_a$ of
$\lag^{+++}$,
which we will refer to as the fundamental $\bar l_a$ representation.
The fundamental weights of $\lag^{+++}$ are defined as usual by
$${2(\alpha_a,l_b)\over (\alpha_a,\alpha_a)}=\delta_{ab}
\eqno(2.2)$$
\par
Here the inner product is taken in the weight lattice of
$\lag^{+++}$. 
The fundamental weights give rise to highest weight representations
with highest weight $l_a$. For the lowest weight representations we
are considering we have ${\bar l_a}=-l_a$. For computational purposes
it is more convenient to consider the lowest weight representations
$\bar l_a$ and their $\lah$ representation content is identical to that
of the highest weight representation $l_a$.
We also note that $(-1,\star)$ also forms a
representation, namely just the highest weight $l_a$ representation.
We first consider decomposing the algebra $\lag^{+++a}$ by deleting  the
node $0$. Following [13] we can write the simple
roots as
$$\alpha_0=y- l_a, \ , \alpha_c,\ \alpha_i
\eqno(2.3)$$
where $\alpha_i, \alpha_c$ are the simple roots of $G^{+++}$ and $y$ is
orthogonal to these in the inner product of the extended algebra 
$\lag^{+++a}$ as can be seen by taking the inner product with any 
simple $\alpha_b$ of $\lag^{+++}$ and using the definition of the 
fundamental weights. As we will be dealing with a number of 
consecutive subalgebras of $\lag^{+++a}$, we have to fix an 
embedding of the different weight lattices in the weight lattice 
of $\lag^{+++a}$. We do this horizontally, {\sl i.e.} by filling 
in zeros in the simple root expansion of any element for those 
components which require embedding. Using this convention we do 
not need to put labels on the different inner products to distinguish
them and it should be clear from the context which space we are
working in.
From (2.3) we see that $y^2+l_a^2=2$ since $\alpha_0^2=2$.
\par
We can further decompose the algebra by deleting the node labelled
$c$. 
Here we assume for simplicity that $a$ is not attached to $c$, but 
an analogous construction works in this case, for example using the
formalism of [13,16]. Taking  into account the
possiblility of starting from non-simply laced algebras $\lag$, we may 
write the 
simple
roots of $\lag^{+++}$ as
$$\alpha_c=x-\nu, \ \alpha_i
\eqno(2.4)$$
where $\alpha_i$ are the roots of $H$ and
$$\nu=-\sum_j A^{\lag^{+++}}_{cj}\lambda_j {(\alpha_c,\alpha_c)\over
(\alpha_j,\alpha_j)}
\eqno(2.5)$$
where $\lambda_j$ are the fundamental weights of $\lah$. The
fundamental weights of $\lag^{+++}$ are given by
$$l_i=\lambda_i+{x\over x^2}(\nu,\lambda_i),\  l_c={x\over x^2}
\eqno(2.6)$$
We note that $x^2+\nu^2=\alpha_c^2$. One can show for the simply
laced case that
$\det A^{\lag^{+++}} =-2\det A^\lag=x^2 \det A^\lah$ [13]. Here and 
throughout the paper $A^\lax$ denotes the generalised Cartan matrix 
of the Kac-Moody algebra $\lax$, or without superscript it will refer
to the preferred $A$ type subalgebra.
\par
As a result of the above decompositions we can write the  roots of
$\lag^{+++a}$ given in  equation (2.1) as
$$\beta=n_0 y +x(n_c-n_0 {(\nu,\lambda_a)\over x^2}) -\Lambda
\eqno(2.7)$$
where
$$\Lambda=n_0\lambda_a+ n_c\nu-\sum_i n_i\alpha_i
\eqno(2.8)$$
We note that $\Lambda$ belongs to the weight lattice of $\lah$. As explained in
[6,17,16], 
if  the adjoint representation of $\lag^{+++a}$ contains generators in
the representation of
$\lah$ with highest weight $\sum_jp_j\lambda_j$,  $p_j$ being the Dynkin
labels, then the following condition must hold
$$\sum_jp_j\lambda_j=\Lambda=n_0\lambda_a+ n_c\nu-\sum_i n_i\alpha_i
\eqno(2.9)$$
where $p_j,n_0,n_c$ and $n_i$ are all non-negative integers. Taking the 
scalar
product with
$\lambda_k$ this condition becomes
$$\sum_j  p_j (\lambda_j,\lambda_k)=
n_0(\lambda_a,\lambda_k)+n_c^{}(\nu,\lambda_k)-n_k
\eqno(2.10)$$
In principle, there should be a factor of $2\over (\alpha_k,\alpha_k)$
with the term $n_k$ but in the cases of interest to us, the subalgebra
will be of simply laced type and so this factor is one.
We may invert this equation for a general Kac-Moody algebra $\lax$ 
in terms of the inverse Cartan matrix by using the relation
$$(A^{\lax})^{-1}_{ab}={2\over (\alpha_a,\alpha_a)}(\lambda_i,\lambda_b)
\eqno(2.11)$$
\par
The scalar product of the roots of $\lag^{+++i}$ take the form
$$\beta^2=n_0^2 y^2+x^2 (n_c-n_0 {(\nu,\lambda_a)\over x^2})^2 
+\sum_{ij} p_i
(\lambda_i,\lambda_j) p_j=2,1,0,\ldots
\eqno(2.12)$$
In the second equation we have inserted the constraint that the roots
of a Kac-Moody algebra must be have length squared which are integers 2 
or
less [18], except for $G_2^{+++}$ where  short roots have length squared 
$2/3$. For a Kac-Moody algebra with symmetric Cartan matrix we can get 
only even integers.
\par
Equations (2.10) and (2.12) are necessary conditions  for the
representation  of $\lag$
with Dynkin indices $p_j$ to belong to the adjoint representation of
$\lag^{+++a}$.  However, they are not as strong as implementing all the
Serre relations and it can happen that they possess
some solutions that  do not actually occur in the adjoint representation.
In fact, this possibility does occur, but it seems rather exceptional,
as the reader can verify by looking at the tables in appendix B and C,
see also 
comments in [11,17,16]. All
the statements made below are assumed to be modulo this exception. We 
also note that for each fixed level $(n_c,n_0)$ there are only a 
finite number of possible solutions to this equation. (In the tables we
have also imposed the additional constraint 
that $\beta$ is an actual root of $\lag^{+++a}$ with non-vanishing 
multiplicity.) The outer multiplicity $\mu$ with which a solution to (2.10) 
and (2.12) appears can be computed recursively using knowledge about 
the multiplicities of the roots of $\lag^{+++a}$, obtained from
computations based on the denominator and Peterson formula [18].
\par
Before beginning the detailed analysis, we make some general remarks. 
Given a finite dimensional semi-simple Lie algebra, its adjoint
representation is a highest weight representation and it is contained in
the tensor products of the fundamental representations. Also, any
highest weight representation of a finite dimensional algebra also has
a lowest weight. However, for
Kac-Moody algebras things are very different. The adjoint representation
is not a highest weight representation (nor a lowest weight
representation) and it is not contained in the
tensor product of the fundamental representations. This is easily seen by
considering the decomposition of a Kac-Moody algebra.
For simplicity, let us consider a Lorentzian Kac-Moody algebra such that
one can delete one node from its Dynkin diagram to find a finite
dimensional semi-simple Lie algebra [13]. The algebra possesses  a root
decomposition  that contains roots at all positive and negative levels and
so there is no highest weight state. A fundamental representation has the
decomposition of equation (2.6).  A crucial factor  in the analysis 
of the representation depends on the sign of $x^2$ which is positive for
finite dimensional semi-simple Lie algebras  and negative for very extended
Kac-Moody algebras . As a result, the highest weight state has
a positive level for a finite dimensional semi-simple Lie algebra and a
negative level for a very extended Kac-Moody algebra.  Therefore, what
we are considering for positive $n_0$ are lowest weight representations.
The root string of the fundamental 
highest weight representation $l_a$  contains  a weight of the form 
$l_a-n_c\alpha_c-\sum_i n_i\alpha_i$ and so  the
level of the weights in the representation is always less than that of 
 the highest weight states. As such, for the Kac-Moody case, the states
in the  fundamental highest weight representation have only negative
levels and so by taking their tensor product one can never obtain the
positive level states  found in the adjoint representation. One way that
one might  find  the adjoint representation is to take tensor products of
the highest and lowest weight representations. Indeed, we have checked,
admittedly  at rather low levels, that the $A_{9}$ representations
found in the adjoint representation of
$E_8^{++}$ are contained in the tensor product of the $l_1$ and the 
$l_1^*$ representation at the appropriate level. 
\par
In contrast, for  finite dimensional semi-simple Lie algebras, the
fundamental representations have  a positive level and, as the level
decreases as one contructs the root string, one finds in the representation
states of positive and negative levels consistent with the fact that
tensor products of these representations contain the adjoint
representation. 
\par
Now we return to our discussion of the representations contained in
$\lag^{+++}$ and its fundamental representations $\bar l_a$.
The roots in the adjoint representation of $\lag^{+++}$ are just
those for which $n_0=0$ and so they will  possess
lowest weight representation of $\lah$, with Dynkin indices $p_j^{(0)}$,
if
$$\sum_j  p_j^{(0)} (\lambda_j,\lambda_k)
=n_c^{(0)}(\nu,\lambda_k)-n_k^{(0)}
\eqno(2.13)$$
The $\bar l_a$ representation of $\lag^{+++}$ is given by just those 
elements in the 
adjoint representation of $\lag^{+++a}$ with $n_0=1$. Therefore 
they will contain  the highest
weight representation of $\lah$, with Dynkin indices $p_j^{(1)}$, if
$$\sum_j  p_j^{(1)} (\lambda_j,\lambda_k)
=(\lambda_a,\lambda_k)+n_c^{(1)}(\nu,\lambda_k)-n_k^{(1)}
\eqno(2.14)$$
We note that the lowest  $\lah$ representation in the fundamental
$\bar l_a$
representation occurs at level $n_c=0$ and has Dynkin index
$p_a^{(1)}=1$,  all the other Dynkin indices being zero. It corresponds
to the root of  $\lag^{+++a}$  with components
$(1,0,\ldots ,0)$.
\par
Let us consider a  representation of $\lah$ with Dynkin indices
$p_j^{(1)}$ that occurs in the fundamental representation
$\bar l_a$  of $\lag^{+++}$,
subtracting the two above equations, we find that  this $\lah$
representation will  also  occur in the adjoint representation of
$\lag^{+++}$ provided  that the equation
$$(\lambda_a+(n_c^{(1)}-n_c^{(0)})\nu,\lambda_k)
=n_k^{(1)}-n_k^{(0)}
\eqno(2.15)$$
holds for  positive $n_k^{(0)}$. Here $n_c^{(1)}-n_c^{(0)}$ can be interpreted as a shift in level. The converse result holds if
$n_k^{(1)}$ is positive.  As a result, we conclude that
$(\lambda_a+(n_c^{(1)}-n_c^{(0)})\nu)$ is in the dual of the weight
lattice $(\Lambda^\lah)^*$ and so is in the orignal root lattice of
$\lah$, $\Lambda^\lah$. As we noted above the the lowest $H$
representation in the fundamental representation $\bar l_a$ has Dynkin indices 
$p_j^{(0)}=\delta_{aj}$. Using equation (2.13), the condition for such a
representation to  occur in the adjoint representation at level $n_c$ is 
$(\lambda_a-n_c^{}\nu,\lambda_k)
=-n_k^{}$. This in agreement with the condition of equation (2.15), but
it also tells us that if we find the lowest component of the $\bar l_a$ 
representation in the adjoint we also may find all the higher $\bar l_a$ states
in the adjoint representation, in a way that is consistent with the
action of $\lag^{+++}$ on the algebra and on the fundamental
representation. 
\par
Let us consider
$E_8^{+++}$,   for which $\lah=A_{10}$, and
$$\nu=\lambda_8,\  x^2=-{2\over 11}
\eqno(2.16)$$
Equation (2.15) tells us that an $A_{10}$ representation that occurs in
the fundamental representation $\bar l_a$ of $E_8^{+++}$ will occur in the
adjoint representation of $E_8^{+++}$ if
$$A^{-1}_{ak}+(n_c^{(1)}-n_c^{(0)}) A^{-1}_{8k}=n_k^{(1)}-n_k^{(0)}
\eqno(2.17)$$
It is straightforward to analyse this equation and for odd
$a$ one finds the following solution to the equation
$$n_c^{(0)}-n_c^{(1)}={(11+3a)\over 2}
\eqno(2.18)$$
The root vectors being related for $a\le 7$ by
$$ n_k^{(0)}-n_k^{(1)}=\cases{k{(a+1)\over 2}, \quad k\le a\cr
{k\over 2}(a+3)-a,\quad k\ge a,\ k\le 8\cr
(11-k)(a+4),\quad k\ge 8\cr}
\eqno(2.19)$$
and for $a=9$ by
$$ n_k^{(0)}-n_k^{(1)}=\cases{5k, \quad k\le 8\cr
26, \quad k=9}
\eqno(2.20)$$
As we noted the lowest $A_{10}$ representation in $\bar l_a$ has highest
weight $\lambda_a$  and one can read off the
precise roots in $\lag^{+++a}$ to which it corresponds. For the case of
$a=1$
we find the lowest representation has Dynkin index $p_1=1$, i.e. it is 
of
the form of a momentum generator $P_a$, and it occurs at level
$n_c^{(0)}=7$ with a root $\beta$ in $E_{12}=E_8^{+++1}$ corresponding to
$ (0,1,3,5,7,9,11,13,15,10,5,7)$. But there are also solutions
occurring higher up in the algebra, on levels $n_c=18,29,\dots$.
Later we will show that there are
actually solutions for all $\bar l_a$ in this case, which will imply that all
the fundamental representations potentially sit within the adjoint.
\par
We now consider  $A_{n-3}^{+++}$ for which the node $c$ is the
node  labelled $n$, $\lah=A_{n-1}$ and
$$\nu=\lambda_3+\lambda_{n-1},\  x^2=-2{(n-2)\over n}
\eqno(2.21)$$
Equation (2.15) tells us that an $A_{n-1}$ representation that occurs in
the fundamental representation $\bar l_a$ of $A_{n-3}^{+++}$ will occur in 
the
adjoint representation of $A_{n-3}^{+++}$ if
$$A^{-1}_{ak}+(n_c^{(1)}-n_c^{(0)})( A^{-1}_{3k}+
A^{-1}_{n-1 k})=n_k^{(1)}-n_k^{(0)}
\eqno(2.22)$$
Let us first consider the $\bar l_1$ representation then one finds that
$$n(n_k^{(1)}-n_k^{(0)})=\cases{ n(1+k(n_c^{(1)}-n_c^{(0)}))
-k(1+2(n_c^{(1)})-n_c^{(0)}),\quad k\le 3 \cr
n(1+3(n_c^{(1)}-n_c^{(0)}))
-k(1+2(n_c^{(1)})-n_c^{(0)}),\quad k\ge 3 \cr}
\eqno(2.23)$$
We therefore conlcude that
$$n_c^{(0)}-n_c^{(1)}={(pn+1)\over 2}
\eqno(2.24)$$
where $p$ is a positive integer. Clearly, if $n$ is even this can never
be the case, but for $n$ odd we find there is always a solution
by choosing $p=1$. In fact one can use equation (2.23) to search for any
representation of $A_{n-1}$ in the adjoint of $A_{n-3}^{+++}$ and one
finds that for $n$ even, despite the infinite number of representations
present, no such  representation ever occurs. Hence, it is not the case
that the $\lah$ representations in the $\bar l_1$ representation of
$\lag^{+++}$ always belong in the adjoint representation and indeed it can
happen that the adjoint representation contains no generator that can be
identified with the space-time translations.
\par
Analysing equation (2.22) one finds that a
$A_{n-1}$ representation that occurs in
the fundamental representation $\bar l_a$ of $A_{n-3}^{+++}$ will occur in 
the
adjoint representation of $A_{n-3}^{+++}$ if
$$n_c^{(0)}-n_c^{(1)}={pn+a\over 2}
\eqno(2.25)$$
where $p$ is a positive integer. 
\par
Now we consider $G_2^{+++}$, for which the node c is the node
labelled 5 and $\lah=A_4$. We normalise the roots such that
$\alpha_c^2={2\over 3}$, $\alpha_4^2=2$ and
$(\alpha_c,\alpha_4)=-1$. For this case
$$\nu=\lambda_4,\  x^2=-{2\over 15}
\eqno(2.26)
$$
Equation (2.15) tells us that an $A_{4}$ representation that occurs in
the fundamental lowest weight 
representation $\bar l_a$ of $G_2^{+++}$ will occur in the
adjoint representation of $G_2^{+++}$ if
$$A^{-1}_{ak}+(n_c^{(1)}-n_c^{(0)}) A^{-1}_{4k}=n_k^{(1)}-n_k^{(0)}
\eqno(2.27)$$
Examining $\bar l_a$ we find there is a solution for
$$n_c^{(0)}-n_c^{(1)}=5-a,\ n_k^{(0)}-n_k^{(1)}=\cases{0,\quad k\le a\cr
k-a,\quad k\ge a\cr}
\eqno(2.28)$$
In particular for $l_1$  the lowest level generator, $P_a$ has a
$\beta$ in
$G_2^{+++1}$ corresponding to $(0,0,1,2,3,4)$.
\par
Although for the case of the Dynkin diagram $E_8^{+++}$ one can delete
the  node labelled 11 to obtain a sub-Dynkin diagram corresponding to
$A_{10}$, it is not the case for all very extended algebras that one
finds a subalgebra  $A_p$ for a particular $p$. However, in the 
application
to  non-linear realisations we have in mind there is always a preferred 
$A$
type subalgebra that is associated with the gravity sector of the theory 
and
it is the decompostion to this subalgebra that is of interest. As a
result, for some very extended algebras it is desirable to delete yet 
one
more node,  labelled d, which lives on the Dynkin diagram of $\lah$.
In these cases the roots in $\lag^{+++a}$ have three levels
$(n_c,n_d,n_0)$. For example,  for the case of $D_{n-3}^{+++}$ the node
$c$ corresponds to the node labelled
$n$ and the remaining subalgebra is $D_{n-1}$. We then delete the node
labelled $n-1$ to find the subalgebra $A_{n-2}$ which is the algebra
that controls the gravity sub-sector of the non-linear realisation.
We now find the analogue of the above constraints after this
decomposition. We write the simple roots of $\lah$ as
$$\alpha_d=x-\tau, \ \alpha_i, i\not=  d
\eqno(2.29)$$
where $\alpha_i$ are the roots of $A_p$ and
$$\tau=-\sum_j A^{H}_{dj}\ \mu_j {(\alpha_d,\alpha_d)\over
(\alpha_j,\alpha_j)}
\eqno(2.30)$$
where $\mu_j$ are the fundamental weights of $A_p$. The
fundamental weights of $\lah$ are given by
$$\lambda_i=\mu_i+{z\over z^2}(\tau,\mu_i),\  l_d={z\over z^2}
\eqno(2.31)$$
and so
$$\nu=-\sum_j A^{G^{+++}}_{cj}(\mu_j+{z\over z^2}(\tau,\mu_a))=\tilde \nu
-\sum_j {z\over z^2} A^{G^{+++}}_{cj} (\tau,\mu_a)
\eqno(2.32)$$
Following the same arguments as before, we find that
  the adjoint representation of $\lag^{+++}$ contains  generators in the
lowest weight representation of $A_p$, with Dynkin indices $ 
q_j^{(0)}$,
if
$$\sum_j  q_j^{(0)} (\mu_j,\mu_k)
=n_c^{(0)}(\tilde \nu,\mu_k)  +n_d^{(0)} (\tau,\mu_k)-n_k^{(0)}
\eqno(2.33)$$
On the other hand, the $\bar l_a$ representation of
$\lag^{+++}$  will contain  generators in the lowest weight representation
of $A_p$, with Dynkin indices $ q_j^{(1)}$, if
$$\sum_j  q_j^{(1)} (\mu_j,\mu_k)=
(\mu_a,\mu_k)+n_c^{(1)}(\tilde\nu,\mu_k) +n_d^{(1)} (\tau,\mu_k)-n_k^{(1)}
\eqno(2.34)$$
Consequently, a  representation of $A_p$ with Dynkin indices
$p_j^{(1)}$ that occurs in the fundamental lowest weight representation
$\bar l_a$  of $\lag^{+++}$, also occurs in the adjoint representation of
$\lag^{+++}$ provided  that
$$(\mu_a+(n_c^{(1)}-n_c^{(0)})\nu +(n_d^{(1)}-n_d^{(0)})\tau,\mu_k)
=n_k^{(1)}-n_k^{(0)}
\eqno(2.35)$$
holds for $n_k^{(0)}$  positive.
\par
Let us consider $E_7^{+++}$. The nodes $c$ and $d$ are the nodes
labelled 9 and 10 respectively, the subalgebra we obtain is $A_8$ and
$$\tilde \nu=\mu_6, \ \tau=\mu_8
\eqno(2.36)$$
Equation (2.35) becomes
$$
A^{-1}_{ak}+(n_c^{(1)}-n_c^{(0)}) A^{-1}_{6k}+
(n_d^{(1)}-n_d^{(0)}) A^{-1}_{8k}=n_k^{(1)}-n_k^{(0)}
\eqno(2.37)$$
Considering the $l_1$ representation we find a soution with
$$
n_c^{(0)}-n_c^{(1)}=2,\quad n_d^{(0)}-n_d^{(1)}=2,\quad \
n_k^{(0)}-n_k^{(1)}=\cases{k-1,\quad k\le 6,
\cr 11-k,\quad k\ge 6\cr}
\eqno(2.38)$$
The lowest level generator $P_a$ in $\bar l_1$ that occurs in $E_7^{+++}$ 
appears at level $(2,2)$ and has 
$\beta$ corresponding to $(0,0,1,2,3,4,5,4,3,2,2)$.
\par
For $D_{n-3}^{+++}$ the  nodes $c$ and $d$ are the nodes
labelled $n$ and $n-1$ respectively, the subalgebra we obtain is
$A_{n-2}$ and
$$\tilde \nu=\mu_4, \ \tau=\mu_{n-3}.
\eqno(2.39)$$
Given any    $A_{n-2}$ representation which occurs in the $\bar l_a$ of
$D_{n-3}^{+++}$ it will also satisfy (2.10) and (2.12) for 
the adjoint representation of
$D_{n-3}^{+++}$ if
$$A^{-1}_{ak}+(n_c^{(1)}-n_c^{(0)}) A^{-1}_{4k}+
(n_d^{(1)}-n_d^{(0)}) A^{-1}_{n-3,k}=n_k^{(1)}-n_k^{(0)}
\eqno(2.40)$$
Analysing this equation for the $\bar l_1$ representation we find a solution
if
$$2(n_c^{(0)}-n_c^{(1)})-(n_d^{(0)}-n_d^{(1)})={(p(n-1)+1)\over 2}
\eqno(2.41)$$
where $p$ is a positive integer. Clearly, this is impossible if $n$ is
odd, but for $n$ even we can always find a solution with $p=1$.
\par
Carrying out the same analysis for the $\bar l_a$ representation of 
$B_{n-3}^{+++}$ we have to analyse
$$A^{-1}_{ak}+(n_c^{(1)}-n_c^{(0)}) A^{-1}_{4k}+
(n_d^{(1)}-n_d^{(0)}) A^{-1}_{n-2,k}=n_k^{(1)}-n_k^{(0)}
\eqno(2.42)$$
For the particular case of $\bar l_1$ we find that this equation always has
a solution.
\par
One could similarly deal with $C_{n}^{+++}$ but here the process has
to be repeated many times and the analysis is not very illuminating in
general. One finds that there is a solution which looks like $P_a$ in
all cases.
\par
Let us summarise the findings of this section. We have shown that the
necessary conditions (2.10) and (2.12) for a representation of the
subalgebra $\lah$ of $\lag^{+++}$ to be
present in the  adjoint is often equivalent to the necessary condition
for the same representation to be present in the fundamental $\bar l_a$
representation. On the other hand, we have shown that for a certain
number of cases some representations of $\lah$ contained in $\bar l_a$ 
cannot occur in the adjoint, like the momentum representation $\bar l_1$
for half the $D$ and half the $A$ cases.
In this analysis we have made use of an auxiliary
algebra $\lag^{+++a}$. The conditions (2.10) and (2.12) are only
necessary conditions and the actual outer mulitplicity $\mu$ of a
given representation requires additional computations, which we have
checked on a case by case basis.
The reader may verify our statements by
examining the tables in 
appendix B where we list the decompositions of $\lag^{+++1}$ for a
number of 
cases at low levels, for which we have also computed the outer 
multiplicities. In appendix B we also include the cases 
of $E_6^{+++}$ and $F_4^{+++}$ where we have applied a modified
procedure to deal with the fact that there are the two nodes which we
delete are joined by a line in the original Dynkin diagram.

\medskip
{\bf {2.2 Fundamentals in the adjoint representation?}}
\medskip
In the above we considered the fundamental and
adjoint representations of 
$\lag^{+++}$ and  answered the question of when the generators of one
representation, when decomposed with respect to the  subalgebra $\lah$,
occur in the other.  To do this we considered the enlarged algebra
$\lag^{+++a}$ as a technical tool to find the generators in both
representations.  In particular, we allowed the relations between the two
levels $n_c^{(1)}$ and $n_c^{(0)}$ to be arbitary. 
\par 
We  now ask a more restrictive question that the states in the lowest
weight 
representation $\bar l_a$ of $\lag^{+++}$ arise in the adjoint at the same
level $n_c$ as
they occur in the lowest weight representation $\bar l_a$ of
$\lag^{+++}$. For simplicity, in this section
we consider $\lag^{+++}$ to have a  symmetric Cartan matrix.  We note from
equation (2.6) that the lowest state in the
$\bar l_a$ representation occurs at level 
$-{(\nu,\lambda_a) \over x^2}$. The higher states occur at level
$n_c^{(1)}-{{(\nu,\lambda_a)}\over{x^2}}$.
From equation (2.7) we find that the $\bar l_a$ representation
can also be found at level  
$\lag^{+++a}$ with $n_0=1$ and we see its states occur at level 
$n^{(1)}_c-{(\nu,\lambda_a) \over x^2}$. 
We now demand that the corresponding representation arises in the adjoint
representation at the same level 
$$n^{(0)}_c=n^{(1)}_c-{(\nu,\lambda_a) \over x^2}
\eqno(2.43)$$
Substituting this into equation (2.15) we find it can be written as 
$$n_k^{(0)}-n_k^{(1)}=-(A^{\lag^{+++}})^{-1}_{k a}
\eqno(2.44)$$
Hence we find the desired correspondence at all levels if
$(A^{\lag^{+++}})^{-1}_{k a}$ is  a negative integer for all $k$. This is
the case for
$E_{10}=E_8^{++}$ as being self-dual its inverse Cartan matrix is
integer valued and being hyperbolic it is also negative. One can verify
that it is also satisfied for $E_8^{+++}$ if $a$ is even. It also
works for $a=9$.
\par
Let us denote the length
squared  of the roots in 
$\lag^{+++a}$ corresponding  to the $\bar l_a$ representation and the adjoint
representation by $\beta^{(1)}$  and  $\beta^{(0)}$
respectively. Using equation (2.12) we find that 
$$(\beta^{(1)})^2=2+(A^{\lag^{+++}})^{-1}_{aa}+(\beta^{(0)})^2\le 2
\eqno{(2.45)}$$
One must check this condition case by case.
From (2.44) it is evident that
$(A^{\lag^{+++}})^{-1}_{ka}$ should be integral for all $k$. Analysing
the case of $E_8^{+++}$ where for even $a$ the relevant row of the
inverse Cartan matrix is integral we find the values $0,-4,-12,-24,-2$
for $a=2,4,6,8,10$ respectively. This means that 
all states in the $\bar l_a$ representation
of $E_8^{+++}$ occur as solutions of the adjoint equations with length
squared less than $2-(A^{\lag^{+++}})^{-1}_{aa}$. In particular, for
$a=2$ we have $(A^{\lag^{+++}})^{-1}_{aa}=0$ and so all fields
deriving from imaginary roots $\beta^{(0)}$ of the adjoint of $E_8^{+++}$
potentially are states of the $l_2$ representation. A similar argument
with appropriate restrictions on the norm of $\beta^{(1)}$ we retrieve
(some) fields of the $\bar l_a$ representation within the adjoint. This
shows how to retrieve the remaining fundamental representations of
$E_8^{+++}$ within the adjoint of $E_8^{+++}$.
\par
We stress that for general indefinite Kac-Moody algebras, unlike the
finite dimensional, affine and hyperbolic case, it is not
true that any (positive or negative) element $\beta$ which has length
squared less than or equal to $2$ is a root of the algebra. The roots
of a Kac-Moody algebra can be characterised as the union of the set of
real roots and the set of imaginary roots. The real roots are those which
are related to the simple roots by a Weyl transformation, and the
imaginary roots are Weyl conjugate to elements in the fundamental Weyl
chamber which have connected support on the Dynkin diagram [18]. This
is to say that if one marks all points on the Dynkin diagram for which
the components in the simple root expansion are non-zero, then this
constitutes a connected subdiagram of the Dynkin diagram.
\par
In
appendix C we also give an example of the $\bar l_1$ representation of the
hyperbolic $E_{10}=E_8^{++}$, where $E_8^{++1}=E_8^{+++}$ as already
noted in [19].  We see that the identification of fields in the
adjoint coming from imaginary roots with $\beta^2\le 0$ with fields in
the $\bar l_1$ representation works well up to level $(12,0)$ 
where for the first time we get a
field in $E_8^{++}$ which is not in the $\bar l_1$ representation. This
is due to the subtlety of the roots of a general indefinite Kac-Moody
algebra described above. As mentioned before, equations (2.10) and
(2.12), or for that purpose (2.45) do not make any statement 
about the actual outer
multiplicity $\mu$ of a given representation. It would be extremely
informative to have a relation for these quantities as well,
augmenting the relations (2.10) and (2.12). We stress
again that the empirical data shows that the outer multiplicity is
rarely vanishing, for instance within $E_8^{++}$ we note that
on the first twelve levels of $E_8^{++}$  only $[1,0,0,0,0,0,0,0,0]$,
$[0,0,0,0,0,0,0,1,0]$ and $[0,0,0,0,1,0,0,0,0]$ (interpreted as $P_a$,
$Z^{a_1a_2}$ and $Z^{a_1\ldots a_5}$) and their multiples
have vanishing outer multiplicity.
\par
We remark that it cannot be the case that the adjoint representation
contains the  irreducible fundamental representation of $\lag^{+++}$
within it as a sub-representation,  but it might be the case that the
fundamental representations  of $\lag^{+++}$, as they  appear in the
adjoint representation, are 
 irreducible representations of a Borel subalgebra of
$\lag^{+++}$.
\medskip
{\bf {3. Interpretation of the $\bar l_1$ representation as "brane" charges}}
\medskip
Much work has been carried out on the solutions of the ten and eleven
dimensional supergravity theories  which
preserve a given fraction of supersymmetry (for a review see
e.g. [20]). 
Those that preserve some 
of the 32 supersymmetries have topological charges that are related to  
 the central charges that occur in the corresponding supersymmetry
algebra [21].  For the p-branes that preserve half the supersymmetry the
relation is particularly simple and they couple electrically or
magnetically to the rank $p-1$  topological central charges.  
The effective actions that describe the dynamics
of the branes which preserve half the supersymmetry are required by
supersymmetry to contain a coupling between the current corresponding to
the central charge and background fields of the supergravity theory.
Indeed, knowing the  fields of a background supergravity theory one can
read off the topological charges and so the 1/2 supersymmetric  branes
that occur in the  given theory.  
\par
It was shown [9] that the $l_1$ representation of $E_{11}=E_8^{+++}$
contained the space-time translations and the two and five form central
charges of the supersymmetry algebra together with an infinite number of
higher level objects. This suggests that the extended
theory, which possess an
$E_8^{+++}$ symmetry, should contain an infinite number of BPS solutions
each of which has as its source a field in the theory and that 
the objects occuring in the
$l_1$ representation are the corresponding charges. As such, one would
expect that there should be a particular 
relationship between the fields of the
theory  and the objects in the $l_1$ representation. 
As every field in the theory is a Goldstone boson of $E_8^{+++}$, we expect
that there is  a correspondence between the generators  in the adjoint
representation of $E_8^{+++}$ associated with  the
positive root space, and the objects in $l_1$. In this section we show
that there is indeed such a correspondence and it is of the expected 
form. 
\par
In this section, for simpicity,  we consider  a $\lag$ which is simply 
laced and for which $\lah$ is of $A$ type. We will show that
given a 
$\lah$ representation, with Dynkin indices $p_j^{(0)}$, that occurs in the
adjoint representation of $\lag^{+++}$ with $p_{l-1}^{(1)}\not= 0$, for some
$l$, then there exist a representation of
$\lah$ in the $\bar l_1$ representation of $\lag^{+++}$  with Dynkin indices 
$p_j^{1}$ given by  
$$p_j^{(1)}-p_j^{(0)}=\cases{1,\quad j=l\cr
-1,\quad j=l-1\cr
0,\quad j\not =l,l-1\cr}
\eqno(3.1)$$
 For the simple case for which $p_l^{(1)}=1$ 
 we find $p_{l-1}^{(0)}= 1$, with  all other Dynkin indices
being zero. This means that if 
$\lag^{+++}$ contains the object $R^{a_1\ldots a_{l}}$ with
 corresponding field $A_{a_1\ldots a_l}$, then the $\bar l_1$
representation of $\lag^{+++}$
contains a generator
$Z^{a_1\ldots a_{l-1}}$. 
\par
The existence of the two representations of interest corresponds to the 
solutions of equations (2.10) and (2.12). Subtracting them we find that the
condition for the $\lah$ representation in $\bar l_1$ 
to appear in the adjoint is
given by 
$$\sum_j(\lambda_{k}, \lambda_{j})(p_j^{(1)}-p_j^{(0)})=
(\lambda_{1}, \lambda_{k})-(n_k^{(1)}-n_k^{(0)})
\eqno(3.2)$$
Using the correspondence of equation (3.1) we find that the above equation
becomes 
$$(\lambda_{l}-\lambda_{l-1}-\lambda_{1}, \lambda_{k})=n_k^{(0)}-n_k^{(1)}
\eqno(3.3)$$
Recognising the $\lambda_{l-1}+\lambda_{1}-\lambda_{l}$ is just 
$\alpha_1+\ldots +\alpha_{l-1}$, we find that it is always true provided  
$$n_k^{(1)}-n_k^{(0)}=\cases{1,\quad k\le l-1,\cr
0,\quad k\ge l,\cr}
\eqno(3.4)$$
We note that if $n_k^{(0)}$ is positive then $n_k^{(1)}$ is automatically
positive and so the representation always occurs. Here we can also say
something about the outer multiplicity: One can see that any
$\lah$-representation within $\lag^{+++}$ lifts to a corresponding one
in $\lag^{+++1}$ under the $\lah$ extended by one node to the
left. This representation then will reduce to the original one of
$\lag^{+++}$ and one with one box filled in with a $0$, that is with
the rank decreased by one.  Therefore it is a corollary of the
result that the very-extended $\lag^{+++}$ theories contain at
least the fields of the oxidised theories [16], that, if augmented by the
$\bar l_1$ representation, they also contain all the correct central
charges for these fields.
\par
We now show independently that the additional condition of equation (2.12) is
satisfied for the $\bar l_1$ representation, $n^{(0)}=1$, if it is satisfied
for the adjoint representation, $n^{(0)}=1$. Let us denote the length
squared  of the roots in 
$\lag^{+++1}$ corresponding  to the $\bar l_1$ representation and the adjoint
representation by $\beta^{(1)}$  and  $\beta^{(0)}$
respectively. Using equation (2.12) we find that 
$$(\beta^{(1)})^2-(\beta^{(0)})^2=y^2+{(\nu,\lambda_1)^2\over x^2}
-2n_c (\nu,\lambda_1)+\sum_{ij}p_i^{(1)}A^{-1}_{ij}p_j^{(1)}
-\sum_{ij}p_i^{(0)}A^{-1}_{ij}p_j^{(0)}
\eqno(3.5)$$ 
Using equation (2.9) and the relation $y^2+l_1^2=2$ we find that 
$$(\beta^{(1)})^2-(\beta^{(0)})^2=2-n_k^{(1)}-n_k^{(0)}
-\sum_j(n_j^{(1)}-n_j^{(0)})(p_j^{(1)}+p_j^{(0)})
\eqno(3.6)$$ 
Finally, using equations (3.1) and (3.4) we conclude that the left-hand
side is a negative integer and so if $(\beta^{(0)})^2=2,0,\ldots$ then
$\beta^{(1)}$ has an acceptable length squared. 

\medskip
{\bf {4. Where is   space-time ?}}
\medskip

As we have explained in the introduction  there is
growing evidence for the conjecture [4] that an
extension of eleven dimensional supergravity possesses an
$E_{8}^{+++}$ symmetry
 which is non-linearly realised.
Reference [4] did not address the way space-time should enter
the construction, but  there are now essentially two different proposals
as to how this should arise. 
\par
It has been proposed [6] that the fields of
the non-linear realisation of $E_{10}$  should depend only on time and the
spatial dependence should occur through the equations of motion of the
non-linear realisation, {\sl i.e.} the spatial dependence  should essentially
arise as a result of the properties of the adjoint representation of
$E_{8}^{++}$. Alternatively, it has been proposed [9] that one should form
the non-linear realisation of the semi-direct product of $E_{8}^{+++}$
with the $\bar l_1$ representation. In this case the fields would depend on the
coordinates introduced into the non-linear realisation by the $\bar l_1$
representation. These would include space-time, coordinates for the
central charges of the eleven dimensional supersymmetry algebra and an
infinite number of other coordinates.  There is also  the suggestion of
reference [10] in which a non-linear realisation of $E_{8}^{+++}$ is
considered, but the fields depend on an auxiliary parameter and it is
hoped that the space-time dependence occurs by a mechanism similar to
that proposed in reference [6]. These authors  also  suggested that the
level seven generator in
$E_8^{+++}$, which has the correct
$A_{10}$ properties, {\sl i.e.} $p_1=1$,  be identified with the space-time
translations.  
\par 
In [6] it was shown that for the 3-form, 6-form and dual
graviton representations at levels 1, 2 and 3 of $E_8^{++}$ there are
tensors with the index structures of a $k$-th spatial derivative at levels
$1+3k$, $2+3k$ and $3+3k$ respectively.
 It is
hoped [6] that the dynamics will imply that the fields corresponding
to the latter generators will  turn out to be the spatial derivatives
of fields correponding to the former generators. In fact,  the  roots
giving rise to these $k$-th derivative fields are consistent with the
action of a  generator with
$p_1=1$ at level $3$ which satifies the constraints of equations (2.10) and
(2.12). This element at level three is just the fundamental lowest 
weight of the  over
extended node and so identical with the affine root of the $E_8^+$ 
subalgebra. As discussed in section (2.2),  this is just the first
object in the infinite tower of the
$\bar l_1$ representation. One might interpret this level $3$ generator as
a space-time generator, but as it  has outer
multiplicity zero it  does not appear in the $E_{8}^{++}$ algebra. 
\par
It was  conjectured [10] that the adjoint representation for all 
$\lag^{+++}$, contains a generator that belongs to  the correct $A$
representation,  i.e. $p_1=1,\ p_j=0,\ j\ge2$, to be identified as 
space-time  translations. However, we have seen that this is not the case
for several important examples, namely $D_{n-3}^{+++}$ for odd $n$ and
$A_{n-3}^{+++}$ for even $n$.\footnote{$^1$}{This result was
communicated to the authors of reference [10] who have subsequently changed
their paper to take account of these results.}.  As was explained in
[5,16], there exist two alternative choices for a regular $A_9$
subalgebra of $E_8^{+++}$ and the decompositions there give rise to
the fields of the two type II theories at lowest levels. It is not
hard to see that taking the IIB decomposition of $E_8^{+++}$ there can
be no object with the structure of a space-time translation operator in the
spectrum. For these reason,
the  mechanism
suggested  in [10] can not apply to all $\lag^{+++}$ and also fails in
particular for the interesting cases of type IIB theory and the
bosonic string $D_{24}^{+++}$.  
\par
Furthermore, even when a
generator with   the correct 
$A$ representation property does occur it is usually not unique. For
example, as we found in section two,  for the case of
$E_8^{+++}$,  such a generator occurs at levels $7+11s$ for any
non-negative integer 
$s$  and so one can wonder which of these is the real space-time
translation generator. A similar result occurs for $E_8^{++}$, where
potential space-time operators  occur at level $3+10r$ for any
non-negative integer 
$r$. To make the situation more confusing one finds that some of these
operators occur with outer multiplicity greater than one and so even at a
given level there is often more than one choice. For $E_8^{++}$ the
corresponding statement is that there are $P_a$-like operators at
levels $3+10r$ with $r$ a non-negative integer, and again for $r\ge 1$
it seems that they have non-vanishing outer multiplicity [11].
\par
Using the analysis of section two, one can show that  
given a representation of $A_{10}$ 
with  Dynkin indices $p_j$,  at level $n_c$, in the adjoint representation
of 
$E_{8}^{+++}$, then the adjoint representation also contains a  
representation with Dynkin indicies $p_1+m, p_j, j\ge 2$ at level 
$n_c+m(7+11s)$. Clearly, the field associated to the latter generator
 in the non-linear realisation 
has the correct $A_{10}$ character to be identified with the space-time
derivatives of the  field associated with the former generator. 
A similar result holds for $E_8^{++}$ and the
possible space momentum generators mentioned above. Hence from this
perspective the possible momentum generators are on an equal footing. 
\par
Even if a generator has the correct $A$ properties to be identified
with
a space-time generator $P_a$ one can wonder if its commutator vanishes.
For the level seven candidate of $E_8^{+++}$, suggested in reference [10], 
this is unlikely to be the case as at level 14 there exist a $p_2=1$
representation that has a root of precisely the correct type 
to occur on the right-hand side of the $[P_a,P_b]$ commutator. This would
be inconsistent with our usual understanding of what is meant by a
space-time translation operator, and it would imply  that
the theory has some kind of non-commutative structure. We note that this
does not apply to the level $3$ operator of $E_8^{++}$ mentioned above
as its outer multiplicitly is zero and so does not actually occur in the
theory. 
\par
As we show in section two when a generator with Dynkin indices
$p_1=1,\ p_j=0,\ j\ge2$ does occur in the adjoint representation of
$\lag^{+++}$, 
it can usually be thought of as part of an infinite tower
associated with the $\bar l_1$ representation. For the case of $E_8^{+++}$ the
next object, $Z^{ab}$, in the tower has $p_9=2,\ p_j=0,\ j\not=9$ and
occurs at levels $n_c=8+11k$ ($k\ge 0$, 
for $k=0$ it has outer multiplicity $5$). Given a
representation  Dynkin indices $p_j$  at level $n_c$ in the adjoint
representation of 
$E_{8}^{+++}$, then the adjoint representation also contains a  
solution to (2.10) and (2.12) with Dynkin indices 
$p_9+m, p_j, j\ne 9$ at level 
$n_c+s(11k+8)$, for non-negative integer $s$. 
The fields associated with the latter generator have the correct
structure to be interpreted as derivatives with repect to $Z^{ab}$.
One might hope that this is also a consequence of the dynamics. Hence,
one has in effect a theory where the fields depend on space-time and additional
coordinates associated with the central charges.
 The same is likely to hold
for all the objects of the $\bar l_1$ representation and as such  even if
the derivatives arise from within $E_8^{+++}$ the $\bar l_1$ representation is
likely to play an important role.  A similar result also holds for
$E_8^{++}$.
\par 
The introduction of space-time by considering the semi-direct product
of 
 $E_8^{++}$ with the $\bar l_1$ representation may appear less elegant,
however, as we have explained the $\bar l_1$ representation 
arises naturally in the context of $E_{12}=E^{+++1}_8$ and the roots with 
level $n_0=1$. There may also be reasons associated with the conformal
symmetry found in reference [3,9] to also introduce the level  $n_0=-1$
roots which is the dual highest weight $l_1$ representation. The levels
$n_0=-1,0,1$ form what is known as the local subalgebra of
$\lag^{+++a}$ and the lowest $n_c$ level fields just form an algebra
resembling the conformal group with the only difference being that the
Lorentz generators $J_{ab}$ are traded in for the generators $K^a{}_b$
of the general
linear group (which has its Cartan invariant subalgebra has the Lorentz
group).\footnote{$^2$}{In [4] the local subalgebra taken for $E_8^{+++}$
was that which was invariant under the Cartan involution
$E_a\leftrightarrow -F_a, H_a\to -H_a$. Considering the gravity line
$SL(n)$ one finds that the corresponding local subalgebra is $SO(n)$ and not
$SO(1,n-1)$.  As a result, to arrive at Minkowski space one  must carry
out a Wick rotation.  To arrive directly in Minkowski
space one can take the above involution except for $E_1\leftrightarrow
F_1$. A modification along these lines was presented in [10], but the
involution taken appears  not  to be the same.} In the supersymmetric
extension one knows [3] that 
the conformal group gets enlarged to
an algebra which must contain $Osp(1|64)$ [3] which in turn has 
$SL(32)$ as a subalgebra. This
$SL(32)$ is contained in the Cartan invariant subalgebra of $E_8^{+++}$
as a truncation,  but not as a subalgebra [9] and hence it is
understandable that the conformal algebra gets enlarged in the way
suggested by adding the $l_1$ representation. We also note that if
we apply the diagram extension to the semi-direct product of the Lorentz
group with a momentum generator, {\sl i.e.} the Poincar\'e group,  then
one finds the  conformal algebra.
In fact, the momentum generators and all the
central charges commute in this local part of $\lag^{+++1}$. If one
considers the whole of
$\lag^{+++1}$, then the momentum generators will still commute but this
is then no longer necessarily true for the higher charges. However, 
if one also
includes the $\bar l_2$ representation, even the $P_a$ operators no longer
need to commute. Hence, it is possible that the semi-direct product is
part of some larger algebra.  
\par
It is clear that how space-time enters is intricately connected with the 
representation theory of $E_{11}$ and in this paper we have tried to
illuminate this connection. Clearly, further work is required on this
fascinating problem.

\medskip
{\bf Acknowledgements}

AK would like to thank the Studienstiftung des deutschen Volkes for
financial support and King's College, London, for hospitality. This
research was supported in part by the PPARC grants 
PPA/G/O/2000/00451  and PPA/G/S4/1998/00613.

\eject

{\bf Appendix A: Dynkin diagrams for the very extended algebras}

\vbox{\centerline{\epsfbox{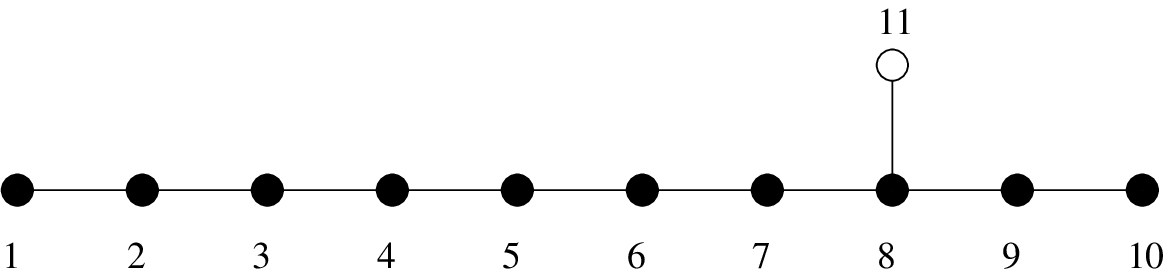}}\centerline{$E_8^{+++}$}}
\vbox{\centerline{\epsfbox{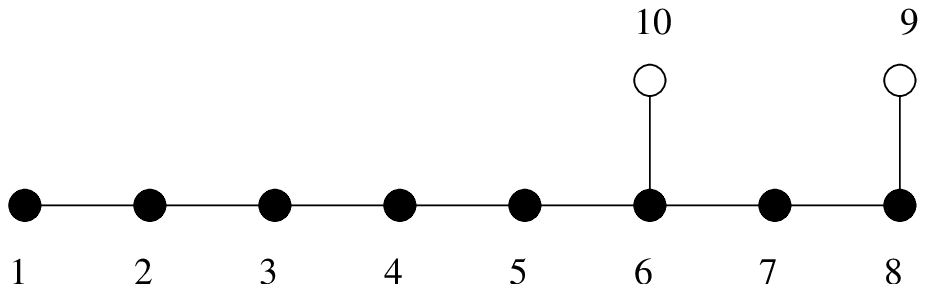}}\centerline{$E_7^{+++}$}}
\vbox{\centerline{\epsfbox{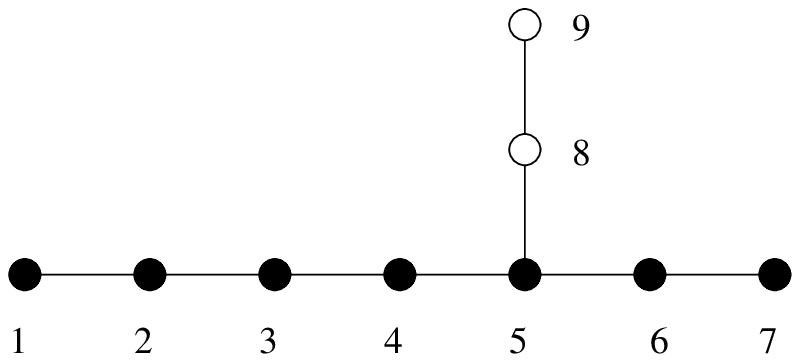}}\centerline{$E_6^{+++}$}}
\vbox{\centerline{\epsfbox{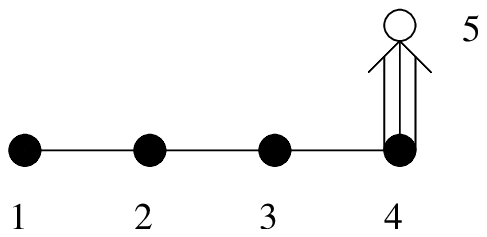}}\centerline{$G_2^{+++}$}}
\vbox{\centerline{\epsfbox{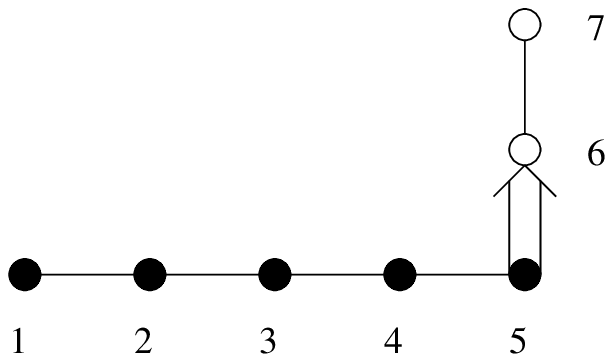}}\centerline{$F_4^{+++}$}}
\vbox{\centerline{\epsfbox{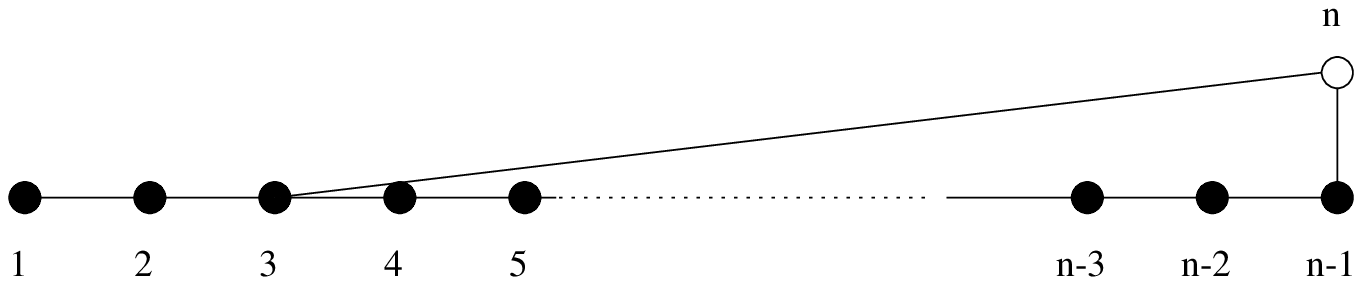}}\centerline{$A_{n-3}^{+++}$}}
\vbox{\centerline{\epsfbox{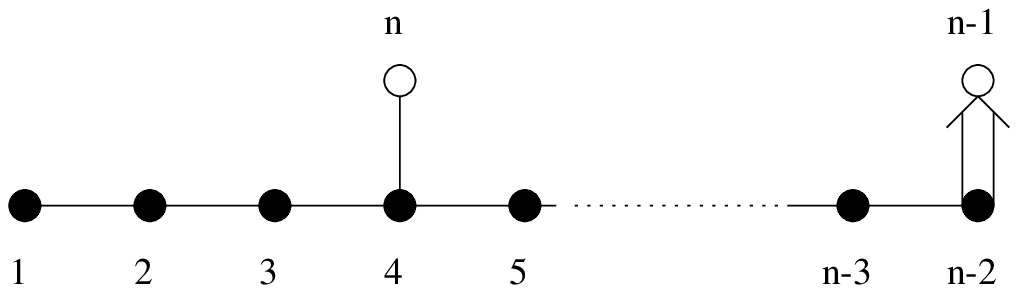}}\centerline{$B_{n-3}^{+++}$}}
\vbox{\centerline{\epsfbox{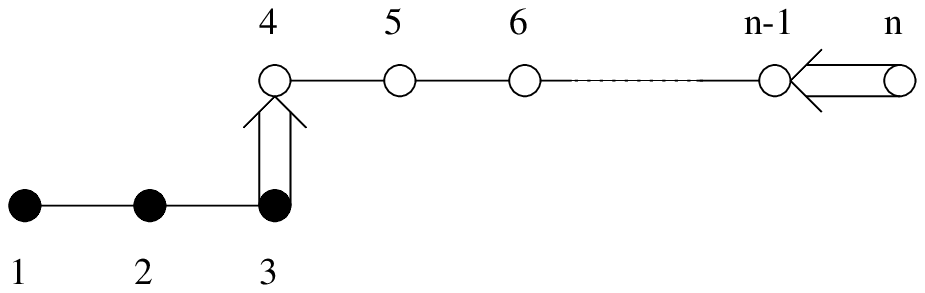}}\centerline{$C_{n-3}^{+++}$}}
\vbox{\centerline{\epsfbox{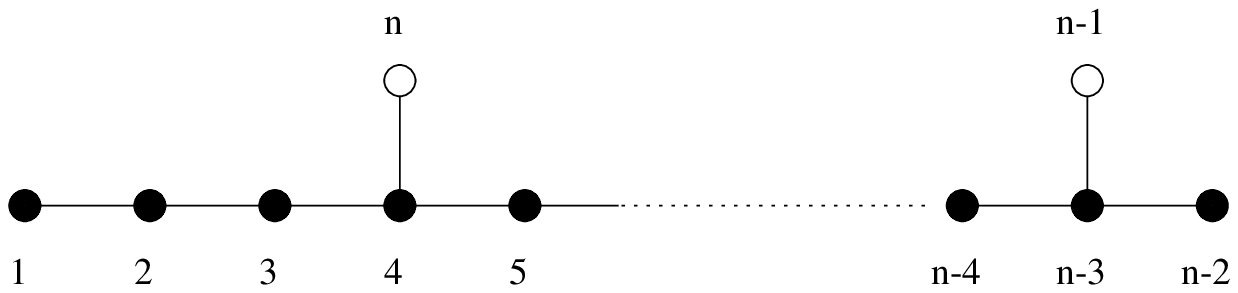}}\centerline{$D_{n-3}^{+++}$}}

\eject

{\bf Appendix B: Tables for the $\bar l_1$ representation of various
$\lag^{+++}$}
\medskip
{\bf B1: $E_8^{+++1}$}
\medskip

\vbox{\settabs\+(8,0)\quad\quad&[0,0,0,0,0,0,0,0,1,0]\quad&(0,2,4,6,8,10,12,14,16,10,5,8)\quad&-10\quad\quad&95\quad\quad&5\cr
\+$(n_c,n_0)$&$p_i$&$\beta$&$\beta^2$&$ht(\beta)$&$\mu$\cr
\hrule
\+(1,0)&[0,0,0,0,0,0,0,1,0,0]&(0,0,0,0,0,0,0,0,0,0,0,1)&2&1&1\cr
\+(2,0)&[0,0,0,0,1,0,0,0,0,0]&(0,0,0,0,0,0,1,2,3,2,1,2)&2&11&1\cr
\+(3,0)&[0,0,1,0,0,0,0,0,0,1]&(0,0,0,0,1,2,3,4,5,3,1,3)&2&22&1\cr
\+(3,0)&[0,1,0,0,0,0,0,0,0,0]&(0,0,0,1,2,3,4,5,6,4,2,3)&0&30&0\cr
\+(4,0)&[0,1,0,0,0,0,0,1,0,0]&(0,0,0,1,2,3,4,5,6,4,2,4)&2&31&1\cr
\+(4,0)&[1,0,0,0,0,0,0,0,0,2]&(0,0,1,2,3,4,5,6,7,4,1,4)&2&37&1\cr
\+(4,0)&[1,0,0,0,0,0,0,0,1,0]&(0,0,1,2,3,4,5,6,7,4,2,4)&0&38&0\cr
\+(4,0)&[0,0,0,0,0,0,0,0,0,1]&(0,1,2,3,4,5,6,7,8,5,2,4)&-2&47&1\cr
\+(5,0)&[1,0,0,0,0,0,1,0,0,1]&(0,0,1,2,3,4,5,6,8,5,2,5)&2&41&1\cr
\+(5,0)&[0,1,0,0,1,0,0,0,0,0]&(0,0,0,1,2,3,5,7,9,6,3,5)&2&41&1\cr
\+(5,0)&[1,0,0,0,0,1,0,0,0,0]&(0,0,1,2,3,4,5,7,9,6,3,5)&0&45&0\cr
\+(5,0)&[0,0,0,0,0,0,0,1,0,1]&(0,1,2,3,4,5,6,7,8,5,2,5)&0&48&1\cr
\+(5,0)&[0,0,0,0,0,0,1,0,0,0]&(0,1,2,3,4,5,6,7,9,6,3,5)&-2&51&1\cr
\+(6,0)&[1,0,0,0,1,0,0,0,1,0]&(0,0,1,2,3,4,6,8,10,6,3,6)&2&49&1\cr
\+(6,0)&[0,1,1,0,0,0,0,0,0,1]&(0,0,0,1,3,5,7,9,11,7,3,6)&2&52&1\cr
\+(6,0)&[0,0,0,0,0,0,1,1,0,0]&(0,1,2,3,4,5,6,7,9,6,3,6)&2&52&1\cr
\+(6,0)&[0,0,0,0,0,1,0,0,0,2]&(0,1,2,3,4,5,6,8,10,6,2,6)&2&53&1\cr
\+(6,0)&[1,0,0,1,0,0,0,0,0,1]&(0,0,1,2,3,5,7,9,11,7,3,6)&0&54&1\cr
\+(6,0)&[0,0,0,0,0,1,0,0,1,0]&(0,1,2,3,4,5,6,8,10,6,3,6)&0&54&0\cr
\+(6,0)&[0,0,0,0,1,0,0,0,0,1]&(0,1,2,3,4,5,7,9,11,7,3,6)&-2&58&2\cr
\+(6,0)&[0,2,0,0,0,0,0,0,0,0]&(0,0,0,2,4,6,8,10,12,8,4,6)&0&60&0\cr
\+(6,0)&[1,0,1,0,0,0,0,0,0,0]&(0,0,1,2,4,6,8,10,12,8,4,6)&-2&61&1\cr
\+(6,0)&[0,0,0,1,0,0,0,0,0,0]&(0,1,2,3,4,6,8,10,12,8,4,6)&-4&64&1\cr
\+(7,0)&[1,0,0,1,0,0,1,0,0,0]&(0,0,1,2,3,5,7,9,12,8,4,7)&2&58&1\cr
\+(7,0)&[0,0,0,0,1,0,0,1,0,1]&(0,1,2,3,4,5,7,9,11,7,3,7)&2&59&1\cr
\+(7,0)&[1,0,1,0,0,0,0,0,1,1]&(0,0,1,2,4,6,8,10,12,7,3,7)&2&60&1\cr
\+(7,0)&[0,2,0,0,0,0,0,1,0,0]&(0,0,0,2,4,6,8,10,12,8,4,7)&2&61&1\cr
\+(7,0)&[1,0,1,0,0,0,0,1,0,0]&(0,0,1,2,4,6,8,10,12,8,4,7)&0&62&1\cr
\+(7,0)&[0,0,0,0,1,0,1,0,0,0]&(0,1,2,3,4,5,7,9,12,8,4,7)&0&62&1\cr
\+(7,0)&[0,0,0,1,0,0,0,0,1,1]&(0,1,2,3,4,6,8,10,12,7,3,7)&0&63&1\cr
\+(7,0)&[0,0,0,1,0,0,0,1,0,0]&(0,1,2,3,4,6,8,10,12,8,4,7)&-2&65&2\cr
\+(7,0)&[1,1,0,0,0,0,0,0,0,2]&(0,0,1,3,5,7,9,11,13,8,3,7)&0&67&1\cr
\+(7,0)&[1,1,0,0,0,0,0,0,1,0]&(0,0,1,3,5,7,9,11,13,8,4,7)&-2&68&2\cr
\+(7,0)&[0,0,1,0,0,0,0,0,0,2]&(0,1,2,3,5,7,9,11,13,8,3,7)&-2&69&3\cr
\+(7,0)&[0,0,1,0,0,0,0,0,1,0]&(0,1,2,3,5,7,9,11,13,8,4,7)&-4&70&3\cr
\+(7,0)&[2,0,0,0,0,0,0,0,0,1]&(0,0,2,4,6,8,10,12,14,9,4,7)&-4&76&1\cr
\+(7,0)&[0,1,0,0,0,0,0,0,0,1]&(0,1,2,4,6,8,10,12,14,9,4,7)&-6&77&4\cr
\+(7,0)&[1,0,0,0,0,0,0,0,0,0]&(0,1,3,5,7,9,11,13,15,10,5,7)&-8&86&1\cr}
\vbox{\settabs\+(8,0)\quad\quad&[0,0,0,0,0,0,0,0,1,0]\quad&(0,2,4,6,8,10,12,14,16,10,5,8)\quad&-10\quad\quad&95\quad\quad&5\cr
\+$(n_c,n_0)$&$p_i$&$\beta$&$\beta^2$&$ht(\beta)$&$\mu$\cr
\hrule
\+(8,0)&[1,0,1,0,0,1,0,0,0,1]&(0,0,1,2,4,6,8,11,14,9,4,8)&2&67&1\cr
\+(8,0)&[0,0,0,1,0,0,1,0,1,0]&(0,1,2,3,4,6,8,10,13,8,4,8)&2&67&1\cr
\+(8,0)&[1,1,0,0,0,0,0,1,1,0]&(0,0,1,3,5,7,9,11,13,8,4,8)&2&69&1\cr
\+(8,0)&[0,0,0,0,2,0,0,0,0,1]&(0,1,2,3,4,5,8,11,14,9,4,8)&2&69&1\cr
\+(8,0)&[0,0,1,0,0,0,0,1,0,2]&(0,1,2,3,5,7,9,11,13,8,3,8)&2&70&1\cr
\+(8,0)&[0,0,0,1,0,1,0,0,0,1]&(0,1,2,3,4,6,8,11,14,9,4,8)&0&70&1\cr
\+(8,0)&[1,1,0,0,0,0,1,0,0,1]&(0,0,1,3,5,7,9,11,14,9,4,8)&0&71&2\cr
\+(8,0)&[1,0,0,2,0,0,0,0,0,0]&(0,0,1,2,3,6,9,12,15,10,5,8)&2&71&1\cr
\+(8,0)&[0,2,0,0,1,0,0,0,0,0]&(0,0,0,2,4,6,9,12,15,10,5,8)&2&71&1\cr
\+(8,0)&[0,0,1,0,0,0,0,1,1,0]&(0,1,2,3,5,7,9,11,13,8,4,8)&0&71&1\cr
\+(8,0)&[1,0,1,0,1,0,0,0,0,0]&(0,0,1,2,4,6,9,12,15,10,5,8)&0&72&1\cr
\+(8,0)&[0,0,1,0,0,0,1,0,0,1]&(0,1,2,3,5,7,9,11,14,9,4,8)&-2&73&4\cr
\+(8,0)&[2,0,0,0,0,0,0,0,1,2]&(0,0,2,4,6,8,10,12,14,8,3,8)&2&75&1\cr
\+(8,0)&[1,1,0,0,0,1,0,0,0,0]&(0,0,1,3,5,7,9,12,15,10,5,8)&-2&75&2\cr
\+(8,0)&[0,0,0,1,1,0,0,0,0,0]&(0,1,2,3,4,6,9,12,15,10,5,8)&-2&75&2\cr
\+(8,0)&[2,0,0,0,0,0,0,0,2,0]&(0,0,2,4,6,8,10,12,14,8,4,8)&0&76&0\cr
\+(8,0)&[0,1,0,0,0,0,0,0,1,2]&(0,1,2,4,6,8,10,12,14,8,3,8)&0&76&1\cr
\+(8,0)&[2,0,0,0,0,0,0,1,0,1]&(0,0,2,4,6,8,10,12,14,9,4,8)&-2&77&2\cr
\+(8,0)&[0,1,0,0,0,0,0,0,2,0]&(0,1,2,4,6,8,10,12,14,8,4,8)&-2&77&2\cr
\+(8,0)&[0,0,1,0,0,1,0,0,0,0]&(0,1,2,3,5,7,9,12,15,10,5,8)&-4&77&3\cr
\+(8,0)&[0,1,0,0,0,0,0,1,0,1]&(0,1,2,4,6,8,10,12,14,9,4,8)&-4&78&6\cr
\+(8,0)&[2,0,0,0,0,0,1,0,0,0]&(0,0,2,4,6,8,10,12,15,10,5,8)&-4&80&2\cr
\+(8,0)&[0,1,0,0,0,0,1,0,0,0]&(0,1,2,4,6,8,10,12,15,10,5,8)&-6&81&7\cr
\+(8,0)&[1,0,0,0,0,0,0,0,0,3]&(0,1,3,5,7,9,11,13,15,9,3,8)&-2&84&2\cr
\+(8,0)&[1,0,0,0,0,0,0,0,1,1]&(0,1,3,5,7,9,11,13,15,9,4,8)&-6&85&7\cr
\+(8,0)&[1,0,0,0,0,0,0,1,0,0]&(0,1,3,5,7,9,11,13,15,10,5,8)&-8&87&6\cr
\+(8,0)&[0,0,0,0,0,0,0,0,0,2]&(0,2,4,6,8,10,12,14,16,10,4,8)&-8&94&3\cr
\+(8,0)&[0,0,0,0,0,0,0,0,1,0]&(0,2,4,6,8,10,12,14,16,10,5,8)&-10&95&5\cr
\+(0,1)&[1,0,0,0,0,0,0,0,0,0]&(1,0,0,0,0,0,0,0,0,0,0,0)&2&1&1\cr
\+(1,1)&[0,0,0,0,0,0,0,0,1,0]&(1,1,1,1,1,1,1,1,1,0,0,1)&2&10&1\cr
\+(2,1)&[0,0,0,0,0,1,0,0,0,0]&(1,1,1,1,1,1,1,2,3,2,1,2)&2&17&1\cr
\+(3,1)&[0,0,0,1,0,0,0,0,0,1]&(1,1,1,1,1,2,3,4,5,3,1,3)&2&26&1\cr
\+(3,1)&[0,0,1,0,0,0,0,0,0,0]&(1,1,1,1,2,3,4,5,6,4,2,3)&0&33&1\cr
\+(4,1)&[0,0,1,0,0,0,0,1,0,0]&(1,1,1,1,2,3,4,5,6,4,2,4)&2&34&1\cr
\+(4,1)&[0,1,0,0,0,0,0,0,0,2]&(1,1,1,2,3,4,5,6,7,4,1,4)&2&39&1\cr
\+(4,1)&[0,1,0,0,0,0,0,0,1,0]&(1,1,1,2,3,4,5,6,7,4,2,4)&0&40&1\cr
\+(4,1)&[1,0,0,0,0,0,0,0,0,1]&(1,1,2,3,4,5,6,7,8,5,2,4)&-2&48&2\cr
\+(4,1)&[0,0,0,0,0,0,0,0,0,0]&(1,2,3,4,5,6,7,8,9,6,3,4)&-4&58&1\cr
\+(5,1)&[0,1,0,0,0,0,1,0,0,1]&(1,1,1,2,3,4,5,6,8,5,2,5)&2&43&1\cr
\+(5,1)&[0,0,1,0,1,0,0,0,0,0]&(1,1,1,1,2,3,5,7,9,6,3,5)&2&44&1\cr
\+(5,1)&[0,1,0,0,0,1,0,0,0,0]&(1,1,1,2,3,4,5,7,9,6,3,5)&0&47&1\cr
\+(5,1)&[1,0,0,0,0,0,0,1,0,1]&(1,1,2,3,4,5,6,7,8,5,2,5)&0&49&2\cr
\+(5,1)&[1,0,0,0,0,0,1,0,0,0]&(1,1,2,3,4,5,6,7,9,6,3,5)&-2&52&2\cr}
\vbox{\settabs\+(8,0)\quad\quad&[0,0,0,0,0,0,0,0,1,0]\quad&(0,2,4,6,8,10,12,14,16,10,5,8)\quad&-10\quad\quad&95\quad\quad&5\cr
\+$(n_c,n_0)$&$p_i$&$\beta$&$\beta^2$&$ht(\beta)$&$\mu$\cr
\hrule
\+(5,1)&[0,0,0,0,0,0,0,0,0,3]&(1,2,3,4,5,6,7,8,9,5,1,5)&2&56&1\cr
\+(5,1)&[0,0,0,0,0,0,0,0,1,1]&(1,2,3,4,5,6,7,8,9,5,2,5)&-2&57&2\cr
\+(5,1)&[0,0,0,0,0,0,0,1,0,0]&(1,2,3,4,5,6,7,8,9,6,3,5)&-4&59&3\cr
\+(6,1)&[0,1,0,0,1,0,0,0,1,0]&(1,1,1,2,3,4,6,8,10,6,3,6)&2&51&1\cr
\+(6,1)&[1,0,0,0,0,0,1,1,0,0]&(1,1,2,3,4,5,6,7,9,6,3,6)&2&53&1\cr
\+(6,1)&[1,0,0,0,0,1,0,0,0,2]&(1,1,2,3,4,5,6,8,10,6,2,6)&2&54&1\cr
\+(6,1)&[1,0,0,0,0,1,0,0,1,0]&(1,1,2,3,4,5,6,8,10,6,3,6)&0&55&1\cr
\+(6,1)&[0,0,2,0,0,0,0,0,0,1]&(1,1,1,1,3,5,7,9,11,7,3,6)&2&55&1\cr
\+(6,1)&[0,1,0,1,0,0,0,0,0,1]&(1,1,1,2,3,5,7,9,11,7,3,6)&0&56&2\cr
\+(6,1)&[1,0,0,0,1,0,0,0,0,1]&(1,1,2,3,4,5,7,9,11,7,3,6)&-2&59&4\cr
\+(6,1)&[0,0,0,0,0,0,1,0,0,2]&(1,2,3,4,5,6,7,8,10,6,2,6)&0&60&2\cr
\+(6,1)&[0,0,0,0,0,0,0,2,0,0]&(1,2,3,4,5,6,7,8,9,6,3,6)&0&60&1\cr
\+(6,1)&[0,0,0,0,0,0,1,0,1,0]&(1,2,3,4,5,6,7,8,10,6,3,6)&-2&61&3\cr
\+(6,1)&[0,1,1,0,0,0,0,0,0,0]&(1,1,1,2,4,6,8,10,12,8,4,6)&-2&63&2\cr
\+(6,1)&[0,0,0,0,0,1,0,0,0,1]&(1,2,3,4,5,6,7,9,11,7,3,6)&-4&64&5\cr
\+(6,1)&[1,0,0,1,0,0,0,0,0,0]&(1,1,2,3,4,6,8,10,12,8,4,6)&-4&65&3\cr
\+(6,1)&[0,0,0,0,1,0,0,0,0,0]&(1,2,3,4,5,6,8,10,12,8,4,6)&-6&69&5\cr
\+(7,1)&[1,0,0,0,1,0,0,1,0,1]&(1,1,2,3,4,5,7,9,11,7,3,7)&2&60&1\cr
\+(7,1)&[0,1,0,1,0,0,1,0,0,0]&(1,1,1,2,3,5,7,9,12,8,4,7)&2&60&1\cr
\+(7,1)&[0,1,1,0,0,0,0,0,1,1]&(1,1,1,2,4,6,8,10,12,7,3,7)&2&62&1\cr
\+(7,1)&[1,0,0,0,1,0,1,0,0,0]&(1,1,2,3,4,5,7,9,12,8,4,7)&0&63&2\cr
\+(7,1)&[1,0,0,1,0,0,0,0,1,1]&(1,1,2,3,4,6,8,10,12,7,3,7)&0&64&2\cr
\+(7,1)&[0,1,1,0,0,0,0,1,0,0]&(1,1,1,2,4,6,8,10,12,8,4,7)&0&64&2\cr
\+(7,1)&[0,0,0,0,0,0,2,0,0,1]&(1,2,3,4,5,6,7,8,11,7,3,7)&2&64&1\cr
\+(7,1)&[0,0,0,0,0,1,0,1,0,1]&(1,2,3,4,5,6,7,9,11,7,3,7)&0&65&2\cr
\+(7,1)&[1,0,0,1,0,0,0,1,0,0]&(1,1,2,3,4,6,8,10,12,8,4,7)&-2&66&4\cr
\+(7,1)&[0,0,0,0,1,0,0,0,0,3]&(1,2,3,4,5,6,8,10,12,7,2,7)&2&67&1\cr
\+(7,1)&[0,0,0,0,1,0,0,0,1,1]&(1,2,3,4,5,6,8,10,12,7,3,7)&-2&68&4\cr
\+(7,1)&[0,0,0,0,0,1,1,0,0,0]&(1,2,3,4,5,6,7,9,12,8,4,7)&-2&68&2\cr
\+(7,1)&[0,2,0,0,0,0,0,0,0,2]&(1,1,1,3,5,7,9,11,13,8,3,7)&0&69&1\cr
\+(7,1)&[1,0,1,0,0,0,0,0,0,2]&(1,1,2,3,5,7,9,11,13,8,3,7)&-2&70&5\cr
\+(7,1)&[0,2,0,0,0,0,0,0,1,0]&(1,1,1,3,5,7,9,11,13,8,4,7)&-2&70&3\cr
\+(7,1)&[0,0,0,0,1,0,0,1,0,0]&(1,2,3,4,5,6,8,10,12,8,4,7)&-4&70&7\cr
\+(7,1)&[1,0,1,0,0,0,0,0,1,0]&(1,1,2,3,5,7,9,11,13,8,4,7)&-4&71&7\cr
\+(7,1)&[0,0,0,1,0,0,0,0,0,2]&(1,2,3,4,5,7,9,11,13,8,3,7)&-4&73&7\cr
\+(7,1)&[0,0,0,1,0,0,0,0,1,0]&(1,2,3,4,5,7,9,11,13,8,4,7)&-6&74&11\cr
\+(7,1)&[1,1,0,0,0,0,0,0,0,1]&(1,1,2,4,6,8,10,12,14,9,4,7)&-6&78&8\cr
\+(7,1)&[0,0,1,0,0,0,0,0,0,1]&(1,2,3,4,6,8,10,12,14,9,4,7)&-8&80&15\cr
\+(7,1)&[2,0,0,0,0,0,0,0,0,0]&(1,1,3,5,7,9,11,13,15,10,5,7)&-8&87&2\cr
\+(7,1)&[0,1,0,0,0,0,0,0,0,0]&(1,2,3,5,7,9,11,13,15,10,5,7)&-10&88&8\cr}

\eject
{\bf B2: $E_7^{+++1}$}

\vbox{\settabs\+(4,2,1)\quad\quad\quad&[0,0,0,0,0,0,1,0,1,-1,-1]\quad&(1,2,3,4,5,6,7,5,3,2,4)\quad&-4\quad\quad&42\quad\quad&11\cr
\+$(n_c,n_d,n_0)$&$p_i$&$\beta$&$\beta^2$&$ht(\beta)$&$\mu$\cr
\hrule
\+(1,0,0)&[0,0,0,0,0,1,0,0]&(0,0,0,0,0,0,0,0,0,0,1)&2&1&1\cr
\+(2,0,0)&[0,0,1,0,0,0,0,0]&(0,0,0,0,1,2,3,2,1,0,2)&2&11&1\cr
\+(3,0,0)&[1,0,0,0,0,0,0,1]&(0,0,1,2,3,4,5,3,1,0,3)&2&22&1\cr
\+(3,0,0)&[0,0,0,0,0,0,0,0]&(0,1,2,3,4,5,6,4,2,0,3)&0&30&0\cr
\+(4,0,0)&[0,0,0,0,0,1,0,0]&(0,1,2,3,4,5,6,4,2,0,4)&2&31&1\cr
\+(5,0,0)&[0,0,1,0,0,0,0,0]&(0,1,2,3,5,7,9,6,3,0,5)&2&41&1\cr
\+(0,1,0)&[0,0,0,0,0,0,0,1]&(0,0,0,0,0,0,0,0,0,1,0)&2&1&1\cr
\+(1,1,0)&[0,0,0,0,1,0,0,0]&(0,0,0,0,0,0,1,1,1,1,1)&2&5&1\cr
\+(2,1,0)&[0,0,1,0,0,0,0,1]&(0,0,0,0,1,2,3,2,1,1,2)&2&12&1\cr
\+(2,1,0)&[0,1,0,0,0,0,0,0]&(0,0,0,1,2,3,4,3,2,1,2)&0&18&1\cr
\+(3,1,0)&[0,1,0,0,0,1,0,0]&(0,0,0,1,2,3,4,3,2,1,3)&2&19&1\cr
\+(3,1,0)&[1,0,0,0,0,0,0,2]&(0,0,1,2,3,4,5,3,1,1,3)&2&23&1\cr
\+(3,1,0)&[1,0,0,0,0,0,1,0]&(0,0,1,2,3,4,5,3,2,1,3)&0&24&1\cr
\+(3,1,0)&[0,0,0,0,0,0,0,1]&(0,1,2,3,4,5,6,4,2,1,3)&-2&31&2\cr
\+(4,1,0)&[1,0,0,0,1,0,0,1]&(0,0,1,2,3,4,6,4,2,1,4)&2&27&1\cr
\+(4,1,0)&[0,1,1,0,0,0,0,0]&(0,0,0,1,3,5,7,5,3,1,4)&2&29&1\cr
\+(4,1,0)&[1,0,0,1,0,0,0,0]&(0,0,1,2,3,5,7,5,3,1,4)&0&31&1\cr
\+(4,1,0)&[0,0,0,0,0,1,0,1]&(0,1,2,3,4,5,6,4,2,1,4)&0&32&2\cr
\+(4,1,0)&[0,0,0,0,1,0,0,0]&(0,1,2,3,4,5,7,5,3,1,4)&-2&35&2\cr
\+(5,1,0)&[1,0,1,0,0,0,1,0]&(0,0,1,2,4,6,8,5,3,1,5)&2&35&1\cr
\+(5,1,0)&[0,0,0,0,1,1,0,0]&(0,1,2,3,4,5,7,5,3,1,5)&2&36&1\cr
\+(5,1,0)&[0,0,0,1,0,0,0,2]&(0,1,2,3,4,6,8,5,2,1,5)&2&37&1\cr
\+(5,1,0)&[0,0,0,1,0,0,1,0]&(0,1,2,3,4,6,8,5,3,1,5)&0&38&1\cr
\+(5,1,0)&[1,1,0,0,0,0,0,1]&(0,0,1,3,5,7,9,6,3,1,5)&0&40&2\cr
\+(5,1,0)&[0,0,1,0,0,0,0,1]&(0,1,2,3,5,7,9,6,3,1,5)&-2&42&4\cr
\+(5,1,0)&[2,0,0,0,0,0,0,0]&(0,0,2,4,6,8,10,7,4,1,5)&-2&47&1\cr
\+(5,1,0)&[0,1,0,0,0,0,0,0]&(0,1,2,4,6,8,10,7,4,1,5)&-4&48&3\cr
\+(2,2,0)&[0,1,0,0,0,0,0,1]&(0,0,0,1,2,3,4,3,2,2,2)&2&19&1\cr
\+(2,2,0)&[1,0,0,0,0,0,0,0]&(0,0,1,2,3,4,5,4,3,2,2)&0&26&0\cr
\+(3,2,0)&[0,1,0,0,1,0,0,0]&(0,0,0,1,2,3,5,4,3,2,3)&2&23&1\cr
\+(3,2,0)&[1,0,0,0,0,0,1,1]&(0,0,1,2,3,4,5,3,2,2,3)&2&25&1\cr
\+(3,2,0)&[1,0,0,0,0,1,0,0]&(0,0,1,2,3,4,5,4,3,2,3)&0&27&1\cr
\+(3,2,0)&[0,0,0,0,0,0,0,2]&(0,1,2,3,4,5,6,4,2,2,3)&0&32&1\cr
\+(3,2,0)&[0,0,0,0,0,0,1,0]&(0,1,2,3,4,5,6,4,3,2,3)&-2&33&2\cr
\+(4,2,0)&[1,0,0,0,1,0,1,0]&(0,0,1,2,3,4,6,4,3,2,4)&2&29&1\cr
\+(4,2,0)&[0,1,1,0,0,0,0,1]&(0,0,0,1,3,5,7,5,3,2,4)&2&30&1\cr
\+(4,2,0)&[1,0,0,1,0,0,0,1]&(0,0,1,2,3,5,7,5,3,2,4)&0&32&2\cr
\+(4,2,0)&[0,0,0,0,0,1,0,2]&(0,1,2,3,4,5,6,4,2,2,4)&2&33&1\cr}
\vbox{\settabs\+(4,2,1)\quad\quad\quad&[0,0,0,0,0,0,1,0,1,-1,-1]\quad&(1,2,3,4,5,6,7,5,3,2,4)\quad&-4\quad\quad&42\quad\quad&11\cr
\+$(n_c,n_d,n_0)$&$p_i$&$\beta$&$\beta^2$&$ht(\beta)$&$\mu$\cr
\hrule
\+(4,2,0)&[0,0,0,0,0,1,1,0]&(0,1,2,3,4,5,6,4,3,2,4)&0&34&1\cr
\+(4,2,0)&[0,2,0,0,0,0,0,0]&(0,0,0,2,4,6,8,6,4,2,4)&0&36&1\cr
\+(4,2,0)&[0,0,0,0,1,0,0,1]&(0,1,2,3,4,5,7,5,3,2,4)&-2&36&4\cr
\+(4,2,0)&[1,0,1,0,0,0,0,0]&(0,0,1,2,4,6,8,6,4,2,4)&-2&37&3\cr
\+(4,2,0)&[0,0,0,1,0,0,0,0]&(0,1,2,3,4,6,8,6,4,2,4)&-4&40&3\cr
\+(0,0,1)&[1,0,0,0,0,0,0,0]&(1,0,0,0,0,0,0,0,0,0,0)&2&1&1\cr
\+(1,0,1)&[0,0,0,0,0,0,1,0]&(1,1,1,1,1,1,1,0,0,0,1)&2&8&1\cr
\+(2,0,1)&[0,0,0,1,0,0,0,0]&(1,1,1,1,1,2,3,2,1,0,2)&2&15&1\cr
\+(3,0,1)&[0,1,0,0,0,0,0,1]&(1,1,1,2,3,4,5,3,1,0,3)&2&24&1\cr
\+(3,0,1)&[1,0,0,0,0,0,0,0]&(1,1,2,3,4,5,6,4,2,0,3)&0&31&1\cr
\+(4,0,1)&[1,0,0,0,0,1,0,0]&(1,1,2,3,4,5,6,4,2,0,4)&2&32&1\cr
\+(4,0,1)&[0,0,0,0,0,0,0,2]&(1,2,3,4,5,6,7,4,1,0,4)&2&37&1\cr
\+(4,0,1)&[0,0,0,0,0,0,1,0]&(1,2,3,4,5,6,7,4,2,0,4)&0&38&1\cr
\+(5,0,1)&[0,0,0,0,1,0,0,1]&(1,2,3,4,5,6,8,5,2,0,5)&2&41&1\cr
\+(5,0,1)&[1,0,1,0,0,0,0,0]&(1,1,2,3,5,7,9,6,3,0,5)&2&42&1\cr
\+(5,0,1)&[0,0,0,1,0,0,0,0]&(1,2,3,4,5,7,9,6,3,0,5)&0&45&1\cr
\+(0,1,1)&[0,0,0,0,0,0,0,0]&(1,1,1,1,1,1,1,1,1,1,0)&2&10&1\cr
\+(1,1,1)&[0,0,0,0,0,1,0,0]&(1,1,1,1,1,1,1,1,1,1,1)&2&11&1\cr
\+(2,1,1)&[0,0,0,1,0,0,0,1]&(1,1,1,1,1,2,3,2,1,1,2)&2&16&1\cr
\+(2,1,1)&[0,0,1,0,0,0,0,0]&(1,1,1,1,2,3,4,3,2,1,2)&0&21&2\cr
\+(3,1,1)&[0,0,1,0,0,1,0,0]&(1,1,1,1,2,3,4,3,2,1,3)&2&22&1\cr
\+(3,1,1)&[0,1,0,0,0,0,0,2]&(1,1,1,2,3,4,5,3,1,1,3)&2&25&1\cr
\+(3,1,1)&[0,1,0,0,0,0,1,0]&(1,1,1,2,3,4,5,3,2,1,3)&0&26&2\cr
\+(3,1,1)&[1,0,0,0,0,0,0,1]&(1,1,2,3,4,5,6,4,2,1,3)&-2&32&4\cr
\+(3,1,1)&[0,0,0,0,0,0,0,0]&(1,2,3,4,5,6,7,5,3,1,3)&-4&40&3\cr
\+(4,1,1)&[0,1,0,0,1,0,0,1]&(1,1,1,2,3,4,6,4,2,1,4)&2&29&1\cr
\+(4,1,1)&[0,0,2,0,0,0,0,0]&(1,1,1,1,3,5,7,5,3,1,4)&2&32&1\cr
\+(4,1,1)&[1,0,0,0,0,1,0,1]&(1,1,2,3,4,5,6,4,2,1,4)&0&33&3\cr
\+(4,1,1)&[0,1,0,1,0,0,0,0]&(1,1,1,2,3,5,7,5,3,1,4)&0&33&2\cr
\+(4,1,1)&[1,0,0,0,1,0,0,0]&(1,1,2,3,4,5,7,5,3,1,4)&-2&36&4\cr
\+(4,1,1)&[0,0,0,0,0,0,0,3]&(1,2,3,4,5,6,7,4,1,1,4)&2&38&1\cr
\+(4,1,1)&[0,0,0,0,0,0,1,1]&(1,2,3,4,5,6,7,4,2,1,4)&-2&39&4\cr
\+(4,1,1)&[0,0,0,0,0,1,0,0]&(1,2,3,4,5,6,7,5,3,1,4)&-4&41&7\cr
\+(2,2,1)&[0,0,1,0,0,0,0,1]&(1,1,1,1,2,3,4,3,2,2,2)&2&22&1\cr
\+(2,2,1)&[0,1,0,0,0,0,0,0]&(1,1,1,2,3,4,5,4,3,2,2)&0&28&1\cr
\+(3,2,1)&[0,0,1,0,1,0,0,0]&(1,1,1,1,2,3,5,4,3,2,3)&2&26&1\cr
\+(3,2,1)&[0,1,0,0,0,0,1,1]&(1,1,1,2,3,4,5,3,2,2,3)&2&27&1\cr
\+(3,2,1)&[0,1,0,0,0,1,0,0]&(1,1,1,2,3,4,5,4,3,2,3)&0&29&2\cr
\+(3,2,1)&[1,0,0,0,0,0,0,2]&(1,1,2,3,4,5,6,4,2,2,3)&0&33&2\cr}
\vbox{\settabs\+(4,2,1)\quad\quad\quad&[0,0,0,0,0,0,1,0,1,-1,-1]\quad&(1,2,3,4,5,6,7,5,3,2,4)\quad&-4\quad\quad&42\quad\quad&11\cr
\+$(n_c,n_d,n_0)$&$p_i$&$\beta$&$\beta^2$&$ht(\beta)$&$\mu$\cr
\hrule
\+(3,2,1)&[1,0,0,0,0,0,1,0]&(1,1,2,3,4,5,6,4,3,2,3)&-2&34&4\cr
\+(3,2,1)&[0,0,0,0,0,0,0,1]&(1,2,3,4,5,6,7,5,3,2,3)&-4&41&5\cr
\+(4,2,1)&[0,1,0,0,1,0,1,0]&(1,1,1,2,3,4,6,4,3,2,4)&2&31&1\cr
\+(4,2,1)&[0,0,2,0,0,0,0,1]&(1,1,1,1,3,5,7,5,3,2,4)&2&33&1\cr
\+(4,2,1)&[1,0,0,0,0,1,0,2]&(1,1,2,3,4,5,6,4,2,2,4)&2&34&1\cr
\+(4,2,1)&[0,1,0,1,0,0,0,1]&(1,1,1,2,3,5,7,5,3,2,4)&0&34&3\cr
\+(4,2,1)&[1,0,0,0,0,1,1,0]&(1,1,2,3,4,5,6,4,3,2,4)&0&35&2\cr
\+(4,2,1)&[1,0,0,0,1,0,0,1]&(1,1,2,3,4,5,7,5,3,2,4)&-2&37&7\cr
\+(4,2,1)&[0,1,1,0,0,0,0,0]&(1,1,1,2,4,6,8,6,4,2,4)&-2&39&5\cr
\+(4,2,1)&[0,0,0,0,0,0,1,2]&(1,2,3,4,5,6,7,4,2,2,4)&0&40&2\cr
\+(4,2,1)&[1,0,0,1,0,0,0,0]&(1,1,2,3,4,6,8,6,4,2,4)&-4&41&8\cr
\+(4,2,1)&[0,0,0,0,0,0,2,0]&(1,2,3,4,5,6,7,4,3,2,4)&-2&41&3\cr
\+(4,2,1)&[0,0,0,0,0,1,0,1]&(1,2,3,4,5,6,7,5,3,2,4)&-4&42&11\cr
\+(4,2,1)&[0,0,0,0,1,0,0,0]&(1,2,3,4,5,6,8,6,4,2,4)&-6&45&13\cr}

\medskip
{\bf B3: $E_6^{+++1}$}

\vbox{\settabs\+(3,5,1)\quad\quad\quad&[0,1,0,0,0,0,0]\quad&(1,2,3,5,7,9,6,3,5,3)\quad&-10\quad\quad&44\quad\quad&32\quad\quad\cr
\+$(n_d,n_c,n_0)$&$p_i$&$\beta$&$\beta^2$&$ht(\beta)$&$\mu$\cr
\hrule
\+(1,0,0)&[0,0,0,0,0,0,0]&(0,0,0,0,0,0,0,0,0,1)&2&1&1\cr
\+(0,1,0)&[0,0,0,0,1,0,0]&(0,0,0,0,0,0,0,0,1,0)&2&1&1\cr
\+(1,1,0)&[0,0,0,0,1,0,0]&(0,0,0,0,0,0,0,0,1,1)&2&2&1\cr
\+(0,2,0)&[0,1,0,0,0,0,0]&(0,0,0,1,2,3,2,1,2,0)&2&11&1\cr
\+(1,2,0)&[0,0,1,0,0,0,1]&(0,0,0,0,1,2,1,0,2,1)&2&7&1\cr
\+(1,2,0)&[0,1,0,0,0,0,0]&(0,0,0,1,2,3,2,1,2,1)&0&12&1\cr
\+(2,2,0)&[0,1,0,0,0,0,0]&(0,0,0,1,2,3,2,1,2,2)&2&13&1\cr
\+(0,3,0)&[0,0,0,0,0,0,1]&(0,1,2,3,4,5,3,1,3,0)&2&22&1\cr
\+(1,3,0)&[0,1,0,0,1,0,0]&(0,0,0,1,2,3,2,1,3,1)&2&13&1\cr
\+(1,3,0)&[1,0,0,0,0,0,2]&(0,0,1,2,3,4,2,0,3,1)&2&16&1\cr
\+(1,3,0)&[1,0,0,0,0,1,0]&(0,0,1,2,3,4,2,1,3,1)&0&17&1\cr
\+(1,3,0)&[0,0,0,0,0,0,1]&(0,1,2,3,4,5,3,1,3,1)&-2&23&2\cr
\+(2,3,0)&[0,1,0,0,1,0,0]&(0,0,0,1,2,3,2,1,3,2)&2&14&1\cr
\+(2,3,0)&[1,0,0,0,0,0,2]&(0,0,1,2,3,4,2,0,3,2)&2&17&1\cr
\+(2,3,0)&[1,0,0,0,0,1,0]&(0,0,1,2,3,4,2,1,3,2)&0&18&1\cr
\+(2,3,0)&[0,0,0,0,0,0,1]&(0,1,2,3,4,5,3,1,3,2)&-2&24&2\cr
\+(3,3,0)&[0,0,0,0,0,0,1]&(0,1,2,3,4,5,3,1,3,3)&2&25&1\cr}
\vbox{\settabs\+(3,5,1)\quad\quad\quad&[0,1,0,0,0,0,0]\quad&(1,2,3,5,7,9,6,3,5,3)\quad&-10\quad\quad&44\quad\quad&32\quad\quad\cr
\+$(n_d,n_c,n_0)$&$p_i$&$\beta$&$\beta^2$&$ht(\beta)$&$\mu$\cr
\hrule
\+(1,4,0)&[1,0,0,1,0,0,1]&(0,0,1,2,3,5,3,1,4,1)&2&20&1\cr
\+(1,4,0)&[0,2,0,0,0,0,0]&(0,0,0,2,4,6,4,2,4,1)&2&23&1\cr
\+(1,4,0)&[1,0,1,0,0,0,0]&(0,0,1,2,4,6,4,2,4,1)&0&24&1\cr
\+(1,4,0)&[0,0,0,0,1,0,1]&(0,1,2,3,4,5,3,1,4,1)&0&24&2\cr
\+(1,4,0)&[0,0,0,1,0,0,0]&(0,1,2,3,4,6,4,2,4,1)&-2&27&2\cr
\+(2,4,0)&[1,0,0,0,1,1,0]&(0,0,1,2,3,4,2,1,4,2)&2&19&1\cr
\+(2,4,0)&[0,1,1,0,0,0,1]&(0,0,0,1,3,5,3,1,4,2)&2&19&1\cr
\+(2,4,0)&[1,0,0,1,0,0,1]&(0,0,1,2,3,5,3,1,4,2)&0&21&2\cr
\+(2,4,0)&[0,0,0,0,0,1,2]&(0,1,2,3,4,5,2,0,4,2)&2&23&1\cr
\+(2,4,0)&[0,2,0,0,0,0,0]&(0,0,0,2,4,6,4,2,4,2)&0&24&1\cr
\+(2,4,0)&[0,0,0,0,0,2,0]&(0,1,2,3,4,5,2,1,4,2)&0&24&0\cr
\+(2,4,0)&[1,0,1,0,0,0,0]&(0,0,1,2,4,6,4,2,4,2)&-2&25&3\cr
\+(2,4,0)&[0,0,0,0,1,0,1]&(0,1,2,3,4,5,3,1,4,2)&-2&25&4\cr
\+(2,4,0)&[0,0,0,1,0,0,0]&(0,1,2,3,4,6,4,2,4,2)&-4&28&3\cr
\+(3,4,0)&[1,0,0,1,0,0,1]&(0,0,1,2,3,5,3,1,4,3)&2&22&1\cr
\+(3,4,0)&[0,2,0,0,0,0,0]&(0,0,0,2,4,6,4,2,4,3)&2&25&1\cr
\+(3,4,0)&[1,0,1,0,0,0,0]&(0,0,1,2,4,6,4,2,4,3)&0&26&1\cr
\+(3,4,0)&[0,0,0,0,1,0,1]&(0,1,2,3,4,5,3,1,4,3)&0&26&2\cr
\+(3,4,0)&[0,0,0,1,0,0,0]&(0,1,2,3,4,6,4,2,4,3)&-2&29&2\cr
\+(1,5,0)&[1,1,0,0,0,1,0]&(0,0,1,3,5,7,4,2,5,1)&2&28&1\cr
\+(1,5,0)&[0,0,0,1,1,0,0]&(0,1,2,3,4,6,4,2,5,1)&2&28&1\cr
\+(1,5,0)&[0,0,1,0,0,0,2]&(0,1,2,3,5,7,4,1,5,1)&2&29&1\cr
\+(1,5,0)&[0,0,1,0,0,1,0]&(0,1,2,3,5,7,4,2,5,1)&0&30&1\cr
\+(1,5,0)&[2,0,0,0,0,0,1]&(0,0,2,4,6,8,5,2,5,1)&0&33&1\cr
\+(1,5,0)&[0,1,0,0,0,0,1]&(0,1,2,4,6,8,5,2,5,1)&-2&34&4\cr
\+(1,5,0)&[1,0,0,0,0,0,0]&(0,1,3,5,7,9,6,3,5,1)&-4&40&2\cr
\+(2,5,0)&[1,0,1,0,0,1,1]&(0,0,1,2,4,6,3,1,5,2)&2&24&1\cr
\+(2,5,0)&[1,0,0,2,0,0,0]&(0,0,1,2,3,6,4,2,5,2)&2&25&1\cr
\+(2,5,0)&[0,2,0,0,1,0,0]&(0,0,0,2,4,6,4,2,5,2)&2&25&1\cr
\+(2,5,0)&[1,0,1,0,1,0,0]&(0,0,1,2,4,6,4,2,5,2)&0&26&2\cr
\+(2,5,0)&[0,0,0,0,2,0,1]&(0,1,2,3,4,5,3,1,5,2)&2&26&1\cr
\+(2,5,0)&[0,0,0,1,0,1,1]&(0,1,2,3,4,6,3,1,5,2)&0&27&2\cr
\+(2,5,0)&[1,1,0,0,0,0,2]&(0,0,1,3,5,7,4,1,5,2)&0&28&2\cr
\+(2,5,0)&[1,1,0,0,0,1,0]&(0,0,1,3,5,7,4,2,5,2)&-2&29&5\cr
\+(2,5,0)&[0,0,0,1,1,0,0]&(0,1,2,3,4,6,4,2,5,2)&-2&29&4\cr
\+(2,5,0)&[0,0,1,0,0,0,2]&(0,1,2,3,5,7,4,1,5,2)&-2&30&5\cr
\+(2,5,0)&[0,0,1,0,0,1,0]&(0,1,2,3,5,7,4,2,5,2)&-4&31&7\cr
\+(2,5,0)&[2,0,0,0,0,0,1]&(0,0,2,4,6,8,5,2,5,2)&-4&34&4\cr
\+(2,5,0)&[0,1,0,0,0,0,1]&(0,1,2,4,6,8,5,2,5,2)&-6&35&13\cr
\+(2,5,0)&[1,0,0,0,0,0,0]&(0,1,3,5,7,9,6,3,5,2)&-8&41&6\cr
\+(3,5,0)&[1,0,1,0,0,1,1]&(0,0,1,2,4,6,3,1,5,3)&2&25&1\cr
\+(3,5,0)&[1,0,0,2,0,0,0]&(0,0,1,2,3,6,4,2,5,3)&2&26&1\cr
\+(3,5,0)&[0,2,0,0,1,0,0]&(0,0,0,2,4,6,4,2,5,3)&2&26&1\cr
\+(3,5,0)&[1,0,1,0,1,0,0]&(0,0,1,2,4,6,4,2,5,3)&0&27&2\cr}
\vbox{\settabs\+(3,5,1)\quad\quad\quad&[0,1,0,0,0,0,0]\quad&(1,2,3,5,7,9,6,3,5,3)\quad&-10\quad\quad&44\quad\quad&32\quad\quad\cr
\+$(n_d,n_c,n_0)$&$p_i$&$\beta$&$\beta^2$&$ht(\beta)$&$\mu$\cr
\hrule
\+(3,5,0)&[0,0,0,0,2,0,1]&(0,1,2,3,4,5,3,1,5,3)&2&27&1\cr
\+(3,5,0)&[0,0,0,1,0,1,1]&(0,1,2,3,4,6,3,1,5,3)&0&28&2\cr
\+(3,5,0)&[1,1,0,0,0,0,2]&(0,0,1,3,5,7,4,1,5,3)&0&29&2\cr
\+(3,5,0)&[1,1,0,0,0,1,0]&(0,0,1,3,5,7,4,2,5,3)&-2&30&5\cr
\+(3,5,0)&[0,0,0,1,1,0,0]&(0,1,2,3,4,6,4,2,5,3)&-2&30&4\cr
\+(3,5,0)&[0,0,1,0,0,0,2]&(0,1,2,3,5,7,4,1,5,3)&-2&31&5\cr
\+(3,5,0)&[0,0,1,0,0,1,0]&(0,1,2,3,5,7,4,2,5,3)&-4&32&7\cr
\+(3,5,0)&[2,0,0,0,0,0,1]&(0,0,2,4,6,8,5,2,5,3)&-4&35&4\cr
\+(3,5,0)&[0,1,0,0,0,0,1]&(0,1,2,4,6,8,5,2,5,3)&-6&36&13\cr
\+(3,5,0)&[1,0,0,0,0,0,0]&(0,1,3,5,7,9,6,3,5,3)&-8&42&6\cr
\+(4,5,0)&[1,1,0,0,0,1,0]&(0,0,1,3,5,7,4,2,5,4)&2&31&1\cr
\+(4,5,0)&[0,0,0,1,1,0,0]&(0,1,2,3,4,6,4,2,5,4)&2&31&1\cr
\+(4,5,0)&[0,0,1,0,0,0,2]&(0,1,2,3,5,7,4,1,5,4)&2&32&1\cr
\+(4,5,0)&[0,0,1,0,0,1,0]&(0,1,2,3,5,7,4,2,5,4)&0&33&1\cr
\+(4,5,0)&[2,0,0,0,0,0,1]&(0,0,2,4,6,8,5,2,5,4)&0&36&1\cr
\+(4,5,0)&[0,1,0,0,0,0,1]&(0,1,2,4,6,8,5,2,5,4)&-2&37&4\cr
\+(4,5,0)&[1,0,0,0,0,0,0]&(0,1,3,5,7,9,6,3,5,4)&-4&43&2\cr
\+(0,0,1)&[1,0,0,0,0,0,0]&(1,0,0,0,0,0,0,0,0,0)&2&1&1\cr
\+(0,1,1)&[0,0,0,0,0,1,0]&(1,1,1,1,1,1,0,0,1,0)&2&7&1\cr
\+(1,1,1)&[0,0,0,0,0,1,0]&(1,1,1,1,1,1,0,0,1,1)&2&8&1\cr
\+(0,2,1)&[0,0,1,0,0,0,0]&(1,1,1,1,2,3,2,1,2,0)&2&14&1\cr
\+(1,2,1)&[0,0,0,1,0,0,1]&(1,1,1,1,1,2,1,0,2,1)&2&11&1\cr
\+(1,2,1)&[0,0,1,0,0,0,0]&(1,1,1,1,2,3,2,1,2,1)&0&15&2\cr
\+(2,2,1)&[0,0,1,0,0,0,0]&(1,1,1,1,2,3,2,1,2,2)&2&16&1\cr
\+(0,3,1)&[1,0,0,0,0,0,1]&(1,1,2,3,4,5,3,1,3,0)&2&23&1\cr
\+(0,3,1)&[0,0,0,0,0,0,0]&(1,2,3,4,5,6,4,2,3,0)&0&30&1\cr
\+(1,3,1)&[0,0,1,0,1,0,0]&(1,1,1,1,2,3,2,1,3,1)&2&16&1\cr
\+(1,3,1)&[0,1,0,0,0,0,2]&(1,1,1,2,3,4,2,0,3,1)&2&18&1\cr
\+(1,3,1)&[0,1,0,0,0,1,0]&(1,1,1,2,3,4,2,1,3,1)&0&19&2\cr
\+(1,3,1)&[1,0,0,0,0,0,1]&(1,1,2,3,4,5,3,1,3,1)&-2&24&4\cr
\+(1,3,1)&[0,0,0,0,0,0,0]&(1,2,3,4,5,6,4,2,3,1)&-4&31&3\cr
\+(2,3,1)&[0,0,1,0,1,0,0]&(1,1,1,1,2,3,2,1,3,2)&2&17&1\cr
\+(2,3,1)&[0,1,0,0,0,0,2]&(1,1,1,2,3,4,2,0,3,2)&2&19&1\cr
\+(2,3,1)&[0,1,0,0,0,1,0]&(1,1,1,2,3,4,2,1,3,2)&0&20&2\cr
\+(2,3,1)&[1,0,0,0,0,0,1]&(1,1,2,3,4,5,3,1,3,2)&-2&25&4\cr
\+(2,3,1)&[0,0,0,0,0,0,0]&(1,2,3,4,5,6,4,2,3,2)&-4&32&3\cr
\+(3,3,1)&[1,0,0,0,0,0,1]&(1,1,2,3,4,5,3,1,3,3)&2&26&1\cr
\+(3,3,1)&[0,0,0,0,0,0,0]&(1,2,3,4,5,6,4,2,3,3)&0&33&1\cr
\+(0,4,1)&[0,0,0,0,1,0,0]&(1,2,3,4,5,6,4,2,4,0)&2&31&1\cr
\+(1,4,1)&[0,1,0,1,0,0,1]&(1,1,1,2,3,5,3,1,4,1)&2&22&1\cr
\+(1,4,1)&[1,0,0,0,1,0,1]&(1,1,2,3,4,5,3,1,4,1)&0&25&3\cr
\+(1,4,1)&[0,1,1,0,0,0,0]&(1,1,1,2,4,6,4,2,4,1)&0&26&2\cr
\+(1,4,1)&[1,0,0,1,0,0,0]&(1,1,2,3,4,6,4,2,4,1)&-2&28&4\cr}
\vbox{\settabs\+(3,5,1)\quad\quad\quad&[0,1,0,0,0,0,0]\quad&(1,2,3,5,7,9,6,3,5,3)\quad&-10\quad\quad&44\quad\quad&32\quad\quad\cr
\+$(n_d,n_c,n_0)$&$p_i$&$\beta$&$\beta^2$&$ht(\beta)$&$\mu$\cr
\hrule
\+(1,4,1)&[0,0,0,0,0,0,3]&(1,2,3,4,5,6,3,0,4,1)&2&29&1\cr
\+(1,4,1)&[0,0,0,0,0,1,1]&(1,2,3,4,5,6,3,1,4,1)&-2&30&4\cr
\+(1,4,1)&[0,0,0,0,1,0,0]&(1,2,3,4,5,6,4,2,4,1)&-4&32&7\cr
\+(2,4,1)&[0,1,0,0,1,1,0]&(1,1,1,2,3,4,2,1,4,2)&2&21&1\cr
\+(2,4,1)&[0,0,2,0,0,0,1]&(1,1,1,1,3,5,3,1,4,2)&2&22&1\cr
\+(2,4,1)&[0,1,0,1,0,0,1]&(1,1,1,2,3,5,3,1,4,2)&0&23&3\cr
\+(2,4,1)&[1,0,0,0,0,1,2]&(1,1,2,3,4,5,2,0,4,2)&2&24&1\cr
\+(2,4,1)&[1,0,0,0,0,2,0]&(1,1,2,3,4,5,2,1,4,2)&0&25&1\cr
\+(2,4,1)&[1,0,0,0,1,0,1]&(1,1,2,3,4,5,3,1,4,2)&-2&26&7\cr
\+(2,4,1)&[0,1,1,0,0,0,0]&(1,1,1,2,4,6,4,2,4,2)&-2&27&5\cr
\+(2,4,1)&[1,0,0,1,0,0,0]&(1,1,2,3,4,6,4,2,4,2)&-4&29&8\cr
\+(2,4,1)&[0,0,0,0,0,0,3]&(1,2,3,4,5,6,3,0,4,2)&0&30&2\cr
\+(2,4,1)&[0,0,0,0,0,1,1]&(1,2,3,4,5,6,3,1,4,2)&-4&31&9\cr
\+(2,4,1)&[0,0,0,0,1,0,0]&(1,2,3,4,5,6,4,2,4,2)&-6&33&13\cr
\+(3,4,1)&[0,1,0,1,0,0,1]&(1,1,1,2,3,5,3,1,4,3)&2&24&1\cr
\+(3,4,1)&[1,0,0,0,1,0,1]&(1,1,2,3,4,5,3,1,4,3)&0&27&3\cr
\+(3,4,1)&[0,1,1,0,0,0,0]&(1,1,1,2,4,6,4,2,4,3)&0&28&2\cr
\+(3,4,1)&[1,0,0,1,0,0,0]&(1,1,2,3,4,6,4,2,4,3)&-2&30&4\cr
\+(3,4,1)&[0,0,0,0,0,0,3]&(1,2,3,4,5,6,3,0,4,3)&2&31&1\cr
\+(3,4,1)&[0,0,0,0,0,1,1]&(1,2,3,4,5,6,3,1,4,3)&-2&32&4\cr
\+(3,4,1)&[0,0,0,0,1,0,0]&(1,2,3,4,5,6,4,2,4,3)&-4&34&7\cr
\+(4,4,1)&[0,0,0,0,1,0,0]&(1,2,3,4,5,6,4,2,4,4)&2&35&1\cr}

\medskip
{\bf B4: $G_2^{+++1}$}

\vbox{\settabs\+(13,1)\quad\quad&[1,0,0,1,0,4]\quad&(1,3,5,7,10,13)\quad&-76\quad\quad&39\quad\quad&330\cr
\+$(n_c,n_0)$&$p_i$&$\beta$&$\beta^2$&$ht(\beta)$&$\mu$\cr
\hrule
\+(1,0)&[0,0,0,1]&(0,0,0,0,0,1)&2/3&1&1\cr
\+(2,0)&[0,0,1,0]&(0,0,0,0,1,2)&2/3&3&1\cr
\+(3,0)&[0,0,1,1]&(0,0,0,0,1,3)&2&4&1\cr
\+(3,0)&[0,1,0,0]&(0,0,0,1,2,3)&0&6&0\cr
\+(4,0)&[0,1,0,1]&(0,0,0,1,2,4)&2/3&7&1\cr
\+(4,0)&[1,0,0,0]&(0,0,1,2,3,4)&-4/3&10&0\cr
\+(5,0)&[0,1,1,0]&(0,0,0,1,3,5)&2/3&9&1\cr
\+(5,0)&[1,0,0,1]&(0,0,1,2,3,5)&-4/3&11&1\cr
\+(5,0)&[0,0,0,0]&(0,1,2,3,4,5)&-10/3&15&0\cr
\+(6,0)&[0,1,1,1]&(0,0,0,1,3,6)&2&10&1\cr
\+(6,0)&[1,0,0,2]&(0,0,1,2,3,6)&0&12&0\cr
\+(6,0)&[0,2,0,0]&(0,0,0,2,4,6)&0&12&0\cr
\+(6,0)&[1,0,1,0]&(0,0,1,2,4,6)&-2&13&2\cr
\+(6,0)&[0,0,0,1]&(0,1,2,3,4,6)&-4&16&1\cr
\+(7,0)&[0,2,0,1]&(0,0,0,2,4,7)&2/3&13&1\cr
\+(7,0)&[1,0,1,1]&(0,0,1,2,4,7)&-4/3&14&2\cr
\+(7,0)&[1,1,0,0]&(0,0,1,3,5,7)&-10/3&16&2\cr
\+(7,0)&[0,0,0,2]&(0,1,2,3,4,7)&-10/3&17&1\cr
\+(7,0)&[0,0,1,0]&(0,1,2,3,5,7)&-16/3&18&2\cr}
\vbox{\settabs\+(13,1)\quad\quad&[1,0,0,1,0,4]\quad&(1,3,5,7,10,13)\quad&-76\quad\quad&39\quad\quad&330\cr
\+$(n_c,n_0)$&$p_i$&$\beta$&$\beta^2$&$ht(\beta)$&$\mu$\cr
\hrule
\+(0,1)&[1,0,0,0]&(1,0,0,0,0,0)&2&1&1\cr
\+(1,1)&[0,0,0,0]&(1,1,1,1,1,1)&2/3&6&1\cr
\+(2,1)&[0,0,0,1]&(1,1,1,1,1,2)&2/3&7&1\cr
\+(3,1)&[0,0,0,2]&(1,1,1,1,1,3)&2&8&1\cr
\+(3,1)&[0,0,1,0]&(1,1,1,1,2,3)&0&9&1\cr
\+(4,1)&[0,0,1,1]&(1,1,1,1,2,4)&2/3&10&1\cr
\+(4,1)&[0,1,0,0]&(1,1,1,2,3,4)&-4/3&12&1\cr
\+(5,1)&[0,0,2,0]&(1,1,1,1,3,5)&2/3&12&1\cr
\+(5,1)&[0,1,0,1]&(1,1,1,2,3,5)&-4/3&13&2\cr
\+(5,1)&[1,0,0,0]&(1,1,2,3,4,5)&-10/3&16&1\cr
\+(6,1)&[0,0,2,1]&(1,1,1,1,3,6)&2&13&1\cr
\+(6,1)&[0,1,0,2]&(1,1,1,2,3,6)&0&14&1\cr
\+(6,1)&[0,1,1,0]&(1,1,1,2,4,6)&-2&15&3\cr
\+(6,1)&[1,0,0,1]&(1,1,2,3,4,6)&-4&17&3\cr
\+(6,1)&[0,0,0,0]&(1,2,3,4,5,6)&-6&21&1\cr
\+(7,1)&[0,1,1,1]&(1,1,1,2,4,7)&-4/3&16&3\cr
\+(7,1)&[1,0,0,2]&(1,1,2,3,4,7)&-10/3&18&3\cr
\+(7,1)&[0,2,0,0]&(1,1,1,3,5,7)&-10/3&18&3\cr
\+(7,1)&[1,0,1,0]&(1,1,2,3,5,7)&-16/3&19&6\cr
\+(7,1)&[0,0,0,1]&(1,2,3,4,5,7)&-22/3&22&4\cr}

\medskip
{\bf B5: $F_4^{+++1}$}

\vbox{\settabs\+(6,10,1)\quad\quad\quad&[0,0,1,0,0]\quad&(1,2,3,4,6,8,10,6)\quad&-8\quad\quad&40\quad\quad&64\quad\quad\cr
\+$(n_c,n_d,n_0)$&$p_i$&$\beta$&$\beta^2$&$ht(\beta)$&$\mu$\cr
\hrule
\+(1,0,0)&[0,0,0,0,0]&(0,0,0,0,0,0,0,1)&1&1&1\cr
\+(0,1,0)&[0,0,0,0,1]&(0,0,0,0,0,0,1,0)&1&1&1\cr
\+(1,1,0)&[0,0,0,0,1]&(0,0,0,0,0,0,1,1)&1&2&1\cr
\+(0,2,0)&[0,0,0,1,0]&(0,0,0,0,0,1,2,0)&2&3&1\cr
\+(1,2,0)&[0,0,0,1,0]&(0,0,0,0,0,1,2,1)&1&4&1\cr
\+(2,2,0)&[0,0,0,1,0]&(0,0,0,0,0,1,2,2)&2&5&1\cr
\+(1,3,0)&[0,0,1,0,0]&(0,0,0,0,1,2,3,1)&1&7&1\cr
\+(2,3,0)&[0,0,1,0,0]&(0,0,0,0,1,2,3,2)&1&8&1\cr
\+(1,4,0)&[0,1,0,0,0]&(0,0,0,1,2,3,4,1)&1&11&1\cr
\+(2,4,0)&[0,0,1,0,1]&(0,0,0,0,1,2,4,2)&2&9&1\cr
\+(2,4,0)&[0,1,0,0,0]&(0,0,0,1,2,3,4,2)&0&12&1\cr
\+(3,4,0)&[0,1,0,0,0]&(0,0,0,1,2,3,4,3)&1&13&1\cr
\+(1,5,0)&[1,0,0,0,0]&(0,0,1,2,3,4,5,1)&1&16&1\cr
\+(2,5,0)&[0,1,0,0,1]&(0,0,0,1,2,3,5,2)&1&13&1\cr
\+(2,5,0)&[1,0,0,0,0]&(0,0,1,2,3,4,5,2)&-1&17&2\cr
\+(3,5,0)&[0,1,0,0,1]&(0,0,0,1,2,3,5,3)&1&14&1\cr
\+(3,5,0)&[1,0,0,0,0]&(0,0,1,2,3,4,5,3)&-1&18&2\cr
\+(4,5,0)&[1,0,0,0,0]&(0,0,1,2,3,4,5,4)&1&19&1\cr}
\vbox{\settabs\+(6,10,1)\quad\quad\quad&[0,0,1,0,0]\quad&(1,2,3,4,6,8,10,6)\quad&-8\quad\quad&40\quad\quad&64\quad\quad\cr
\+$(n_c,n_d,n_0)$&$p_i$&$\beta$&$\beta^2$&$ht(\beta)$&$\mu$\cr
\hrule
\+(1,6,0)&[0,0,0,0,0]&(0,1,2,3,4,5,6,1)&1&22&1\cr
\+(2,6,0)&[0,1,0,1,0]&(0,0,0,1,2,4,6,2)&2&15&1\cr
\+(2,6,0)&[1,0,0,0,1]&(0,0,1,2,3,4,6,2)&0&18&1\cr
\+(2,6,0)&[0,0,0,0,0]&(0,1,2,3,4,5,6,2)&-2&23&3\cr
\+(3,6,0)&[0,1,0,1,0]&(0,0,0,1,2,4,6,3)&1&16&1\cr
\+(3,6,0)&[1,0,0,0,1]&(0,0,1,2,3,4,6,3)&-1&19&3\cr
\+(3,6,0)&[0,0,0,0,0]&(0,1,2,3,4,5,6,3)&-3&24&3\cr
\+(4,6,0)&[0,1,0,1,0]&(0,0,0,1,2,4,6,4)&2&17&1\cr
\+(4,6,0)&[1,0,0,0,1]&(0,0,1,2,3,4,6,4)&0&20&1\cr
\+(4,6,0)&[0,0,0,0,0]&(0,1,2,3,4,5,6,4)&-2&25&3\cr
\+(5,6,0)&[0,0,0,0,0]&(0,1,2,3,4,5,6,5)&1&26&1\cr
\+(0,0,1)&[1,0,0,0,0]&(1,0,0,0,0,0,0,0)&2&1&1\cr
\+(0,1,1)&[0,0,0,0,0]&(1,1,1,1,1,1,1,0)&1&7&1\cr
\+(1,1,1)&[0,0,0,0,0]&(1,1,1,1,1,1,1,1)&1&8&1\cr
\+(0,2,1)&[0,0,0,0,1]&(1,1,1,1,1,1,2,0)&2&8&1\cr
\+(1,2,1)&[0,0,0,0,1]&(1,1,1,1,1,1,2,1)&1&9&1\cr
\+(2,2,1)&[0,0,0,0,1]&(1,1,1,1,1,1,2,2)&2&10&1\cr
\+(1,3,1)&[0,0,0,1,0]&(1,1,1,1,1,2,3,1)&1&11&1\cr
\+(2,3,1)&[0,0,0,1,0]&(1,1,1,1,1,2,3,2)&1&12&1\cr
\+(1,4,1)&[0,0,1,0,0]&(1,1,1,1,2,3,4,1)&1&14&1\cr
\+(2,4,1)&[0,0,0,1,1]&(1,1,1,1,1,2,4,2)&2&13&1\cr
\+(2,4,1)&[0,0,1,0,0]&(1,1,1,1,2,3,4,2)&0&15&2\cr
\+(3,4,1)&[0,0,1,0,0]&(1,1,1,1,2,3,4,3)&1&16&1\cr
\+(1,5,1)&[0,1,0,0,0]&(1,1,1,2,3,4,5,1)&1&18&1\cr
\+(2,5,1)&[0,0,1,0,1]&(1,1,1,1,2,3,5,2)&1&16&1\cr
\+(2,5,1)&[0,1,0,0,0]&(1,1,1,2,3,4,5,2)&-1&19&3\cr
\+(3,5,1)&[0,0,1,0,1]&(1,1,1,1,2,3,5,3)&1&17&1\cr
\+(3,5,1)&[0,1,0,0,0]&(1,1,1,2,3,4,5,3)&-1&20&3\cr
\+(4,5,1)&[0,1,0,0,0]&(1,1,1,2,3,4,5,4)&1&21&1\cr
\+(1,6,1)&[1,0,0,0,0]&(1,1,2,3,4,5,6,1)&1&23&1\cr
\+(2,6,1)&[0,0,1,1,0]&(1,1,1,1,2,4,6,2)&2&18&1\cr
\+(2,6,1)&[0,1,0,0,1]&(1,1,1,2,3,4,6,2)&0&20&2\cr
\+(2,6,1)&[1,0,0,0,0]&(1,1,2,3,4,5,6,2)&-2&24&4\cr
\+(3,6,1)&[0,0,1,1,0]&(1,1,1,1,2,4,6,3)&1&19&1\cr
\+(3,6,1)&[0,1,0,0,1]&(1,1,1,2,3,4,6,3)&-1&21&4\cr
\+(3,6,1)&[1,0,0,0,0]&(1,1,2,3,4,5,6,3)&-3&25&6\cr
\+(4,6,1)&[0,0,1,1,0]&(1,1,1,1,2,4,6,4)&2&20&1\cr
\+(4,6,1)&[0,1,0,0,1]&(1,1,1,2,3,4,6,4)&0&22&2\cr
\+(4,6,1)&[1,0,0,0,0]&(1,1,2,3,4,5,6,4)&-2&26&4\cr
\+(5,6,1)&[1,0,0,0,0]&(1,1,2,3,4,5,6,5)&1&27&1\cr}

\eject
{\bf B6: $D_8^{+++1}$}

\vbox{\settabs\+(2,2,2)\quad\quad\quad&[-1,0,0,0,0,0,0,1,0,1,-1,2]\quad&(2,3,4,5,6,5,4,3,3,1,2,2)\quad&-4\quad\quad&40\quad\quad&13\cr
\+$(n_c,n_d,n_0)$&$p_i$&$\beta$&$\beta^2$&$ht(\beta)$&$\mu$\cr
\hrule
\+(1,0,0)&[0,0,0,1,0,0,0,0,0]&(0,0,0,0,0,0,0,0,0,0,0,1)&2&1&1\cr
\+(2,0,0)&[1,0,0,0,0,0,1,0,0]&(0,0,1,2,3,2,1,0,0,0,0,2)&2&11&1\cr
\+(2,0,0)&[0,0,0,0,0,0,0,1,0]&(0,1,2,3,4,3,2,1,0,0,0,2)&0&18&0\cr
\+(0,1,0)&[0,0,0,0,0,0,0,1,0]&(0,0,0,0,0,0,0,0,0,0,1,0)&2&1&1\cr
\+(1,1,0)&[0,0,1,0,0,0,0,0,1]&(0,0,0,0,1,1,1,1,1,0,1,1)&2&7&1\cr
\+(1,1,0)&[0,1,0,0,0,0,0,0,0]&(0,0,0,1,2,2,2,2,2,1,1,1)&0&14&1\cr
\+(2,1,0)&[1,0,0,0,0,1,0,0,1]&(0,0,1,2,3,2,1,1,1,0,1,2)&2&14&1\cr
\+(2,1,0)&[0,1,0,1,0,0,0,0,0]&(0,0,0,1,2,2,2,2,2,1,1,2)&2&15&1\cr
\+(2,1,0)&[1,0,0,0,1,0,0,0,0]&(0,0,1,2,3,2,2,2,2,1,1,2)&0&18&1\cr
\+(2,1,0)&[0,0,0,0,0,0,1,0,1]&(0,1,2,3,4,3,2,1,1,0,1,2)&0&20&1\cr
\+(2,1,0)&[0,0,0,0,0,1,0,0,0]&(0,1,2,3,4,3,2,2,2,1,1,2)&-2&23&2\cr
\+(1,2,0)&[0,1,0,0,0,0,0,1,0]&(0,0,0,1,2,2,2,2,2,1,2,1)&2&15&1\cr
\+(1,2,0)&[1,0,0,0,0,0,0,0,1]&(0,0,1,2,3,3,3,3,3,1,2,1)&0&22&1\cr
\+(1,2,0)&[0,0,0,0,0,0,0,0,0]&(0,1,2,3,4,4,4,4,4,2,2,1)&-2&31&1\cr
\+(2,2,0)&[1,0,0,0,1,0,0,1,0]&(0,0,1,2,3,2,2,2,2,1,2,2)&2&19&1\cr
\+(2,2,0)&[0,1,1,0,0,0,0,0,1]&(0,0,0,1,3,3,3,3,3,1,2,2)&2&21&1\cr
\+(2,2,0)&[1,0,0,1,0,0,0,0,1]&(0,0,1,2,3,3,3,3,3,1,2,2)&0&23&2\cr
\+(2,2,0)&[0,0,0,0,0,1,0,0,2]&(0,1,2,3,4,3,2,2,2,0,2,2)&2&23&1\cr
\+(2,2,0)&[0,0,0,0,0,1,0,1,0]&(0,1,2,3,4,3,2,2,2,1,2,2)&0&24&1\cr
\+(2,2,0)&[0,0,0,0,1,0,0,0,1]&(0,1,2,3,4,3,3,3,3,1,2,2)&-2&27&3\cr
\+(2,2,0)&[0,2,0,0,0,0,0,0,0]&(0,0,0,2,4,4,4,4,4,2,2,2)&0&28&1\cr
\+(2,2,0)&[1,0,1,0,0,0,0,0,0]&(0,0,1,2,4,4,4,4,4,2,2,2)&-2&29&3\cr
\+(2,2,0)&[0,0,0,1,0,0,0,0,0]&(0,1,2,3,4,4,4,4,4,2,2,2)&-4&32&3\cr
\+(1,3,0)&[1,0,0,0,0,0,1,0,0]&(0,0,1,2,3,3,3,3,4,2,3,1)&2&25&1\cr
\+(1,3,0)&[0,0,0,0,0,0,0,1,0]&(0,1,2,3,4,4,4,4,4,2,3,1)&0&32&1\cr
\+(2,3,0)&[1,0,0,1,0,0,1,0,0]&(0,0,1,2,3,3,3,3,4,2,3,2)&2&26&1\cr
\+(2,3,0)&[0,0,0,0,1,0,0,1,1]&(0,1,2,3,4,3,3,3,3,1,3,2)&2&28&1\cr
\+(2,3,0)&[1,0,1,0,0,0,0,0,2]&(0,0,1,2,4,4,4,4,4,1,3,2)&2&29&1\cr
\+(2,3,0)&[0,2,0,0,0,0,0,1,0]&(0,0,0,2,4,4,4,4,4,2,3,2)&2&29&1\cr
\+(2,3,0)&[1,0,1,0,0,0,0,1,0]&(0,0,1,2,4,4,4,4,4,2,3,2)&0&30&2\cr
\+(2,3,0)&[0,0,0,0,1,0,1,0,0]&(0,1,2,3,4,3,3,3,4,2,3,2)&0&30&1\cr
\+(2,3,0)&[0,0,0,1,0,0,0,0,2]&(0,1,2,3,4,4,4,4,4,1,3,2)&0&32&1\cr
\+(2,3,0)&[0,0,0,1,0,0,0,1,0]&(0,1,2,3,4,4,4,4,4,2,3,2)&-2&33&4\cr
\+(2,3,0)&[1,1,0,0,0,0,0,0,1]&(0,0,1,3,5,5,5,5,5,2,3,2)&-2&36&4\cr
\+(2,3,0)&[0,0,1,0,0,0,0,0,1]&(0,1,2,3,5,5,5,5,5,2,3,2)&-4&38&6\cr
\+(2,3,0)&[2,0,0,0,0,0,0,0,0]&(0,0,2,4,6,6,6,6,6,3,3,2)&-4&44&2\cr
\+(2,3,0)&[0,1,0,0,0,0,0,0,0]&(0,1,2,4,6,6,6,6,6,3,3,2)&-6&45&6\cr
\+(0,0,1)&[1,0,0,0,0,0,0,0,0]&(1,0,0,0,0,0,0,0,0,0,0,0)&2&1&1\cr
\+(1,0,1)&[0,0,0,0,1,0,0,0,0]&(1,1,1,1,1,0,0,0,0,0,0,1)&2&6&1\cr
\+(2,0,1)&[0,1,0,0,0,0,1,0,0]&(1,1,1,2,3,2,1,0,0,0,0,2)&2&13&1\cr
\+(2,0,1)&[1,0,0,0,0,0,0,1,0]&(1,1,2,3,4,3,2,1,0,0,0,2)&0&19&1\cr
\+(2,0,1)&[0,0,0,0,0,0,0,0,1]&(1,2,3,4,5,4,3,2,1,0,0,2)&-2&27&1\cr}
\vbox{\settabs\+(2,2,2)\quad\quad\quad&[-1,0,0,0,0,0,0,1,0,1,-1,2]\quad&(2,3,4,5,6,5,4,3,3,1,2,2)\quad&-4\quad\quad&40\quad\quad&13\cr
\+$(n_c,n_d,n_0)$&$p_i$&$\beta$&$\beta^2$&$ht(\beta)$&$\mu$\cr
\hrule
\+(0,1,1)&[0,0,0,0,0,0,0,0,1]&(1,1,1,1,1,1,1,1,1,0,1,0)&2&10&1\cr
\+(1,1,1)&[0,0,0,1,0,0,0,0,1]&(1,1,1,1,1,1,1,1,1,0,1,1)&2&11&1\cr
\+(1,1,1)&[0,0,1,0,0,0,0,0,0]&(1,1,1,1,2,2,2,2,2,1,1,1)&0&17&2\cr
\+(2,1,1)&[0,1,0,0,0,1,0,0,1]&(1,1,1,2,3,2,1,1,1,0,1,2)&2&16&1\cr
\+(2,1,1)&[0,0,1,1,0,0,0,0,0]&(1,1,1,1,2,2,2,2,2,1,1,2)&2&18&1\cr
\+(2,1,1)&[0,1,0,0,1,0,0,0,0]&(1,1,1,2,3,2,2,2,2,1,1,2)&0&20&2\cr
\+(2,1,1)&[1,0,0,0,0,0,1,0,1]&(1,1,2,3,4,3,2,1,1,0,1,2)&0&21&2\cr
\+(2,1,1)&[1,0,0,0,0,1,0,0,0]&(1,1,2,3,4,3,2,2,2,1,1,2)&-2&24&4\cr
\+(2,1,1)&[0,0,0,0,0,0,0,1,1]&(1,2,3,4,5,4,3,2,1,0,1,2)&-2&28&2\cr
\+(2,1,1)&[0,0,0,0,0,0,1,0,0]&(1,2,3,4,5,4,3,2,2,1,1,2)&-4&30&5\cr
\+(1,2,1)&[0,0,1,0,0,0,0,1,0]&(1,1,1,1,2,2,2,2,2,1,2,1)&2&18&1\cr
\+(1,2,1)&[0,1,0,0,0,0,0,0,1]&(1,1,1,2,3,3,3,3,3,1,2,1)&0&24&2\cr
\+(1,2,1)&[1,0,0,0,0,0,0,0,0]&(1,1,2,3,4,4,4,4,4,2,2,1)&-2&32&2\cr
\+(2,2,1)&[0,1,0,0,1,0,0,1,0]&(1,1,1,2,3,2,2,2,2,1,2,2)&2&21&1\cr
\+(2,2,1)&[1,0,0,0,0,1,0,0,2]&(1,1,2,3,4,3,2,2,2,0,2,2)&2&24&1\cr
\+(2,2,1)&[0,0,2,0,0,0,0,0,1]&(1,1,1,1,3,3,3,3,3,1,2,2)&2&24&1\cr
\+(2,2,1)&[1,0,0,0,0,1,0,1,0]&(1,1,2,3,4,3,2,2,2,1,2,2)&0&25&2\cr
\+(2,2,1)&[0,1,0,1,0,0,0,0,1]&(1,1,1,2,3,3,3,3,3,1,2,2)&0&25&3\cr
\+(2,2,1)&[1,0,0,0,1,0,0,0,1]&(1,1,2,3,4,3,3,3,3,1,2,2)&-2&28&6\cr
\+(2,2,1)&[0,0,0,0,0,0,0,2,1]&(1,2,3,4,5,4,3,2,1,0,2,2)&2&29&1\cr
\+(2,2,1)&[0,0,0,0,0,0,1,0,2]&(1,2,3,4,5,4,3,2,2,0,2,2)&0&30&0\cr
\+(2,2,1)&[0,1,1,0,0,0,0,0,0]&(1,1,1,2,4,4,4,4,4,2,2,2)&-2&31&5\cr
\+(2,2,1)&[0,0,0,0,0,0,1,1,0]&(1,2,3,4,5,4,3,2,2,1,2,2)&-2&31&2\cr
\+(2,2,1)&[1,0,0,1,0,0,0,0,0]&(1,1,2,3,4,4,4,4,4,2,2,2)&-4&33&8\cr
\+(2,2,1)&[0,0,0,0,0,1,0,0,1]&(1,2,3,4,5,4,3,3,3,1,2,2)&-4&33&11\cr
\+(2,2,1)&[0,0,0,0,1,0,0,0,0]&(1,2,3,4,5,4,4,4,4,2,2,2)&-6&37&9\cr
\+(1,3,1)&[0,1,0,0,0,0,1,0,0]&(1,1,1,2,3,3,3,3,4,2,3,1)&2&27&1\cr
\+(1,3,1)&[1,0,0,0,0,0,0,1,0]&(1,1,2,3,4,4,4,4,4,2,3,1)&0&33&2\cr
\+(1,3,1)&[0,0,0,0,0,0,0,0,1]&(1,2,3,4,5,5,5,5,5,2,3,1)&-2&41&2\cr
\+(2,3,1)&[0,1,0,1,0,0,1,0,0]&(1,1,1,2,3,3,3,3,4,2,3,2)&2&28&1\cr
\+(2,3,1)&[1,0,0,0,1,0,0,1,1]&(1,1,2,3,4,3,3,3,3,1,3,2)&2&29&1\cr
\+(2,3,1)&[1,0,0,0,1,0,1,0,0]&(1,1,2,3,4,3,3,3,4,2,3,2)&0&31&2\cr
\+(2,3,1)&[0,1,1,0,0,0,0,0,2]&(1,1,1,2,4,4,4,4,4,1,3,2)&2&31&1\cr
\+(2,3,1)&[0,1,1,0,0,0,0,1,0]&(1,1,1,2,4,4,4,4,4,2,3,2)&0&32&3\cr
\+(2,3,1)&[1,0,0,1,0,0,0,0,2]&(1,1,2,3,4,4,4,4,4,1,3,2)&0&33&2\cr
\+(2,3,1)&[1,0,0,1,0,0,0,1,0]&(1,1,2,3,4,4,4,4,4,2,3,2)&-2&34&7\cr
\+(2,3,1)&[0,0,0,0,0,1,0,1,1]&(1,2,3,4,5,4,3,3,3,1,3,2)&0&34&2\cr
\+(2,3,1)&[0,0,0,0,0,1,1,0,0]&(1,2,3,4,5,4,3,3,4,2,3,2)&-2&36&3\cr
\+(2,3,1)&[0,0,0,0,1,0,0,0,2]&(1,2,3,4,5,4,4,4,4,1,3,2)&-2&37&4\cr
\+(2,3,1)&[0,2,0,0,0,0,0,0,1]&(1,1,1,3,5,5,5,5,5,2,3,2)&-2&38&5\cr
\+(2,3,1)&[0,0,0,0,1,0,0,1,0]&(1,2,3,4,5,4,4,4,4,2,3,2)&-4&38&11\cr
\+(2,3,1)&[1,0,1,0,0,0,0,0,1]&(1,1,2,3,5,5,5,5,5,2,3,2)&-4&39&13\cr
\+(2,3,1)&[0,0,0,1,0,0,0,0,1]&(1,2,3,4,5,5,5,5,5,2,3,2)&-6&42&19\cr
\+(2,3,1)&[1,1,0,0,0,0,0,0,0]&(1,1,2,4,6,6,6,6,6,3,3,2)&-6&46&12\cr
\+(2,3,1)&[0,0,1,0,0,0,0,0,0]&(1,2,3,4,6,6,6,6,6,3,3,2)&-8&48&19\cr}

\medskip
{\bf B7: $B_8^{+++1}$}

\vbox{\settabs\+(2,4,2)\quad\quad\quad&[-1,0,0,0,0,0,0,1,0,1,-2,2]\quad&(2,3,4,5,6,5,4,3,3,3,4,2)\quad&-4\quad\quad&44\quad\quad&14\cr
\+$(n_c,n_d,n_0)$&$p_i$&$\beta$&$\beta^2$&$ht(\beta)$&$\mu$\cr
\hrule
\+(1,0,0)&[0,0,0,1,0,0,0,0,0]&(0,0,0,0,0,0,0,0,0,0,0,1)&2&1&1\cr
\+(2,0,0)&[1,0,0,0,0,0,1,0,0]&(0,0,1,2,3,2,1,0,0,0,0,2)&2&11&1\cr
\+(2,0,0)&[0,0,0,0,0,0,0,1,0]&(0,1,2,3,4,3,2,1,0,0,0,2)&0&18&0\cr
\+(3,0,0)&[0,0,0,1,0,0,0,1,0]&(0,1,2,3,4,3,2,1,0,0,0,3)&2&19&1\cr
\+(3,0,0)&[1,1,0,0,0,0,0,0,1]&(0,0,1,3,5,4,3,2,1,0,0,3)&2&22&1\cr
\+(3,0,0)&[0,0,1,0,0,0,0,0,1]&(0,1,2,3,5,4,3,2,1,0,0,3)&0&24&0\cr
\+(3,0,0)&[2,0,0,0,0,0,0,0,0]&(0,0,2,4,6,5,4,3,2,1,0,3)&0&30&0\cr
\+(3,0,0)&[0,1,0,0,0,0,0,0,0]&(0,1,2,4,6,5,4,3,2,1,0,3)&-2&31&1\cr
\+(0,1,0)&[0,0,0,0,0,0,0,0,1]&(0,0,0,0,0,0,0,0,0,0,1,0)&1&1&1\cr
\+(1,1,0)&[0,0,1,0,0,0,0,0,0]&(0,0,0,0,1,1,1,1,1,1,1,1)&1&8&1\cr
\+(2,1,0)&[1,0,0,0,0,1,0,0,0]&(0,0,1,2,3,2,1,1,1,1,1,2)&1&15&1\cr
\+(2,1,0)&[0,0,0,0,0,0,1,0,0]&(0,1,2,3,4,3,2,1,1,1,1,2)&-1&21&1\cr
\+(3,1,0)&[0,0,0,1,0,0,1,0,0]&(0,1,2,3,4,3,2,1,1,1,1,3)&1&22&1\cr
\+(3,1,0)&[1,1,0,0,0,0,0,1,0]&(0,0,1,3,5,4,3,2,1,1,1,3)&1&24&1\cr
\+(3,1,0)&[0,0,1,0,0,0,0,1,0]&(0,1,2,3,5,4,3,2,1,1,1,3)&-1&26&1\cr
\+(3,1,0)&[2,0,0,0,0,0,0,0,1]&(0,0,2,4,6,5,4,3,2,1,1,3)&-1&31&1\cr
\+(3,1,0)&[0,1,0,0,0,0,0,0,1]&(0,1,2,4,6,5,4,3,2,1,1,3)&-3&32&2\cr
\+(3,1,0)&[1,0,0,0,0,0,0,0,0]&(0,1,3,5,7,6,5,4,3,2,1,3)&-5&40&2\cr
\+(0,2,0)&[0,0,0,0,0,0,0,1,0]&(0,0,0,0,0,0,0,0,0,1,2,0)&2&3&1\cr
\+(1,2,0)&[0,0,1,0,0,0,0,0,1]&(0,0,0,0,1,1,1,1,1,1,2,1)&2&9&1\cr
\+(1,2,0)&[0,1,0,0,0,0,0,0,0]&(0,0,0,1,2,2,2,2,2,2,2,1)&0&16&1\cr
\+(2,2,0)&[1,0,0,0,0,1,0,0,1]&(0,0,1,2,3,2,1,1,1,1,2,2)&2&16&1\cr
\+(2,2,0)&[0,1,0,1,0,0,0,0,0]&(0,0,0,1,2,2,2,2,2,2,2,2)&2&17&1\cr
\+(2,2,0)&[1,0,0,0,1,0,0,0,0]&(0,0,1,2,3,2,2,2,2,2,2,2)&0&20&1\cr
\+(2,2,0)&[0,0,0,0,0,0,1,0,1]&(0,1,2,3,4,3,2,1,1,1,2,2)&0&22&1\cr
\+(2,2,0)&[0,0,0,0,0,1,0,0,0]&(0,1,2,3,4,3,2,2,2,2,2,2)&-2&25&3\cr
\+(3,2,0)&[1,0,1,0,0,1,0,0,0]&(0,0,1,2,4,3,2,2,2,2,2,3)&2&23&1\cr
\+(3,2,0)&[0,0,0,1,0,0,1,0,1]&(0,1,2,3,4,3,2,1,1,1,2,3)&2&23&1\cr
\+(3,2,0)&[1,1,0,0,0,0,0,1,1]&(0,0,1,3,5,4,3,2,1,1,2,3)&2&25&1\cr
\+(3,2,0)&[0,0,0,1,0,1,0,0,0]&(0,1,2,3,4,3,2,2,2,2,2,3)&0&26&2\cr
\+(3,2,0)&[1,1,0,0,0,0,1,0,0]&(0,0,1,3,5,4,3,2,2,2,2,3)&0&27&2\cr
\+(3,2,0)&[0,0,1,0,0,0,0,1,1]&(0,1,2,3,5,4,3,2,1,1,2,3)&0&27&1\cr
\+(3,2,0)&[0,0,1,0,0,0,1,0,0]&(0,1,2,3,5,4,3,2,2,2,2,3)&-2&29&4\cr
\+(3,2,0)&[2,0,0,0,0,0,0,0,2]&(0,0,2,4,6,5,4,3,2,1,2,3)&0&32&1\cr
\+(3,2,0)&[2,0,0,0,0,0,0,1,0]&(0,0,2,4,6,5,4,3,2,2,2,3)&-2&33&3\cr
\+(3,2,0)&[0,1,0,0,0,0,0,0,2]&(0,1,2,4,6,5,4,3,2,1,2,3)&-2&33&2\cr
\+(3,2,0)&[0,1,0,0,0,0,0,1,0]&(0,1,2,4,6,5,4,3,2,2,2,3)&-4&34&6\cr
\+(3,2,0)&[1,0,0,0,0,0,0,0,1]&(0,1,3,5,7,6,5,4,3,2,2,3)&-6&41&7\cr
\+(3,2,0)&[0,0,0,0,0,0,0,0,0]&(0,2,4,6,8,7,6,5,4,3,2,3)&-8&50&3\cr
\+(1,3,0)&[0,1,0,0,0,0,0,0,1]&(0,0,0,1,2,2,2,2,2,2,3,1)&1&17&1\cr
\+(1,3,0)&[1,0,0,0,0,0,0,0,0]&(0,0,1,2,3,3,3,3,3,3,3,1)&-1&25&1\cr}
\vbox{\settabs\+(2,4,2)\quad\quad\quad&[-1,0,0,0,0,0,0,1,0,1,-2,2]\quad&(2,3,4,5,6,5,4,3,3,3,4,2)\quad&-4\quad\quad&44\quad\quad&14\cr
\+$(n_c,n_d,n_0)$&$p_i$&$\beta$&$\beta^2$&$ht(\beta)$&$\mu$\cr
\hrule
\+(2,3,0)&[1,0,0,0,1,0,0,0,1]&(0,0,1,2,3,2,2,2,2,2,3,2)&1&21&1\cr
\+(2,3,0)&[0,1,1,0,0,0,0,0,0]&(0,0,0,1,3,3,3,3,3,3,3,2)&1&24&1\cr
\+(2,3,0)&[1,0,0,1,0,0,0,0,0]&(0,0,1,2,3,3,3,3,3,3,3,2)&-1&26&2\cr
\+(2,3,0)&[0,0,0,0,0,1,0,0,1]&(0,1,2,3,4,3,2,2,2,2,3,2)&-1&26&2\cr
\+(2,3,0)&[0,0,0,0,1,0,0,0,0]&(0,1,2,3,4,3,3,3,3,3,3,2)&-3&30&3\cr
\+(0,0,1)&[1,0,0,0,0,0,0,0,0]&(1,0,0,0,0,0,0,0,0,0,0,0)&2&1&1\cr
\+(1,0,1)&[0,0,0,0,1,0,0,0,0]&(1,1,1,1,1,0,0,0,0,0,0,1)&2&6&1\cr
\+(2,0,1)&[0,1,0,0,0,0,1,0,0]&(1,1,1,2,3,2,1,0,0,0,0,2)&2&13&1\cr
\+(2,0,1)&[1,0,0,0,0,0,0,1,0]&(1,1,2,3,4,3,2,1,0,0,0,2)&0&19&1\cr
\+(2,0,1)&[0,0,0,0,0,0,0,0,1]&(1,2,3,4,5,4,3,2,1,0,0,2)&-2&27&1\cr
\+(3,0,1)&[1,0,0,1,0,0,0,1,0]&(1,1,2,3,4,3,2,1,0,0,0,3)&2&20&1\cr
\+(3,0,1)&[0,0,0,0,0,1,1,0,0]&(1,2,3,4,5,3,1,0,0,0,0,3)&2&22&1\cr
\+(3,0,1)&[0,2,0,0,0,0,0,0,1]&(1,1,1,3,5,4,3,2,1,0,0,3)&2&24&1\cr
\+(3,0,1)&[0,0,0,0,1,0,0,1,0]&(1,2,3,4,5,3,2,1,0,0,0,3)&0&24&1\cr
\+(3,0,1)&[1,0,1,0,0,0,0,0,1]&(1,1,2,3,5,4,3,2,1,0,0,3)&0&25&1\cr
\+(3,0,1)&[0,0,0,1,0,0,0,0,1]&(1,2,3,4,5,4,3,2,1,0,0,3)&-2&28&2\cr
\+(3,0,1)&[1,1,0,0,0,0,0,0,0]&(1,1,2,4,6,5,4,3,2,1,0,3)&-2&32&2\cr
\+(3,0,1)&[0,0,1,0,0,0,0,0,0]&(1,2,3,4,6,5,4,3,2,1,0,3)&-4&34&2\cr
\+(0,1,1)&[0,0,0,0,0,0,0,0,0]&(1,1,1,1,1,1,1,1,1,1,1,0)&1&11&1\cr
\+(1,1,1)&[0,0,0,1,0,0,0,0,0]&(1,1,1,1,1,1,1,1,1,1,1,1)&1&12&1\cr
\+(2,1,1)&[0,1,0,0,0,1,0,0,0]&(1,1,1,2,3,2,1,1,1,1,1,2)&1&17&1\cr
\+(2,1,1)&[1,0,0,0,0,0,1,0,0]&(1,1,2,3,4,3,2,1,1,1,1,2)&-1&22&2\cr
\+(2,1,1)&[0,0,0,0,0,0,0,1,0]&(1,2,3,4,5,4,3,2,1,1,1,2)&-3&29&2\cr
\+(3,1,1)&[1,0,0,1,0,0,1,0,0]&(1,1,2,3,4,3,2,1,1,1,1,3)&1&23&1\cr
\+(3,1,1)&[0,2,0,0,0,0,0,1,0]&(1,1,1,3,5,4,3,2,1,1,1,3)&1&26&1\cr
\+(3,1,1)&[0,0,0,0,0,2,0,0,0]&(1,2,3,4,5,3,1,1,1,1,1,3)&1&26&1\cr
\+(3,1,1)&[1,0,1,0,0,0,0,1,0]&(1,1,2,3,5,4,3,2,1,1,1,3)&-1&27&2\cr
\+(3,1,1)&[0,0,0,0,1,0,1,0,0]&(1,2,3,4,5,3,2,1,1,1,1,3)&-1&27&2\cr
\+(3,1,1)&[0,0,0,1,0,0,0,1,0]&(1,2,3,4,5,4,3,2,1,1,1,3)&-3&30&4\cr
\+(3,1,1)&[1,1,0,0,0,0,0,0,1]&(1,1,2,4,6,5,4,3,2,1,1,3)&-3&33&4\cr
\+(3,1,1)&[0,0,1,0,0,0,0,0,1]&(1,2,3,4,6,5,4,3,2,1,1,3)&-5&35&5\cr
\+(3,1,1)&[2,0,0,0,0,0,0,0,0]&(1,1,3,5,7,6,5,4,3,2,1,3)&-5&41&3\cr
\+(3,1,1)&[0,1,0,0,0,0,0,0,0]&(1,2,3,5,7,6,5,4,3,2,1,3)&-7&42&7\cr
\+(0,2,1)&[0,0,0,0,0,0,0,0,1]&(1,1,1,1,1,1,1,1,1,1,2,0)&2&12&1\cr
\+(1,2,1)&[0,0,0,1,0,0,0,0,1]&(1,1,1,1,1,1,1,1,1,1,2,1)&2&13&1\cr
\+(1,2,1)&[0,0,1,0,0,0,0,0,0]&(1,1,1,1,2,2,2,2,2,2,2,1)&0&19&2\cr
\+(2,2,1)&[0,1,0,0,0,1,0,0,1]&(1,1,1,2,3,2,1,1,1,1,2,2)&2&18&1\cr
\+(2,2,1)&[0,0,1,1,0,0,0,0,0]&(1,1,1,1,2,2,2,2,2,2,2,2)&2&20&1\cr
\+(2,2,1)&[0,1,0,0,1,0,0,0,0]&(1,1,1,2,3,2,2,2,2,2,2,2)&0&22&2\cr
\+(2,2,1)&[1,0,0,0,0,0,1,0,1]&(1,1,2,3,4,3,2,1,1,1,2,2)&0&23&2\cr
\+(2,2,1)&[1,0,0,0,0,1,0,0,0]&(1,1,2,3,4,3,2,2,2,2,2,2)&-2&26&5\cr
\+(2,2,1)&[0,0,0,0,0,0,0,1,1]&(1,2,3,4,5,4,3,2,1,1,2,2)&-2&30&2\cr
\+(2,2,1)&[0,0,0,0,0,0,1,0,0]&(1,2,3,4,5,4,3,2,2,2,2,2)&-4&32&6\cr}
\vbox{\settabs\+(2,4,2)\quad\quad\quad&[-1,0,0,0,0,0,0,1,0,1,-2,2]\quad&(2,3,4,5,6,5,4,3,3,3,4,2)\quad&-4\quad\quad&44\quad\quad&14\cr
\+$(n_c,n_d,n_0)$&$p_i$&$\beta$&$\beta^2$&$ht(\beta)$&$\mu$\cr
\hrule
\+(1,3,1)&[0,0,1,0,0,0,0,0,1]&(1,1,1,1,2,2,2,2,2,2,3,1)&1&20&1\cr
\+(1,3,1)&[0,1,0,0,0,0,0,0,0]&(1,1,1,2,3,3,3,3,3,3,3,1)&-1&27&2\cr
\+(2,3,1)&[0,1,0,0,1,0,0,0,1]&(1,1,1,2,3,2,2,2,2,2,3,2)&1&23&1\cr
\+(2,3,1)&[1,0,0,0,0,1,0,0,1]&(1,1,2,3,4,3,2,2,2,2,3,2)&-1&27&3\cr
\+(2,3,1)&[0,0,2,0,0,0,0,0,0]&(1,1,1,1,3,3,3,3,3,3,3,2)&1&27&1\cr
\+(2,3,1)&[0,1,0,1,0,0,0,0,0]&(1,1,1,2,3,3,3,3,3,3,3,2)&-1&28&3\cr
\+(2,3,1)&[1,0,0,0,1,0,0,0,0]&(1,1,2,3,4,3,3,3,3,3,3,2)&-3&31&6\cr
\+(2,3,1)&[0,0,0,0,0,0,0,1,2]&(1,2,3,4,5,4,3,2,1,1,3,2)&1&31&1\cr
\+(2,3,1)&[0,0,0,0,0,0,1,0,1]&(1,2,3,4,5,4,3,2,2,2,3,2)&-3&33&2\cr
\+(2,3,1)&[0,0,0,0,0,1,0,0,0]&(1,2,3,4,5,4,3,3,3,3,3,2)&-5&36&12\cr}

\medskip
{\bf B8: $A_1^{+++1}$}

\vbox{\settabs\+(4,1)\quad\quad&[-1,2,0,1,2]\quad&(1,1,3,5,4)\quad&-14\quad\quad&14\quad\quad&3\cr
\+$(n_c,n_0)$&$p_i$&$\beta$&$\beta^2$&$ht(\beta)$&$\mu$\cr
\hrule
\+(1,0)&[0,0,2]&(0,0,0,0,1)&2&1&1\cr
\+(1,0)&[0,1,0]&(0,0,0,1,1)&0&2&0\cr
\+(2,0)&[0,1,2]&(0,0,0,1,2)&2&3&1\cr
\+(2,0)&[0,2,0]&(0,0,0,2,2)&0&4&0\cr
\+(2,0)&[1,0,1]&(0,0,1,2,2)&-2&5&0\cr
\+(2,0)&[0,0,0]&(0,1,2,3,2)&-4&8&0\cr
\+(3,0)&[0,2,2]&(0,0,0,2,3)&2&5&1\cr
\+(3,0)&[1,0,3]&(0,0,1,2,3)&0&6&0\cr
\+(3,0)&[0,3,0]&(0,0,0,3,3)&0&6&0\cr
\+(3,0)&[1,1,1]&(0,0,1,3,3)&-4&7&1\cr
\+(3,0)&[2,0,0]&(0,0,2,4,3)&-6&9&0\cr
\+(3,0)&[0,0,2]&(0,1,2,3,3)&-6&9&0\cr
\+(3,0)&[0,1,0]&(0,1,2,4,3)&-8&10&0\cr
\+(4,0)&[0,3,2]&(0,0,0,3,4)&2&7&1\cr
\+(4,0)&[1,1,3]&(0,0,1,3,4)&-2&8&1\cr
\+(4,0)&[0,4,0]&(0,0,0,4,4)&0&8&0\cr
\+(4,0)&[1,2,1]&(0,0,1,4,4)&-6&9&2\cr
\+(4,0)&[2,0,2]&(0,0,2,4,4)&-8&10&1\cr
\+(4,0)&[0,0,4]&(0,1,2,3,4)&-4&10&0\cr
\+(4,0)&[2,1,0]&(0,0,2,5,4)&-10&11&1\cr
\+(4,0)&[0,1,2]&(0,1,2,4,4)&-10&11&1\cr
\+(4,0)&[0,2,0]&(0,1,2,5,4)&-12&12&0\cr
\+(4,0)&[1,0,1]&(0,1,3,5,4)&-14&13&1\cr
\+(4,0)&[0,0,0]&(0,2,4,6,4)&-16&16&0\cr}
\vbox{\settabs\+(4,1)\quad\quad&[-1,2,0,1,2]\quad&(1,1,3,5,4)\quad&-14\quad\quad&14\quad\quad&3\cr
\+$(n_c,n_0)$&$p_i$&$\beta$&$\beta^2$&$ht(\beta)$&$\mu$\cr
\hrule
\+(5,0)&[0,4,2]&(0,0,0,4,5)&2&9&1\cr
\+(5,0)&[1,2,3]&(0,0,1,4,5)&-4&10&2\cr
\+(5,0)&[0,5,0]&(0,0,0,5,5)&0&10&0\cr
\+(5,0)&[2,0,4]&(0,0,2,4,5)&-6&11&2\cr
\+(5,0)&[1,3,1]&(0,0,1,5,5)&-8&11&3\cr
\+(5,0)&[2,1,2]&(0,0,2,5,5)&-12&12&4\cr
\+(5,0)&[0,1,4]&(0,1,2,4,5)&-8&12&1\cr
\+(5,0)&[2,2,0]&(0,0,2,6,5)&-14&13&3\cr
\+(5,0)&[0,2,2]&(0,1,2,5,5)&-14&13&3\cr
\+(5,0)&[3,0,1]&(0,0,3,6,5)&-16&14&2\cr
\+(5,0)&[1,0,3]&(0,1,3,5,5)&-16&14&3\cr
\+(5,0)&[0,3,0]&(0,1,2,6,5)&-16&14&2\cr
\+(5,0)&[1,1,1]&(0,1,3,6,5)&-20&15&5\cr
\+(5,0)&[2,0,0]&(0,1,4,7,5)&-22&17&1\cr
\+(5,0)&[0,0,2]&(0,2,4,6,5)&-22&17&2\cr
\+(5,0)&[0,1,0]&(0,2,4,7,5)&-24&18&1\cr
\+(0,1)&[1,0,0]&(1,0,0,0,0)&2&1&1\cr
\+(1,1)&[0,0,1]&(1,1,1,1,1)&0&5&1\cr
\+(2,1)&[0,0,3]&(1,1,1,1,2)&2&6&1\cr
\+(2,1)&[0,1,1]&(1,1,1,2,2)&-2&7&1\cr
\+(2,1)&[1,0,0]&(1,1,2,3,2)&-4&9&0\cr
\+(3,1)&[0,1,3]&(1,1,1,2,3)&0&8&1\cr
\+(3,1)&[0,2,1]&(1,1,1,3,3)&-4&9&2\cr
\+(3,1)&[1,0,2]&(1,1,2,3,3)&-6&10&1\cr
\+(3,1)&[1,1,0]&(1,1,2,4,3)&-8&11&1\cr
\+(3,1)&[0,0,1]&(1,2,3,4,3)&-10&13&0\cr
\+(4,1)&[0,2,3]&(1,1,1,3,4)&-2&10&2\cr
\+(4,1)&[1,0,4]&(1,1,2,3,4)&-4&11&1\cr
\+(4,1)&[0,3,1]&(1,1,1,4,4)&-6&11&3\cr
\+(4,1)&[1,1,2]&(1,1,2,4,4)&-10&12&5\cr
\+(4,1)&[1,2,0]&(1,1,2,5,4)&-12&13&3\cr
\+(4,1)&[2,0,1]&(1,1,3,5,4)&-14&14&3\cr
\+(4,1)&[0,0,3]&(1,2,3,4,4)&-12&14&1\cr
\+(4,1)&[0,1,1]&(1,2,3,5,4)&-16&15&2\cr
\+(4,1)&[1,0,0]&(1,2,4,6,4)&-18&17&1\cr}

\eject

{\bf Appendix C: The $\bar l_1$ representation of $E_8^{++}$}

\vbox{\settabs\+(11,0)\quad\quad&[0,0,0,0,0,0,1,0,0]\quad&(0,3,6,9,12,15,18,21,14,7,11)\quad&-10\quad\quad&116\quad\quad&12\cr
\+$(n_c,n_0)$&$p_i$&$\beta$&$\beta^2$&$ht(\beta)$&$\mu$\cr
\hrule
\+(1,0)&[0,0,0,0,0,0,1,0,0]&(0,0,0,0,0,0,0,0,0,0,1)&2&1&1\cr
\+(2,0)&[0,0,0,1,0,0,0,0,0]&(0,0,0,0,0,1,2,3,2,1,2)&2&11&1\cr
\+(3,0)&[0,1,0,0,0,0,0,0,1]&(0,0,0,1,2,3,4,5,3,1,3)&2&22&1\cr
\+(3,0)&[1,0,0,0,0,0,0,0,0]&(0,0,1,2,3,4,5,6,4,2,3)&0&30&0\cr
\+(4,0)&[1,0,0,0,0,0,1,0,0]&(0,0,1,2,3,4,5,6,4,2,4)&2&31&1\cr
\+(4,0)&[0,0,0,0,0,0,0,0,2]&(0,1,2,3,4,5,6,7,4,1,4)&2&37&1\cr
\+(4,0)&[0,0,0,0,0,0,0,1,0]&(0,1,2,3,4,5,6,7,4,2,4)&0&38&0\cr
\+(5,0)&[1,0,0,1,0,0,0,0,0]&(0,0,1,2,3,5,7,9,6,3,5)&2&41&1\cr
\+(5,0)&[0,0,0,0,0,1,0,0,1]&(0,1,2,3,4,5,6,8,5,2,5)&2&41&1\cr
\+(5,0)&[0,0,0,0,1,0,0,0,0]&(0,1,2,3,4,5,7,9,6,3,5)&0&45&0\cr
\+(6,0)&[0,0,0,1,0,0,0,1,0]&(0,1,2,3,4,6,8,10,6,3,6)&2&49&1\cr
\+(6,0)&[1,1,0,0,0,0,0,0,1]&(0,0,1,3,5,7,9,11,7,3,6)&2&52&1\cr
\+(6,0)&[0,0,1,0,0,0,0,0,1]&(0,1,2,3,5,7,9,11,7,3,6)&0&54&1\cr
\+(6,0)&[2,0,0,0,0,0,0,0,0]&(0,0,2,4,6,8,10,12,8,4,6)&0&60&0\cr
\+(6,0)&[0,1,0,0,0,0,0,0,0]&(0,1,2,4,6,8,10,12,8,4,6)&-2&61&1\cr
\+(7,0)&[0,0,1,0,0,1,0,0,0]&(0,1,2,3,5,7,9,12,8,4,7)&2&58&1\cr
\+(7,0)&[0,1,0,0,0,0,0,1,1]&(0,1,2,4,6,8,10,12,7,3,7)&2&60&1\cr
\+(7,0)&[2,0,0,0,0,0,1,0,0]&(0,0,2,4,6,8,10,12,8,4,7)&2&61&1\cr
\+(7,0)&[0,1,0,0,0,0,1,0,0]&(0,1,2,4,6,8,10,12,8,4,7)&0&62&1\cr
\+(7,0)&[1,0,0,0,0,0,0,0,2]&(0,1,3,5,7,9,11,13,8,3,7)&0&67&1\cr
\+(7,0)&[1,0,0,0,0,0,0,1,0]&(0,1,3,5,7,9,11,13,8,4,7)&-2&68&2\cr
\+(7,0)&[0,0,0,0,0,0,0,0,1]&(0,2,4,6,8,10,12,14,9,4,7)&-4&76&1\cr
\+(8,0)&[0,1,0,0,1,0,0,0,1]&(0,1,2,4,6,8,11,14,9,4,8)&2&67&1\cr
\+(8,0)&[1,0,0,0,0,0,1,1,0]&(0,1,3,5,7,9,11,13,8,4,8)&2&69&1\cr
\+(8,0)&[2,0,0,1,0,0,0,0,0]&(0,0,2,4,6,9,12,15,10,5,8)&2&71&1\cr
\+(8,0)&[1,0,0,0,0,1,0,0,1]&(0,1,3,5,7,9,11,14,9,4,8)&0&71&2\cr
\+(8,0)&[0,0,2,0,0,0,0,0,0]&(0,1,2,3,6,9,12,15,10,5,8)&2&71&1\cr
\+(8,0)&[0,1,0,1,0,0,0,0,0]&(0,1,2,4,6,9,12,15,10,5,8)&0&72&1\cr
\+(8,0)&[1,0,0,0,1,0,0,0,0]&(0,1,3,5,7,9,12,15,10,5,8)&-2&75&2\cr
\+(8,0)&[0,0,0,0,0,0,0,1,2]&(0,2,4,6,8,10,12,14,8,3,8)&2&75&1\cr
\+(8,0)&[0,0,0,0,0,0,0,2,0]&(0,2,4,6,8,10,12,14,8,4,8)&0&76&0\cr
\+(8,0)&[0,0,0,0,0,0,1,0,1]&(0,2,4,6,8,10,12,14,9,4,8)&-2&77&2\cr
\+(8,0)&[0,0,0,0,0,1,0,0,0]&(0,2,4,6,8,10,12,15,10,5,8)&-4&80&2\cr
\+(9,0)&[1,0,0,0,1,0,1,0,0]&(0,1,3,5,7,9,12,15,10,5,9)&2&76&1\cr
\+(9,0)&[0,1,1,0,0,0,0,1,0]&(0,1,2,4,7,10,13,16,10,5,9)&2&77&1\cr
\+(9,0)&[1,0,0,1,0,0,0,0,2]&(0,1,3,5,7,10,13,16,10,4,9)&2&78&1\cr
\+(9,0)&[1,0,0,1,0,0,0,1,0]&(0,1,3,5,7,10,13,16,10,5,9)&0&79&2\cr
\+(9,0)&[0,0,0,0,0,1,0,1,1]&(0,2,4,6,8,10,12,15,9,4,9)&2&79&1\cr
\+(9,0)&[0,0,0,0,0,1,1,0,0]&(0,2,4,6,8,10,12,15,10,5,9)&0&81&1\cr
\+(9,0)&[2,1,0,0,0,0,0,0,1]&(0,0,2,5,8,11,14,17,11,5,9)&2&82&1\cr}
\vbox{\settabs\+(11,0)\quad\quad&[0,0,0,0,0,0,1,0,0]\quad&(0,3,6,9,12,15,18,21,14,7,11)\quad&-10\quad\quad&116\quad\quad&12\cr
\+$(n_c,n_0)$&$p_i$&$\beta$&$\beta^2$&$ht(\beta)$&$\mu$\cr
\hrule
\+(9,0)&[0,0,0,0,1,0,0,0,2]&(0,2,4,6,8,10,13,16,10,4,9)&0&82&1\cr
\+(9,0)&[0,2,0,0,0,0,0,0,1]&(0,1,2,5,8,11,14,17,11,5,9)&0&83&1\cr
\+(9,0)&[0,0,0,0,1,0,0,1,0]&(0,2,4,6,8,10,13,16,10,5,9)&-2&83&3\cr
\+(9,0)&[1,0,1,0,0,0,0,0,1]&(0,1,3,5,8,11,14,17,11,5,9)&-2&84&4\cr
\+(9,0)&[0,0,0,1,0,0,0,0,1]&(0,2,4,6,8,11,14,17,11,5,9)&-4&87&4\cr
\+(9,0)&[3,0,0,0,0,0,0,0,0]&(0,0,3,6,9,12,15,18,12,6,9)&0&90&0\cr
\+(9,0)&[1,1,0,0,0,0,0,0,0]&(0,1,3,6,9,12,15,18,12,6,9)&-4&91&3\cr
\+(9,0)&[0,0,1,0,0,0,0,0,0]&(0,2,4,6,9,12,15,18,12,6,9)&-6&93&4\cr
\+(10,0)&[1,0,1,0,0,0,1,0,1]&(0,1,3,5,8,11,14,17,11,5,10)&2&85&1\cr
\+(10,0)&[1,0,0,1,1,0,0,0,0]&(0,1,3,5,7,10,14,18,12,6,10)&2&86&1\cr
\+(10,0)&[0,0,0,0,1,1,0,0,1]&(0,2,4,6,8,10,13,17,11,5,10)&2&86&1\cr
\+(10,0)&[0,2,0,0,0,1,0,0,0]&(0,1,2,5,8,11,14,18,12,6,10)&2&87&1\cr
\+(10,0)&[0,0,0,1,0,0,0,2,0]&(0,2,4,6,8,11,14,17,10,5,10)&2&87&1\cr
\+(10,0)&[1,0,1,0,0,1,0,0,0]&(0,1,3,5,8,11,14,18,12,6,10)&0&88&2\cr
\+(10,0)&[0,0,0,1,0,0,1,0,1]&(0,2,4,6,8,11,14,17,11,5,10)&0&88&2\cr
\+(10,0)&[1,1,0,0,0,0,0,1,1]&(0,1,3,6,9,12,15,18,11,5,10)&0&90&3\cr
\+(10,0)&[0,0,0,0,2,0,0,0,0]&(0,2,4,6,8,10,14,18,12,6,10)&0&90&0\cr
\+(10,0)&[3,0,0,0,0,0,1,0,0]&(0,0,3,6,9,12,15,18,12,6,10)&2&91&1\cr
\+(10,0)&[0,0,1,0,0,0,0,0,3]&(0,2,4,6,9,12,15,18,11,4,10)&2&91&1\cr
\+(10,0)&[0,0,0,1,0,1,0,0,0]&(0,2,4,6,8,11,14,18,12,6,10)&-2&91&3\cr
\+(10,0)&[1,1,0,0,0,0,1,0,0]&(0,1,3,6,9,12,15,18,12,6,10)&-2&92&4\cr
\+(10,0)&[0,0,1,0,0,0,0,1,1]&(0,2,4,6,9,12,15,18,11,5,10)&-2&92&4\cr
\+(10,0)&[0,0,1,0,0,0,1,0,0]&(0,2,4,6,9,12,15,18,12,6,10)&-4&94&6\cr
\+(10,0)&[2,0,0,0,0,0,0,0,2]&(0,1,4,7,10,13,16,19,12,5,10)&-2&97&3\cr
\+(10,0)&[2,0,0,0,0,0,0,1,0]&(0,1,4,7,10,13,16,19,12,6,10)&-4&98&4\cr
\+(10,0)&[0,1,0,0,0,0,0,0,2]&(0,2,4,7,10,13,16,19,12,5,10)&-4&98&4\cr
\+(10,0)&[0,1,0,0,0,0,0,1,0]&(0,2,4,7,10,13,16,19,12,6,10)&-6&99&10\cr
\+(10,0)&[1,0,0,0,0,0,0,0,1]&(0,2,5,8,11,14,17,20,13,6,10)&-8&106&8\cr
\+(10,0)&[0,0,0,0,0,0,0,0,0]&(0,3,6,9,12,15,18,21,14,7,10)&-10&115&3\cr
\+(11,0)&[1,1,0,0,0,1,0,1,0]&(0,1,3,6,9,12,15,19,12,6,11)&2&94&1\cr
\+(11,0)&[0,0,0,1,1,0,0,1,0]&(0,2,4,6,8,11,15,19,12,6,11)&2&94&1\cr
\+(11,0)&[1,0,1,1,0,0,0,0,1]&(0,1,3,5,8,12,16,20,13,6,11)&2&95&1\cr
\+(11,0)&[0,0,1,0,0,1,0,0,2]&(0,2,4,6,9,12,15,19,12,5,11)&2&95&1\cr
\+(11,0)&[0,0,1,0,0,0,2,0,0]&(0,2,4,6,9,12,15,18,12,6,11)&2&95&1\cr
\+(11,0)&[0,0,1,0,0,1,0,1,0]&(0,2,4,6,9,12,15,19,12,6,11)&0&96&2\cr
\+(11,0)&[1,1,0,0,1,0,0,0,1]&(0,1,3,6,9,12,16,20,13,6,11)&0&97&3\cr
\+(11,0)&[2,0,0,0,0,0,1,0,2]&(0,1,4,7,10,13,16,19,12,5,11)&2&98&1\cr
\+(11,0)&[0,1,0,0,0,0,0,2,1]&(0,2,4,7,10,13,16,19,11,5,11)&2&98&1\cr
\+(11,0)&[0,0,0,2,0,0,0,0,1]&(0,2,4,6,8,12,16,20,13,6,11)&0&98&1\cr
\+(11,0)&[2,0,0,0,0,0,1,1,0]&(0,1,4,7,10,13,16,19,12,6,11)&0&99&2\cr
\+(11,0)&[0,1,0,0,0,0,1,0,2]&(0,2,4,7,10,13,16,19,12,5,11)&0&99&2\cr
\+(11,0)&[0,0,1,0,1,0,0,0,1]&(0,2,4,6,9,12,16,20,13,6,11)&-2&99&5\cr
\+(11,0)&[0,2,1,0,0,0,0,0,0]&(0,1,2,5,9,13,17,21,14,7,11)&2&100&1\cr}
\vbox{\settabs\+(11,0)\quad\quad&[0,0,0,0,0,0,1,0,0]\quad&(0,3,6,9,12,15,18,21,14,7,11)\quad&-10\quad\quad&116\quad\quad&12\cr
\+$(n_c,n_0)$&$p_i$&$\beta$&$\beta^2$&$ht(\beta)$&$\mu$\cr
\hrule
\+(11,0)&[0,1,0,0,0,0,1,1,0]&(0,2,4,7,10,13,16,19,12,6,11)&-2&100&5\cr
\+(11,0)&[3,0,0,1,0,0,0,0,0]&(0,0,3,6,9,13,17,21,14,7,11)&2&101&1\cr
\+(11,0)&[2,0,0,0,0,1,0,0,1]&(0,1,4,7,10,13,16,20,13,6,11)&-2&101&5\cr
\+(11,0)&[1,0,2,0,0,0,0,0,0]&(0,1,3,5,9,13,17,21,14,7,11)&0&101&1\cr
\+(11,0)&[1,1,0,1,0,0,0,0,0]&(0,1,3,6,9,13,17,21,14,7,11)&-2&102&4\cr
\+(11,0)&[0,1,0,0,0,1,0,0,1]&(0,2,4,7,10,13,16,20,13,6,11)&-4&102&10\cr
\+(11,0)&[0,0,1,1,0,0,0,0,0]&(0,2,4,6,9,13,17,21,14,7,11)&-4&104&5\cr
\+(11,0)&[2,0,0,0,1,0,0,0,0]&(0,1,4,7,10,13,17,21,14,7,11)&-4&105&5\cr
\+(11,0)&[1,0,0,0,0,0,0,1,2]&(0,2,5,8,11,14,17,20,12,5,11)&-2&105&4\cr
\+(11,0)&[1,0,0,0,0,0,0,2,0]&(0,2,5,8,11,14,17,20,12,6,11)&-4&106&5\cr
\+(11,0)&[0,1,0,0,1,0,0,0,0]&(0,2,4,7,10,13,17,21,14,7,11)&-6&106&11\cr
\+(11,0)&[1,0,0,0,0,0,1,0,1]&(0,2,5,8,11,14,17,20,13,6,11)&-6&107&15\cr
\+(11,0)&[1,0,0,0,0,1,0,0,0]&(0,2,5,8,11,14,17,21,14,7,11)&-8&110&14\cr
\+(11,0)&[0,0,0,0,0,0,0,0,3]&(0,3,6,9,12,15,18,21,13,5,11)&-4&113&2\cr
\+(11,0)&[0,0,0,0,0,0,0,1,1]&(0,3,6,9,12,15,18,21,13,6,11)&-8&114&9\cr
\+(11,0)&[0,0,0,0,0,0,1,0,0]&(0,3,6,9,12,15,18,21,14,7,11)&-10&116&12\cr
\+(12,0)&[1,1,0,1,0,0,1,0,0]&(0,1,3,6,9,13,17,21,14,7,12)&2&103&1\cr
\+(12,0)&[0,1,0,0,0,1,1,0,1]&(0,2,4,7,10,13,16,20,13,6,12)&2&103&1\cr
\+(12,0)&[0,0,1,1,0,0,0,1,1]&(0,2,4,6,9,13,17,21,13,6,12)&2&103&1\cr
\+(12,0)&[0,0,1,0,1,1,0,0,0]&(0,2,4,6,9,12,16,21,14,7,12)&2&103&1\cr
\+(12,0)&[2,0,0,0,1,0,0,1,1]&(0,1,4,7,10,13,17,21,13,6,12)&2&104&1\cr
\+(12,0)&[2,0,0,0,0,2,0,0,0]&(0,1,4,7,10,13,16,21,14,7,12)&2&105&1\cr
\+(12,0)&[0,1,0,0,1,0,0,1,1]&(0,2,4,7,10,13,17,21,13,6,12)&0&105&3\cr
\+(12,0)&[0,0,1,1,0,0,1,0,0]&(0,2,4,6,9,13,17,21,14,7,12)&0&105&2\cr
\+(12,0)&[2,0,0,0,1,0,1,0,0]&(0,1,4,7,10,13,17,21,14,7,12)&0&106&2\cr
\+(12,0)&[1,1,1,0,0,0,0,0,2]&(0,1,3,6,10,14,18,22,14,6,12)&2&106&1\cr
\+(12,0)&[0,1,0,0,0,2,0,0,0]&(0,2,4,7,10,13,16,21,14,7,12)&0&106&1\cr
\+(12,0)&[1,1,1,0,0,0,0,1,0]&(0,1,3,6,10,14,18,22,14,7,12)&0&107&3\cr
\+(12,0)&[1,0,0,0,0,0,1,2,0]&(0,2,5,8,11,14,17,20,12,6,12)&2&107&1\cr
\+(12,0)&[0,1,0,0,1,0,1,0,0]&(0,2,4,7,10,13,17,21,14,7,12)&-2&107&6\cr
\+(12,0)&[2,0,0,1,0,0,0,0,2]&(0,1,4,7,10,14,18,22,14,6,12)&0&108&2\cr
\+(12,0)&[1,0,0,0,0,1,0,0,3]&(0,2,5,8,11,14,17,21,13,5,12)&2&108&1\cr
\+(12,0)&[1,0,0,0,0,0,2,0,1]&(0,2,5,8,11,14,17,20,13,6,12)&0&108&2\cr
\+(12,0)&[0,0,2,0,0,0,0,0,2]&(0,2,4,6,10,14,18,22,14,6,12)&0&108&1\cr
\+(12,0)&[2,0,0,1,0,0,0,1,0]&(0,1,4,7,10,14,18,22,14,7,12)&-2&109&6\cr
\+(12,0)&[1,0,0,0,0,1,0,1,1]&(0,2,5,8,11,14,17,21,13,6,12)&-2&109&7\cr
\+(12,0)&[0,1,0,1,0,0,0,0,2]&(0,2,4,7,10,14,18,22,14,6,12)&-2&109&6\cr
\+(12,0)&[0,0,2,0,0,0,0,1,0]&(0,2,4,6,10,14,18,22,14,7,12)&-2&109&4\cr
\+(12,0)&[0,1,0,1,0,0,0,1,0]&(0,2,4,7,10,14,18,22,14,7,12)&-4&110&11\cr
\+(12,0)&[1,0,0,0,0,1,1,0,0]&(0,2,5,8,11,14,17,21,14,7,12)&-4&111&9\cr
\+(12,0)&[3,1,0,0,0,0,0,0,1]&(0,0,3,7,11,15,19,23,15,7,12)&2&112&1\cr
\+(12,0)&[1,0,0,0,1,0,0,0,2]&(0,2,5,8,11,14,18,22,14,6,12)&-4&112&10\cr
\+(12,0)&[1,2,0,0,0,0,0,0,1]&(0,1,3,7,11,15,19,23,15,7,12)&-2&113&4\cr
\+(12,0)&[1,0,0,0,1,0,0,1,0]&(0,2,5,8,11,14,18,22,14,7,12)&-6&113&21\cr}
\vbox{\settabs\+(11,0)\quad\quad&[0,0,0,0,0,0,1,0,0]\quad&(0,3,6,9,12,15,18,21,14,7,11)\quad&-10\quad\quad&116\quad\quad&12\cr
\+$(n_c,n_0)$&$p_i$&$\beta$&$\beta^2$&$ht(\beta)$&$\mu$\cr
\hrule
\+(12,0)&[0,0,0,0,0,0,0,2,2]&(0,3,6,9,12,15,18,21,12,5,12)&2&113&1\cr
\+(12,0)&[2,0,1,0,0,0,0,0,1]&(0,1,4,7,11,15,19,23,15,7,12)&-4&114&9\cr
\+(12,0)&[0,0,0,0,0,0,1,0,3]&(0,3,6,9,12,15,18,21,13,5,12)&0&114&1\cr
\+(12,0)&[0,0,0,0,0,0,0,3,0]&(0,3,6,9,12,15,18,21,12,6,12)&0&114&0\cr
\+(12,0)&[0,1,1,0,0,0,0,0,1]&(0,2,4,7,11,15,19,23,15,7,12)&-6&115&17\cr
\+(12,0)&[0,0,0,0,0,0,1,1,1]&(0,3,6,9,12,15,18,21,13,6,12)&-4&115&7\cr
\+(12,0)&[1,0,0,1,0,0,0,0,1]&(0,2,5,8,11,15,19,23,15,7,12)&-8&117&30\cr
\+(12,0)&[0,0,0,0,0,1,0,0,2]&(0,3,6,9,12,15,18,22,14,6,12)&-6&117&11\cr
\+(12,0)&[0,0,0,0,0,0,2,0,0]&(0,3,6,9,12,15,18,21,14,7,12)&-6&117&7\cr
\+(12,0)&[0,0,0,0,0,1,0,1,0]&(0,3,6,9,12,15,18,22,14,7,12)&-8&118&17\cr
\+(12,0)&[4,0,0,0,0,0,0,0,0]&(0,0,4,8,12,16,20,24,16,8,12)&0&120&0\cr
\+(12,0)&[2,1,0,0,0,0,0,0,0]&(0,1,4,8,12,16,20,24,16,8,12)&-6&121&7\cr
\+(12,0)&[0,0,0,0,1,0,0,0,1]&(0,3,6,9,12,15,19,23,15,7,12)&-10&121&27\cr
\+(12,0)&[0,2,0,0,0,0,0,0,0]&(0,2,4,8,12,16,20,24,16,8,12)&-8&122&8\cr
\+(12,0)&[1,0,1,0,0,0,0,0,0]&(0,2,5,8,12,16,20,24,16,8,12)&-10&123&22\cr
\+(12,0)&[0,0,0,1,0,0,0,0,0]&(0,3,6,9,12,16,20,24,16,8,12)&-12&126&18\cr
\+(0,1)&[1,0,0,0,0,0,0,0,0]&(1,0,0,0,0,0,0,0,0,0,0)&2&1&1\cr
\+(1,1)&[0,0,0,0,0,0,0,1,0]&(1,1,1,1,1,1,1,1,0,0,1)&2&9&1\cr
\+(2,1)&[0,0,0,0,1,0,0,0,0]&(1,1,1,1,1,1,2,3,2,1,2)&2&16&1\cr
\+(3,1)&[0,0,1,0,0,0,0,0,1]&(1,1,1,1,2,3,4,5,3,1,3)&2&25&1\cr
\+(3,1)&[0,1,0,0,0,0,0,0,0]&(1,1,1,2,3,4,5,6,4,2,3)&0&32&1\cr
\+(4,1)&[0,1,0,0,0,0,1,0,0]&(1,1,1,2,3,4,5,6,4,2,4)&2&33&1\cr
\+(4,1)&[1,0,0,0,0,0,0,0,2]&(1,1,2,3,4,5,6,7,4,1,4)&2&38&1\cr
\+(4,1)&[1,0,0,0,0,0,0,1,0]&(1,1,2,3,4,5,6,7,4,2,4)&0&39&1\cr
\+(4,1)&[0,0,0,0,0,0,0,0,1]&(1,2,3,4,5,6,7,8,5,2,4)&-2&47&2\cr
\+(5,1)&[1,0,0,0,0,1,0,0,1]&(1,1,2,3,4,5,6,8,5,2,5)&2&42&1\cr
\+(5,1)&[0,1,0,1,0,0,0,0,0]&(1,1,1,2,3,5,7,9,6,3,5)&2&43&1\cr
\+(5,1)&[1,0,0,0,1,0,0,0,0]&(1,1,2,3,4,5,7,9,6,3,5)&0&46&1\cr
\+(5,1)&[0,0,0,0,0,0,1,0,1]&(1,2,3,4,5,6,7,8,5,2,5)&0&48&2\cr
\+(5,1)&[0,0,0,0,0,1,0,0,0]&(1,2,3,4,5,6,7,9,6,3,5)&-2&51&2\cr
\+(6,1)&[1,0,0,1,0,0,0,1,0]&(1,1,2,3,4,6,8,10,6,3,6)&2&50&1\cr
\+(6,1)&[0,0,0,0,0,1,1,0,0]&(1,2,3,4,5,6,7,9,6,3,6)&2&52&1\cr
\+(6,1)&[0,0,0,0,1,0,0,0,2]&(1,2,3,4,5,6,8,10,6,2,6)&2&53&1\cr
\+(6,1)&[0,2,0,0,0,0,0,0,1]&(1,1,1,3,5,7,9,11,7,3,6)&2&54&1\cr
\+(6,1)&[0,0,0,0,1,0,0,1,0]&(1,2,3,4,5,6,8,10,6,3,6)&0&54&1\cr
\+(6,1)&[1,0,1,0,0,0,0,0,1]&(1,1,2,3,5,7,9,11,7,3,6)&0&55&2\cr
\+(6,1)&[0,0,0,1,0,0,0,0,1]&(1,2,3,4,5,7,9,11,7,3,6)&-2&58&4\cr
\+(6,1)&[1,1,0,0,0,0,0,0,0]&(1,1,2,4,6,8,10,12,8,4,6)&-2&62&2\cr
\+(6,1)&[0,0,1,0,0,0,0,0,0]&(1,2,3,4,6,8,10,12,8,4,6)&-4&64&3\cr
\+(7,1)&[1,0,1,0,0,1,0,0,0]&(1,1,2,3,5,7,9,12,8,4,7)&2&59&1\cr
\+(7,1)&[0,0,0,1,0,0,1,0,1]&(1,2,3,4,5,7,9,11,7,3,7)&2&59&1\cr
\+(7,1)&[1,1,0,0,0,0,0,1,1]&(1,1,2,4,6,8,10,12,7,3,7)&2&61&1\cr
\+(7,1)&[0,0,0,1,0,1,0,0,0]&(1,2,3,4,5,7,9,12,8,4,7)&0&62&2\cr
\+(7,1)&[1,1,0,0,0,0,1,0,0]&(1,1,2,4,6,8,10,12,8,4,7)&0&63&2\cr}
\vbox{\settabs\+(11,0)\quad\quad&[0,0,0,0,0,0,1,0,0]\quad&(0,3,6,9,12,15,18,21,14,7,11)\quad&-10\quad\quad&116\quad\quad&12\cr
\+$(n_c,n_0)$&$p_i$&$\beta$&$\beta^2$&$ht(\beta)$&$\mu$\cr
\hrule
\+(7,1)&[0,0,1,0,0,0,0,1,1]&(1,2,3,4,6,8,10,12,7,3,7)&0&63&2\cr
\+(7,1)&[0,0,1,0,0,0,1,0,0]&(1,2,3,4,6,8,10,12,8,4,7)&-2&65&4\cr
\+(7,1)&[2,0,0,0,0,0,0,0,2]&(1,1,3,5,7,9,11,13,8,3,7)&0&68&1\cr
\+(7,1)&[2,0,0,0,0,0,0,1,0]&(1,1,3,5,7,9,11,13,8,4,7)&-2&69&3\cr
\+(7,1)&[0,1,0,0,0,0,0,0,2]&(1,2,3,5,7,9,11,13,8,3,7)&-2&69&5\cr
\+(7,1)&[0,1,0,0,0,0,0,1,0]&(1,2,3,5,7,9,11,13,8,4,7)&-4&70&7\cr
\+(7,1)&[1,0,0,0,0,0,0,0,1]&(1,2,4,6,8,10,12,14,9,4,7)&-6&77&8\cr
\+(7,1)&[0,0,0,0,0,0,0,0,0]&(1,3,5,7,9,11,13,15,10,5,7)&-8&86&2\cr
\+(8,1)&[0,0,1,0,0,1,0,1,0]&(1,2,3,4,6,8,10,13,8,4,8)&2&67&1\cr
\+(8,1)&[1,1,0,0,1,0,0,0,1]&(1,1,2,4,6,8,11,14,9,4,8)&2&68&1\cr
\+(8,1)&[0,0,0,2,0,0,0,0,1]&(1,2,3,4,5,8,11,14,9,4,8)&2&69&1\cr
\+(8,1)&[2,0,0,0,0,0,1,1,0]&(1,1,3,5,7,9,11,13,8,4,8)&2&70&1\cr
\+(8,1)&[0,1,0,0,0,0,1,0,2]&(1,2,3,5,7,9,11,13,8,3,8)&2&70&1\cr
\+(8,1)&[0,0,1,0,1,0,0,0,1]&(1,2,3,4,6,8,11,14,9,4,8)&0&70&2\cr
\+(8,1)&[0,1,0,0,0,0,1,1,0]&(1,2,3,5,7,9,11,13,8,4,8)&0&71&2\cr
\+(8,1)&[2,0,0,0,0,1,0,0,1]&(1,1,3,5,7,9,11,14,9,4,8)&0&72&2\cr
\+(8,1)&[1,0,2,0,0,0,0,0,0]&(1,1,2,3,6,9,12,15,10,5,8)&2&72&1\cr
\+(8,1)&[1,1,0,1,0,0,0,0,0]&(1,1,2,4,6,9,12,15,10,5,8)&0&73&2\cr
\+(8,1)&[0,1,0,0,0,1,0,0,1]&(1,2,3,5,7,9,11,14,9,4,8)&-2&73&7\cr
\+(8,1)&[0,0,1,1,0,0,0,0,0]&(1,2,3,4,6,9,12,15,10,5,8)&-2&75&4\cr
\+(8,1)&[2,0,0,0,1,0,0,0,0]&(1,1,3,5,7,9,12,15,10,5,8)&-2&76&3\cr
\+(8,1)&[1,0,0,0,0,0,0,1,2]&(1,2,4,6,8,10,12,14,8,3,8)&0&76&2\cr
\+(8,1)&[1,0,0,0,0,0,0,2,0]&(1,2,4,6,8,10,12,14,8,4,8)&-2&77&3\cr
\+(8,1)&[0,1,0,0,1,0,0,0,0]&(1,2,3,5,7,9,12,15,10,5,8)&-4&77&7\cr
\+(8,1)&[1,0,0,0,0,0,1,0,1]&(1,2,4,6,8,10,12,14,9,4,8)&-4&78&11\cr
\+(8,1)&[1,0,0,0,0,1,0,0,0]&(1,2,4,6,8,10,12,15,10,5,8)&-6&81&13\cr
\+(8,1)&[0,0,0,0,0,0,0,0,3]&(1,3,5,7,9,11,13,15,9,3,8)&-2&84&3\cr
\+(8,1)&[0,0,0,0,0,0,0,1,1]&(1,3,5,7,9,11,13,15,9,4,8)&-6&85&10\cr
\+(8,1)&[0,0,0,0,0,0,1,0,0]&(1,3,5,7,9,11,13,15,10,5,8)&-8&87&10\cr
\+(9,1)&[0,1,0,0,1,0,0,1,1]&(1,2,3,5,7,9,12,15,9,4,9)&2&76&1\cr
\+(9,1)&[0,0,1,1,0,0,1,0,0]&(1,2,3,4,6,9,12,15,10,5,9)&2&76&1\cr
\+(9,1)&[2,0,0,0,1,0,1,0,0]&(1,1,3,5,7,9,12,15,10,5,9)&2&77&1\cr
\+(9,1)&[0,1,0,0,0,2,0,0,0]&(1,2,3,5,7,9,11,15,10,5,9)&2&77&1\cr
\+(9,1)&[1,1,1,0,0,0,0,1,0]&(1,1,2,4,7,10,13,16,10,5,9)&2&78&1\cr
\+(9,1)&[0,1,0,0,1,0,1,0,0]&(1,2,3,5,7,9,12,15,10,5,9)&0&78&2\cr
\+(9,1)&[2,0,0,1,0,0,0,0,2]&(1,1,3,5,7,10,13,16,10,4,9)&2&79&1\cr
\+(9,1)&[1,0,0,0,0,0,2,0,1]&(1,2,4,6,8,10,12,14,9,4,9)&2&79&1\cr
\+(9,1)&[2,0,0,1,0,0,0,1,0]&(1,1,3,5,7,10,13,16,10,5,9)&0&80&2\cr
\+(9,1)&[1,0,0,0,0,1,0,1,1]&(1,2,4,6,8,10,12,15,9,4,9)&0&80&3\cr
\+(9,1)&[0,1,0,1,0,0,0,0,2]&(1,2,3,5,7,10,13,16,10,4,9)&0&80&3\cr
\+(9,1)&[0,0,2,0,0,0,0,1,0]&(1,2,3,4,7,10,13,16,10,5,9)&0&80&2\cr
\+(9,1)&[0,1,0,1,0,0,0,1,0]&(1,2,3,5,7,10,13,16,10,5,9)&-2&81&7\cr
\+(9,1)&[1,0,0,0,0,1,1,0,0]&(1,2,4,6,8,10,12,15,10,5,9)&-2&82&6\cr}
\vbox{\settabs\+(11,0)\quad\quad&[0,0,0,0,0,0,1,0,0]\quad&(0,3,6,9,12,15,18,21,14,7,11)\quad&-10\quad\quad&116\quad\quad&12\cr
\+$(n_c,n_0)$&$p_i$&$\beta$&$\beta^2$&$ht(\beta)$&$\mu$\cr
\hrule
\+(9,1)&[1,0,0,0,1,0,0,0,2]&(1,2,4,6,8,10,13,16,10,4,9)&-2&83&6\cr
\+(9,1)&[1,2,0,0,0,0,0,0,1]&(1,1,2,5,8,11,14,17,11,5,9)&0&84&2\cr
\+(9,1)&[1,0,0,0,1,0,0,1,0]&(1,2,4,6,8,10,13,16,10,5,9)&-4&84&13\cr
\+(9,1)&[2,0,1,0,0,0,0,0,1]&(1,1,3,5,8,11,14,17,11,5,9)&-2&85&5\cr
\+(9,1)&[0,0,0,0,0,0,1,0,3]&(1,3,5,7,9,11,13,15,9,3,9)&2&85&1\cr
\+(9,1)&[0,1,1,0,0,0,0,0,1]&(1,2,3,5,8,11,14,17,11,5,9)&-4&86&12\cr
\+(9,1)&[0,0,0,0,0,0,1,1,1]&(1,3,5,7,9,11,13,15,9,4,9)&-2&86&4\cr
\+(9,1)&[1,0,0,1,0,0,0,0,1]&(1,2,4,6,8,11,14,17,11,5,9)&-6&88&24\cr
\+(9,1)&[0,0,0,0,0,1,0,0,2]&(1,3,5,7,9,11,13,16,10,4,9)&-4&88&9\cr
\+(9,1)&[0,0,0,0,0,0,2,0,0]&(1,3,5,7,9,11,13,15,10,5,9)&-4&88&4\cr
\+(9,1)&[0,0,0,0,0,1,0,1,0]&(1,3,5,7,9,11,13,16,10,5,9)&-6&89&16\cr
\+(9,1)&[2,1,0,0,0,0,0,0,0]&(1,1,3,6,9,12,15,18,12,6,9)&-4&92&4\cr
\+(9,1)&[0,0,0,0,1,0,0,0,1]&(1,3,5,7,9,11,14,17,11,5,9)&-8&92&24\cr
\+(9,1)&[0,2,0,0,0,0,0,0,0]&(1,2,3,6,9,12,15,18,12,6,9)&-6&93&6\cr
\+(9,1)&[1,0,1,0,0,0,0,0,0]&(1,2,4,6,9,12,15,18,12,6,9)&-8&94&20\cr
\+(9,1)&[0,0,0,1,0,0,0,0,0]&(1,3,5,7,9,12,15,18,12,6,9)&-10&97&20\cr
\+(10,1)&[0,1,0,1,0,1,0,0,1]&(1,2,3,5,7,10,13,17,11,5,10)&2&84&1\cr
\+(10,1)&[1,0,0,0,1,0,1,1,0]&(1,2,4,6,8,10,13,16,10,5,10)&2&85&1\cr
\+(10,1)&[2,0,1,0,0,0,1,0,1]&(1,1,3,5,8,11,14,17,11,5,10)&2&86&1\cr
\+(10,1)&[0,1,1,0,0,0,0,2,0]&(1,2,3,5,8,11,14,17,10,5,10)&2&86&1\cr
\+(10,1)&[2,0,0,1,1,0,0,0,0]&(1,1,3,5,7,10,14,18,12,6,10)&2&87&1\cr
\+(10,1)&[1,0,0,1,0,0,0,1,2]&(1,2,4,6,8,11,14,17,10,4,10)&2&87&1\cr
\+(10,1)&[1,0,0,0,1,1,0,0,1]&(1,2,4,6,8,10,13,17,11,5,10)&0&87&3\cr
\+(10,1)&[0,1,1,0,0,0,1,0,1]&(1,2,3,5,8,11,14,17,11,5,10)&0&87&3\cr
\+(10,1)&[0,0,2,0,1,0,0,0,0]&(1,2,3,4,7,10,14,18,12,6,10)&2&87&1\cr
\+(10,1)&[1,2,0,0,0,1,0,0,0]&(1,1,2,5,8,11,14,18,12,6,10)&2&88&1\cr
\+(10,1)&[1,0,0,1,0,0,0,2,0]&(1,2,4,6,8,11,14,17,10,5,10)&0&88&2\cr
\+(10,1)&[0,1,0,1,1,0,0,0,0]&(1,2,3,5,7,10,14,18,12,6,10)&0&88&2\cr
\+(10,1)&[2,0,1,0,0,1,0,0,0]&(1,1,3,5,8,11,14,18,12,6,10)&0&89&2\cr
\+(10,1)&[1,0,0,1,0,0,1,0,1]&(1,2,4,6,8,11,14,17,11,5,10)&-2&89&10\cr
\+(10,1)&[0,0,0,0,0,1,1,0,2]&(1,3,5,7,9,11,13,16,10,4,10)&2&89&1\cr
\+(10,1)&[0,1,1,0,0,1,0,0,0]&(1,2,3,5,8,11,14,18,12,6,10)&-2&90&7\cr
\+(10,1)&[0,0,0,0,0,1,1,1,0]&(1,3,5,7,9,11,13,16,10,5,10)&0&90&2\cr
\+(10,1)&[2,1,0,0,0,0,0,1,1]&(1,1,3,6,9,12,15,18,11,5,10)&0&91&3\cr
\+(10,1)&[1,0,0,0,2,0,0,0,0]&(1,2,4,6,8,10,14,18,12,6,10)&-2&91&3\cr
\+(10,1)&[0,2,0,0,0,0,0,0,3]&(1,2,3,6,9,12,15,18,11,4,10)&2&91&1\cr
\+(10,1)&[0,0,0,0,1,0,0,1,2]&(1,3,5,7,9,11,14,17,10,4,10)&0&91&2\cr
\+(10,1)&[1,0,1,0,0,0,0,0,3]&(1,2,4,6,9,12,15,18,11,4,10)&0&92&2\cr
\+(10,1)&[1,0,0,1,0,1,0,0,0]&(1,2,4,6,8,11,14,18,12,6,10)&-4&92&14\cr}
\vbox{\settabs\+(11,0)\quad\quad&[0,0,0,0,0,0,1,0,0]\quad&(0,3,6,9,12,15,18,21,14,7,11)\quad&-10\quad\quad&116\quad\quad&12\cr
\+$(n_c,n_0)$&$p_i$&$\beta$&$\beta^2$&$ht(\beta)$&$\mu$\cr
\hrule
\+(10,1)&[0,2,0,0,0,0,0,1,1]&(1,2,3,6,9,12,15,18,11,5,10)&-2&92&7\cr
\+(10,1)&[0,0,0,0,1,0,0,2,0]&(1,3,5,7,9,11,14,17,10,5,10)&-2&92&4\cr
\+(10,1)&[0,0,0,0,0,2,0,0,1]&(1,3,5,7,9,11,13,17,11,5,10)&-2&92&5\cr
\+(10,1)&[2,1,0,0,0,0,1,0,0]&(1,1,3,6,9,12,15,18,12,6,10)&-2&93&5\cr
\+(10,1)&[1,0,1,0,0,0,0,1,1]&(1,2,4,6,9,12,15,18,11,5,10)&-4&93&19\cr
\+(10,1)&[0,0,0,0,1,0,1,0,1]&(1,3,5,7,9,11,14,17,11,5,10)&-4&93&13\cr
\+(10,1)&[0,2,0,0,0,0,1,0,0]&(1,2,3,6,9,12,15,18,12,6,10)&-4&94&9\cr
\+(10,1)&[1,0,1,0,0,0,1,0,0]&(1,2,4,6,9,12,15,18,12,6,10)&-6&95&30\cr
\+(10,1)&[0,0,0,1,0,0,0,0,3]&(1,3,5,7,9,12,15,18,11,4,10)&-2&95&5\cr
\+(10,1)&[0,0,0,1,0,0,0,1,1]&(1,3,5,7,9,12,15,18,11,5,10)&-6&96&27\cr
\+(10,1)&[0,0,0,0,1,1,0,0,0]&(1,3,5,7,9,11,14,18,12,6,10)&-6&96&15\cr
\+(10,1)&[3,0,0,0,0,0,0,0,2]&(1,1,4,7,10,13,16,19,12,5,10)&-2&98&3\cr
\+(10,1)&[0,0,0,1,0,0,1,0,0]&(1,3,5,7,9,12,15,18,12,6,10)&-8&98&34\cr
\+(10,1)&[3,0,0,0,0,0,0,1,0]&(1,1,4,7,10,13,16,19,12,6,10)&-4&99&5\cr
\+(10,1)&[1,1,0,0,0,0,0,0,2]&(1,2,4,7,10,13,16,19,12,5,10)&-6&99&24\cr
\+(10,1)&[1,1,0,0,0,0,0,1,0]&(1,2,4,7,10,13,16,19,12,6,10)&-8&100&42\cr
\+(10,1)&[0,0,1,0,0,0,0,0,2]&(1,3,5,7,10,13,16,19,12,5,10)&-8&101&31\cr
\+(10,1)&[0,0,1,0,0,0,0,1,0]&(1,3,5,7,10,13,16,19,12,6,10)&-10&102&57\cr
\+(10,1)&[2,0,0,0,0,0,0,0,1]&(1,2,5,8,11,14,17,20,13,6,10)&-10&107&32\cr
\+(10,1)&[0,1,0,0,0,0,0,0,1]&(1,3,5,8,11,14,17,20,13,6,10)&-12&108&58\cr
\+(10,1)&[1,0,0,0,0,0,0,0,0]&(1,3,6,9,12,15,18,21,14,7,10)&-14&116&27\cr}

\medskip
{\bf References}
\medskip

\item{[1]} S. Ferrara, J. Scherk and B. Zumino, {\sl Algebraic
properties of extended supersymmetry}, Nucl. Phys. {\bf B 121} (1977)
393; E. Cremmer, J. Scherk and S. Ferrara, {\sl $SU(4)$ invariant
supergravity theory}, Phys. Lett. {\bf B 74} (1978) 61
\item{[2]} E. Cremmer and B. Julia, {\sl The $N=8$ supergravity
theory. I. The Lagrangian}, Phys. Lett. {\bf B 80} (1978) 48
\item{[3]} P.~C. West, {\sl Hidden superconformal symmetry in {M}
    theory },  JHEP {\bf 08} (2000) 007, {\tt hep-th/0005270}
\item{[4]} P. West, {\sl $E_{11}$ and M Theory}, Class. Quant.
Grav. {\bf 18 } (2001) 4443, {\tt hep-th/0104081}
\item{[5]} I. Schnakenburg and P. West, {\sl Kac-Moody Symmetries of
IIB supergravity}, Phys. Lett. {\bf B 517} (2001) 137-145, {\tt
hep-th/0107181} 
\item{[6]} T. Damour, M. Henneaux and H. Nicolai, {\sl $E_{10}$ and a
``small tension expansion'' of M-theory}, Phys. Rev. Lett. {\bf 89}
(2002) 221601, {\tt hep-th/0207267}
\item{[7]} T. Damour, M. Henneuax and H. Nicolai, {\sl Cosmological
billiards}, Class. Quant. Grav. {\bf 20} (2003) 020, {\tt
hep-th/0212256}
\item{[8]} E. Cremmer, B. Julia, H. Lu and C. N. Pope, {\sl Dualisation
of Dualities I}, Nucl. Phys. {\bf B 523} (1998) 73-144, {\tt
hep-th/9710119} 
\item{[9]} P. West, {\sl $E_{11}$, SL(32) and Central Charges},
Phys. Lett. {\bf B 575} (2003) 333-342, {\tt hep-th/0307098}
\item{[10]} F. Englert and L. Houart, {\sl ${\cal G}^{+++}$ invariant
formulation of gravity and M-theories: Exact BPS solutions}, {\tt
hep-th/0311255} 
\item{[11]}  H. Nicolai and T. Fischbacher, {\sl Low Level
representations of $E_{10}$ and $E_{11}$}, Contribution to the
Proceedings of
the Ramanujan International Symposium on Kac-Moody Algebras and
Applications, ISKMAA-2002, Jan. 28--31, Chennai, India, {\tt
hep-th/0301017}
\item{[12]}  N. Lambert and  P. West, {\sl Coset symmetries in
dimensionally reduced bosonic string theory}, Nucl. Phys. {\bf B 615}  
(2001) 117, {\tt hep-th/0107209}
\item{[13]}  M. R. Gaberdiel, D. I. Olive and P. West, {\sl A
class of Lorentzian Kac-Moody algebras}, Nucl. Phys. {\bf B 645}
(2002) 403-437, {\tt hep-th/0205068}
\item{[14]}  F. Englert, L. Houart, A. Taormina and P. West,
{\sl The Symmetry of M-theories}, JHEP 0309 (2003) 020, {\tt hep-th/0304206}
\item{[15]} F. Englert, L. Houart and P. West, {\sl
  Intersection rules, dynamics and symmetries}, JHEP 0308 (2003) 025, 
{\tt hep-th/0307024}
\item{[16]} A. Kleinschmidt, I. Schnakenburg and P. West, {\sl
Very-extended Kac-Moody algebras and their interpretation at low
levels}, {\tt hep-th/0309198}
\item{[17]} P. West, {\sl Very-extended $E_8$ and $A_8$ at low
levels}, Class. Quant. Grav. {\bf 20} (2003) 2393, {\tt hep-th/0307024}
\item{[18]} V. Kac, {\sl Infinite dimensional algebras}, 3rd edition,
Cambridge University Press (1990)
\item{[19]} A. Kleinschmidt, {\sl $E_{11}$ as $E_{10}$ representation
at low levels}, Nucl. Phys. {\bf B 677} Vol. 3 (2003) 553-586, {\tt
hep-th/0304246} 
\item{[20]} K. Stelle, {\sl BPS branes in supergravity}, Proceedings
of the ICTP summer schools 1996 and 1997, {\tt hep-th/9803116}
\item{[21]} J.A. de Azcarraga, J.P. Gauntlett, J.M. Izquierdo and 
P.K. Townsend, {\sl  Topological Extensions of the supersymmetry
algebra for extended objects}, Phys. Rev. Lett. {\bf 63} (1989) 2443

\end